\documentclass[twoside,12pt,fleqn]{article}
\usepackage{epsfig}

\usepackage{amsmath,amssymb,bm}

\newcommand{\be}{\begin{equation}}
\newcommand{\ee}{\end{equation}}
\newcommand{\bea}{\begin{eqnarray}}
\newcommand{\eea}{\end{eqnarray}}

\newcommand{\PDG}{Nakamura:2010zzi}
\newcommand{\Slash}[1]{\ooalign{\hfil/\hfil\crcr$#1$}}
\newcommand{\bra}[1]{\langle \, #1 \, |}
\newcommand{\ket}[1]{| \, #1 \, \rangle}
\newcommand{\re}{\text{Re }}
\newcommand{\im}{\text{Im }}
\newcommand{\Tr}{\text{Tr}}
\newcommand{\tr}{\text{tr}}
\newcommand{\chiral}{SU(3)$_R\, \times \, $SU(3)$_L$\,}
\newcommand{\iso}[6]{\mbox{$\left( \begin{array}{cc||c} {#1} & {#2} &
{#3} \\ {#4} & {#5} & {#6} \end{array} \right)$}}
\newcommand{\largeN}[1]{\text{``$#1$''}}

\begin{document}

\title{ \vspace{1cm} The nature of the $\Lambda(1405)$ resonance
in chiral dynamics}
\author{Tetsuo\ Hyodo,$^{1}$ Daisuke\ Jido,$^2$\\
\\
$^1$Department of Physics, Tokyo Institute of Technology, \\
Tokyo 152-8551, Japan\\
$^2$Yukawa Institute for Theoretical Physics, Kyoto University, \\
Sakyo, Kyoto 606-8502, Japan }
\maketitle
\begin{abstract} 

The $\Lambda(1405)$ baryon resonance plays an outstanding role in various aspects in hadron and nuclear physics. It has been considered that the $\Lambda(1405)$ resonance is generated by the attractive interaction of the antikaon and the nucleon as a quasi-bound state below its threshold decaying into the $\pi\Sigma$ channel. Thus, the structure of $\Lambda(1405)$ is closely related to the $\bar{K}N$ interaction which is the fundamental ingredient to study few-body systems with antikaon. In this paper, after reviewing the basic properties of the $\Lambda(1405)$ resonance, we introduce the dynamical coupled-channel model which respects chiral symmetry of QCD and the unitarity of the scattering amplitude. We show that the structure of the $\Lambda(1405)$ resonance is dominated by the meson-baryon molecular component and is described as a superposition of two independent states. The meson-baryon nature of $\Lambda(1405)$ leads to various hadronic molecular states in few-body systems with strangeness which are hadron composite systems driven by the hadronic interactions. We summarize the recent progress in the investigation of the $\Lambda(1405)$ structure and future perspective of the physics of the $\Lambda(1405)$ resonance.

\end{abstract}

\eject
\tableofcontents

\section{Introduction}

The $\Lambda(1405)$ resonance is a negative parity baryon resonance with spin 1/2, isospin $I=0$ and strangeness $S=-1$. The resonance is located slightly below the $\bar{K}N$ threshold and decays into the $\pi\Sigma$ channel through the strong interaction. The existence of $\Lambda(1405)$ was theoretically predicted in 1959 by Dalitz and Tuan~\cite{Dalitz:1959dn,Dalitz:1960du}, based on the analysis of the experimental data of the $\bar{K}N$ scattering length. It is shown that the unitarity in coupled-channel $\bar{K}N$-$\pi\Sigma$ system leads to a resonance pole in the $\pi\Sigma$ amplitude. An experimental evidence of this resonance was reported as early as 1961 in the invariant mass spectrum of the $\pi\Sigma$ channel in the $K^-p\to \pi\pi\pi\Sigma$ reaction at 1.15 GeV~\cite{Alston:1961zz}. After a half century, $\Lambda(1405)$ has been well established experimentally and is currently listed as a four-star resonance in the table of the Particle Data Group (PDG)~\cite{\PDG}. In recent years, the structure of $\Lambda(1405)$ has been found to be important in various aspects in the strangeness sector of nonperturbative QCD. At the same time, the experimental information on $\Lambda(1405)$ is being rapidly improved by new data, such as the $\pi\Sigma$ mass spectra in several reactions and the precise measurement of the energy level of the kaonic hydrogen. Thus, it is an important and urgent issue to understand the nature of the $\Lambda(1405)$ resonance.

There is a long-standing discussion on the interpretation of the $\Lambda(1405)$ resonance. It is known to be difficult to describe  $\Lambda(1405)$ as an ordinary three-quark state in simple constituent quark models~\cite{Isgur:1978xj}, because $\Lambda(1405)$ has a lighter mass than the nucleon counterpart, the $N(1535)$ resonance. Moreover, the mass difference from the $\Lambda(1520)$ resonance with $J^P=3/2^-$, which is supposed to be the spin-orbit partner of $\Lambda(1405)$, is too large in comparison with the splitting in the nucleon sector. According to these difficulties of the simple three-quark picture, the meson-baryon quasi-bound picture of $\Lambda(1405)$ attracts much attention. In fact, the $\Lambda(1405)$ resonance can be naturally described as a quasi-bound $\bar{K}N$ state embedded in the $\pi\Sigma$ continuum in coupled-channel meson-baryon scattering models, for instance, by the phenomenological vector-meson exchange potential with flavor SU(3) symmetry~\cite{Dalitz:1967fp}. Strictly speaking, the meson-baryon picture of $\Lambda(1405)$ bears a complementary relationship to the three-quark picture. In reality, the physical $\Lambda(1405)$ resonance should be a mixture of both (and with even more exotic structures), since all possible components with the same quantum number can mix with each other through the strong interaction. It is however meaningful to ask the dominant component in $\Lambda(1405)$, for the understanding of its physical origin and the implication to the $\bar{K}N$ dynamics. 

In this paper, we discuss the nature of the $\Lambda(1405)$ resonance based on the coupled-channel framework with chiral symmetry (chiral unitary approach) which combines the low energy interaction governed by chiral symmetry and the unitarity condition for the coupled-channel scattering amplitude~\cite{Kaiser:1995eg,Oset:1998it,Oller:2000fj,Lutz:2001yb}. This model successfully reproduces the observables in meson-baryon scattering and resonances are dynamically generated in the non-perturbative resummation of the interaction, along the same line with the phenomenological coupled-channel model~\cite{Dalitz:1967fp}. Not only the successful description of data, chiral unitary approach serves as a useful tool to investigate the structure of the resonances. The intensive studies of the $\Lambda(1405)$ resonance reveal the peculiar pole structure in the complex energy plane~\cite{Jido:2003cb}, the meson-baryon nature of the resonance~\cite{Hyodo:2008xr}, the quark structure of the resonance~\cite{Hyodo:2007np,Roca:2008kr}, and the spatial size of the resonance state~\cite{Sekihara:2008qk,Sekihara:2010uz}. 

In addition, chiral unitary approach provides a unique baseline for various applications in low energy QCD. Turning to the strangeness nuclear physics, one of the central issues is the study of the possible bound state of the antikaon in nuclei, the $\bar{K}$ nuclei. Based on the $\bar{K}N$ quasi-bound picture for $\Lambda(1405)$, the $\bar{K}$ nuclei was suggested to exhibit various interesting phenomena~\cite{PL7.288,Akaishi:2002bg}. The interaction of the antikaon and the nucleon is a basic building block of the study of such multi-hadron systems with antikaons. Traditionally, the $\bar{K}$ nuclei has been studied by phenomenological approaches, where the connection with the underlying theory of QCD is not very much clear. At this point it is instructive to recall that the antikaon can be regarded as a flavor partner of the pion, which is the Nambu-Goldstone (NG) boson associated with the spontaneous breakdown of chiral \chiral symmetry in QCD. Chiral unitary approach can shed light on this aspect of the $\bar{K}N$ interaction~\cite{Hyodo:2007jq}, treating the non-perturbative dynamics of the strong attraction in the $\bar{K}N$ channel and its consequence for $\Lambda(1405)$. Such an approach is indispensable to discuss, for instance, the effect of the partial restoration of chiral symmetry on the dynamics of the $\bar{K}$ in nuclear medium. 

The $\Lambda(1405)$ resonance in the meson-baryon picture can be a good candidate of the hadronic molecular state, in which the hadrons are loosely bound by the inter-hadron interaction~\cite{Jido:2008kp}. This is because the antikaon is moderately heavy, and strongly interacts with the nucleon. The binding energy of $\Lambda(1405)$, when regarded as a $\bar{K}N$ bound state, is much smaller than the mass scale of the hadrons, so the antikaon and the nucleon will behave as constituents by keeping their identities in $\Lambda(1405)$. Compared with the quark excitation inside hadrons, the hadronic molecular structure is expected to have a larger spatial size. Physics of the hadronic molecular state is also related to the exotic hadrons and multi-quark states recently observed in heavy quark sector~\cite{Swanson:2006st,Nielsen:2009uh}.

This paper is organized as follows. In the next section, we overview the current status of experimental investigations and theoretical studies of $\Lambda(1405)$. Section~\ref{sec:chiralunitary} introduces the chiral unitary approach in detail from the basic framework. Several recent findings on the nature of the structure of $\Lambda(1405)$ in chiral unitary approach are summarized in Section~\ref{sec:structureL1405}. We then discuss the properties of $\Lambda(1405)$ in various environments in Section~\ref{sec:L1405many} where hadronic molecular states emerge in the few-body systems. The last section is devoted to conclusion of this paper and the future perspective on the studies of $\Lambda(1405)$.

\section{The $\Lambda(1405)$ resonance}

Here we first go through the experimental investigations of the $\pi\Sigma$ spectrum where the resonance peak of $\Lambda(1405)$ is observed. The information on the low energy $K^-p$ scattering data is also presented. We then summarize typical theoretical studies on $\Lambda(1405)$ in the three-quark picture and in the meson-baryon picture.

\subsection{Mass spectrum of the $\pi\Sigma$ channel in bubble chamber experiments}

The $\Lambda(1405)$ resonance exclusively decays into the $\pi\Sigma(I=0)$ channel \textit{via} strong interaction. Since it is very hard to perform the scattering experiment in the $\pi\Sigma$ channel directly, the properties of the resonance have been extracted by analyzing the invariant mass distribution of the $\pi\Sigma$ final state in production experiments. Initiated by Ref.~\cite{Alston:1961zz}, several bubble chamber experiments with hadron induced reactions found the evidence for $\Lambda(1405)$ in the energy region $1382$-$1410$ MeV (a list of early references can be found in PDG~\cite{\PDG}). Among others, it is worth mentioning that Braun \textit{et al.} studied the $\pi^+\Sigma^-$ spectrum in $K^-d\to \pi^+\Sigma^- n$ reaction~\cite{Braun:1977wd} where the resonance energy was found at 1420 MeV. The $\pi^{\pm}\Sigma^{\mp}$ spectrum in the pion induced reaction ($\pi^-p\to K^+\pi\Sigma$) at 1.69 GeV was investigated in Ref.~\cite{Thomas:1973uh}. In this reaction, the mass of the resonance was found to be consistent with 1405 MeV.

The highest-statistics spectrum in bubble chamber experiments is given in Ref.~\cite{Hemingway:1985pz} through the $K^-p\to\pi^-\pi^+(\pi\Sigma)$ reaction at 4.2 GeV in which the $\pi^+(\pi\Sigma)$ final state forms the $\Sigma^+(1660)$ resonance with the sequential decays of $\Sigma^+(1660)\to\pi^+\Lambda(1405)$ and $\Lambda(1405)\to \pi\Sigma$. To enhance the $\Sigma(1660)$ production, events with low momentum transfer $t^{\prime}(p\to \pi\pi\Sigma)<1.0$ GeV were selected, with the additional constraint on the invariant mass of the $\pi\pi\Sigma$ system to be $1.60\leq M(\pi\pi\Sigma)\leq 1.72$ GeV. Two different modes were analyzed: (1) $K^-p\to \pi^-\pi^+(\pi^-\Sigma^+)$ and (2) $K^-p\to \pi^-\pi^+(\pi^+\Sigma^-)$ where the $\pi\Sigma$ pairs in parenthesis were combined to form $\Lambda(1405)$. The $\pi^-\Sigma^+$ spectrum in the process (1) shows a clear peak structure of $\Lambda(1405)$ together with a small amount of $\Lambda(1520)$. The $\pi^+\Sigma^-$ spectrum in the mode (2) was also shown in that paper~\cite{Hemingway:1985pz}. This spectrum is, however, largely contaminated by the nonresonant background contribution, presumably because of the final state interaction between the $\Sigma^-$ and the primary $\pi^+$ in the $\Sigma(1660)$ decay. 

Thus, the cleaner $\pi^-\Sigma^+$ spectrum is frequently shown as a representative of the $\Lambda(1405)$ spectrum and used for the input of theoretical models. In fact, Dalitz and Deloff analyzed this $\pi^-\Sigma^+$ spectrum to extract the mass and width of $\Lambda(1405)$ using several fitting schemes~\cite{Dalitz:1991sq}. This work is the only source of the ``central values" of the mass $M$ and the width $\Gamma$ shown in the current PDG~\cite{\PDG}:
\begin{align}
    M
    &=1406.5\pm 4.0 \text{ MeV},\quad
    \Gamma
    = 50\pm 2 \text{ MeV}.
    \nonumber
\end{align}
It should be nevertheless emphasized that a single charged state of the $\pi\Sigma$ channel may \textit{not} be dominated by the $I=0$ resonant component in view of the strong isospin interference as reported in recent photoproduction experiments discussed below. In addition, since the energy of the $\pi\pi\Sigma$ system is restricted in the range $1.60\leq M(\pi\pi\Sigma)\leq 1.72$ GeV, the higher tail of the $\pi^-\Sigma^+$ spectrum is influenced by the phase space suppression. This effect should be taken into account when the spectral function of the theoretical model is compared with the experimental mass distribution in Ref.~\cite{Hemingway:1985pz}.

\subsection{Mass spectrum of the $\pi\Sigma$ channel in recent experiments}

On top of these classical results, many new $\pi\Sigma$ spectra in recent experiments are becoming available. The first photoproduction experiment was performed by LEPS collaboration at SPring-8 using $\gamma p\to K^+\pi\Sigma$ reaction with the photon energy 1.5 to 2.4 GeV~\cite{Ahn:2003mv}. By excluding the $K^*$ production in the $K^+\pi$ pair, the charged $\pi\Sigma$ invariant mass spectrum was reconstructed, showing a clear peak structure of $\Lambda(1405)$. A remarkable fact is that the ``lineshape" of the $\Lambda(1405)$ resonance is different in $\pi^-\Sigma^+$ and $\pi^+\Sigma^-$ channels, as predicted by the chiral unitary model in Ref.~\cite{Nacher:1998mi}. Neglecting the small $I=2$ component, the $\pi\Sigma$ spectra can be decomposed as
\begin{align}
    \frac{d\sigma(\pi^{+}\Sigma^{-})}{dM_I}
    &\propto \frac{1}{3}|T^{(0)}|^2+\frac{1}{2}|T^{(1)}|^2
    +\frac{2}{\sqrt{6}}\re (T^{(0)}T^{(1)*}) ,
    \label{eq:pipSigmam} \\
    \frac{d\sigma(\pi^{-}\Sigma^{+})}{dM_I}
    &\propto \frac{1}{3}|T^{(0)}|^2+\frac{1}{2}|T^{(1)}|^2
    -\frac{2}{\sqrt{6}}\re (T^{(0)}T^{(1)*}) ,
    \label{eq:pimSigmap} \\
    \frac{d\sigma(\pi^{0}\Sigma^{0})}{dM_I}
    &\propto \frac{1}{3}|T^{(0)}|^2  ,
    \label{eq:pi0Sigma0}
\end{align}
where $T^{(I)}$ represents the $\pi\Sigma$ amplitude with isospin $I$ and $M_{I}$ stands for the invariant mass of the $\pi\Sigma$ pair. It is obvious that the isospin interference term $\re (T^{(0)}T^{(1)*})$ makes the difference of the charged $\pi^{\pm}\Sigma^{\mp}$ spectra.\footnote{The difference of the $\Lambda(1405)$ lineshapes in $\pi^+\Sigma^-$ and $\pi^-\Sigma^+$ spectra primarily comes from the isospin interference effect. In a specific experiment, one should also take into account the final state interaction with the unpaired particles as in Ref.~\cite{Hemingway:1985pz}. It is \textit{not} directly related to the two-pole structure discussed in Section~\ref{subsec:doublepole}.} If the resonant $I=0$ amplitude were much larger than the other isospin components, all the $\pi\Sigma$ spectra would be identical in the resonance energy region. This is not the case. LEPS collaboration further studied the same reaction in Ref.~\cite{Niiyama:2008rt} where the total cross section of the $\Lambda(1405)$ photoproduction was measured for the first time. The production ratios of $\Lambda(1405)$ to $\Sigma(1385)$ were obtained at two regions of the photon energies. At lower energy ($1.5<E_{\gamma}<2.0$ GeV), $\Lambda(1405)$ is produced about one half of $\Sigma(1385)$, while at higher energy ($2.0<E_{\gamma}<2.4$ GeV), the $\Lambda(1405)$ production is much suppressed. The absolute value of the differential cross section $d\sigma/d(\cos\theta)$ was obtained as 0.43 $\mu$b (0.072 $\mu$b)  for the photon energy $1.5<E_{\gamma}<2.0$ GeV ($2.0<E_{\gamma}<2.4$ GeV). The difference in the charged $\pi\Sigma$ spectra was again observed, while the shape of the peak was not consistent with the previous measurement~\cite{Ahn:2003mv}, presumably because of the different kinematical region of the final state pion. Detailed theoretical analysis for the angular dependence of the $\gamma p\to K^+\pi\Sigma$ reaction will make the situation clear.

The neutral $\pi^0\Sigma^0$ spectrum is observed by Crystal Ball Collaboration in the $K^-p\to \pi^0\pi^0\Sigma^0$ reaction at $p_{K^-}=514$-750 MeV~\cite{Prakhov:2004an}. As seen in Eq.~\eqref{eq:pi0Sigma0}, the $\pi^0\Sigma^0$ spectrum is free from the $\Sigma(1385)$ resonance with $I=1$ and the isospin interference term, so it is an ideal channel to study the $\Lambda(1405)$ spectrum. Although it was not explicitly mentioned in Ref.~\cite{Prakhov:2004an}, the peak of the spectrum in the $K^-p\to \pi^0\pi^0\Sigma^0$ reaction locates at 1420 MeV~\cite{Magas:2005vu}. The $pp$ collision experiment was studied at COSY-J\"ulich in the reaction $pp\to pK^+\pi^0\Sigma^0$ with 3.65 GeV proton beam~\cite{Zychor:2007gf}. The $\pi^0\Sigma^0$ spectrum was constructed from the missing mass of $pK^+$ by selecting the events with a $\Lambda$ in the final state and the constraint on the missing mass $MM(pK^+\Lambda)>190$ MeV (charged $\Sigma^{\pm}$ baryons do not decay into the final states with $\Lambda$). The peak position of $\Lambda(1405)$ was found at 1405 MeV. The total cross section of the $pp\to pK^+\Lambda(1405)$ reaction was obtained as 4.5 $\mu$b.

There are also preliminary reports on some ongoing experiments. The result of the photoproduction $\gamma p\to K^+(\pi\Sigma)$ by CLAS collaboration at Jefferson Laboratory was reported in Ref.~\cite{Moriya:2009mx} where all three $\pi\Sigma$ charge states were measured in a single experiment for the first time. The differential cross section $d\sigma/dt$ was also studied. It is remarkable that the interference pattern of the $\pi\Sigma$ spectra by CLAS is different from that in the first data by LEPS~\cite{Ahn:2003mv}; the peak of the $\pi^{+}\Sigma^{-}$ channel is higher (lower) than the peak of the $\pi^{-}\Sigma^{+}$ spectrum in LEPS (CLAS) data. Since the theoretical prediction~\cite{Nacher:1998mi} reproduces the LEPS data~\cite{Ahn:2003mv}, it contradicts with the CLAS new data. We should however note that the angular dependence of the detector acceptance is different from LEPS to CLAS, and the theoretical prediction integrates the angles of the final states. The $p$-wave meson-baryon amplitude then contributes differently, which may cause the difference of the spectra. To consistently understand all the photoproduction results~\cite{Ahn:2003mv,Niiyama:2008rt,Moriya:2009mx}, more refined reaction model should be constructed. The measurement of the $\pi\Sigma$ spectrum in the $pp$ collision at 3.5 GeV was studied by HADES collaboration at GSI~\cite{Siebenson:2010hh}. The finalized results from these experiments, together with the planned experiments by E31 at J-PARC~\cite{JPARCE31} and IKON/KLOE at DA$\Phi$NE~\cite{FilippiTrento} will further shed light on the structure of $\Lambda(1405)$.

\subsection{Low energy $K^-p$ scattering data}

Since $\Lambda(1405)$ is located just below the $\bar{K}N$ threshold, the low energy $K^-p$ scattering data is closely related to the properties of $\Lambda(1405)$. The total cross sections of the $K^-p$ scattering into various meson-baryon channels measured in bubble chamber experiments were shown, for instance, in Refs.~\cite{Abrams:1965zz,Sakitt:1965kh,PRL14.29,kim66,PL16.89,Mast:1975pv,Bangerter:1980px,Ciborowski:1982et,Evans:1983hz}. These data can be used to constrain the global behavior of the meson-baryon amplitude above the $K^-p$ threshold, while some of the data points contradict with each other and the cross sections into the neutral final states ($\pi^0\Sigma^0$, $\pi^0\Lambda$) are relatively poor. The planned new experiment of the low energy $K^-p$ scattering at DA$\Phi$NE~\cite{FilippiTrento} will improve the precision of the total cross section data.

An important quantitative constraint can be given by the threshold branching ratios. They were accurately determined from the $K^-$ capture by hydrogen as~\cite{Tovee:1971ga,Nowak:1978au}
\begin{align}
	\gamma=&\frac{\Gamma(K^{-}p\to\pi^{+}\Sigma^{-})}
	{\Gamma(K^{-}p\to\pi^{-}\Sigma^{+})}
	=  2.36\pm0.04 ,
	\nonumber\\
	R_c=&\frac{\Gamma(K^{-}p\to\text{charged particles})}
	{\Gamma(K^{-}p\to\text{all})}
	= 0.664\pm 0.011 ,
	\label{eq:branchingratio}
	 \\
	R_n=&\frac{\Gamma(K^{-}p\to\pi^{0}\Lambda)}
	{\Gamma(K^{-}p\to\text{neutral particles})}
	= 0.189\pm 0.015 ,
	\nonumber
\end{align}
at the $K^-p$ threshold. The ratio $\gamma$ indicates that the double-charge exchange process $K^-p\to \pi^+\Sigma^-$ is larger than the ordinary process $K^-p\to \pi^-\Sigma^+$.

The $\bar{K}N$ scattering lengths $a^{(I)}(I=0,1)$ are also important threshold quantities. The extrapolation of the low energy $\bar{K}N$ scattering, with the dispersion relations and $M$-matrix method, leads to the scattering lengths $a^{(0)}=-1.70+i0.65$ fm and $a^{(1)}=0.37+i0.60$ fm~\cite{Martin:1980qe}\footnote{In this paper, we adopt the convention where the positive (negative) scattering length corresponds to the attractive (repulsive) interaction. See also Appendix.}. 

The $K^-p$ scattering length, which is the combination $a_{K^-p}=(a^{(0)}+a^{(1)})/2$ in the isospin limit, can be extracted from the measurement of the mass shift by strong interaction $\Delta E$ and the width $\Gamma$ of the $1s$ level of the kaonic hydrogen. In general, the scattering length due to the strong interaction is related to the energy level of the mesic atom \textit{via} Deser-Trueman formula~\cite{Deser:1954vq,Trueman:1961zz}. Systematic improvement of the formula has been achieved by the effective non-relativistic field theory for the kaonic hydrogen~\cite{Meissner:2004jr} and for the kaonic deuterium~\cite{Meissner:2006gx}. There, the isospin breaking correction turns out to be important.

The old experiments of the kaonic hydrogen found attractive energy shifts, which were in contradiction with the analysis of the low energy $\bar{K}N$ scattering~\cite{Martin:1980qe}. The situation became clear by the precise measurement of the kaonic X ray at KEK, which leads to the repulsive shift $\Delta E=-323\pm 63\pm 11$ eV and the width $\Gamma=407\pm 208\pm 100$ eV~\cite{Iwasaki:1997wf}. This result agrees with the negative scattering length and the long-standing puzzle has been solved. However, the quantitative determination is not settled yet, because new data from the DEAR collaboration at DA$\Phi$NE~\cite{Beer:2005qi} deviates from the KEK result. In DEAR experiment, the shift and width were obtained as $\Delta E=-193\pm 37\pm 6$ eV and $\Gamma=249\pm 111\pm 30$ eV. To resolve the discrepancy and further reduce the uncertainties, SIDDHARTA collaboration has performed the newest measurement of the kaonic hydrogen X rays at DA$\Phi$NE~\cite{Bazzi:2011zj}. They found $\Delta E=-283\pm 36\pm 6$ eV and $\Gamma=541\pm 89\pm 22$ eV. The results have the smallest error bars and are closer to the KEK measurement, rather than DEAR. The precise determination of the $K^-p$ scattering length through the measurement of the kaonic hydrogen is an important input for the models of the $S=-1$ meson-baryon scattering.

\subsection{Theoretical studies}

There are plenty of theoretical studies on $\Lambda(1405)$, for which we cannot complete the list of references. Here we pick up typical works in quark-model picture and meson-baryon picture, and summarize the status of lattice QCD simulation.

As mentioned in the introduction, the simple quark model picture~\cite{Isgur:1978xj,Capstick:1986bm} has some difficulties to reproduce $\Lambda(1405)$. To describe negative parity baryons by three quarks, one of the quarks has to be excited to the $l=1$ orbit. This excitation energy is usually normalized in the nucleon sector, where one of the lowest negative parity baryons is $N(1535)$. Thus, the excitation is roughly 600 MeV, which indicates the negative parity $\Lambda^*$ around 1700 MeV. In fact, in a simple three-quark picture, $\Lambda(1405)$ would be heavier than the $N(1535)$ since it contains one strange quark. Moreover, for an $l=1$ state, the spin-orbit partner should appear in the $J^P=3/2^-$ channel. In the nucleon sector, the lowest $J^P=3/2^-$ state is the $N(1520)$ resonance which almost degenerates with $N(1535)$, while $\Lambda(1405)$ and $\Lambda(1520)$ are split more than 100 MeV. However, it is fair to mention that the mass of $\Lambda(1405)$ can be reconciled in the three-quark picture, with the help of certain specific mass splitting operators such as those in the $S_3$ permutation symmetry~\cite{Collins:1998ny} and those in the $1/N_c$ expansions~\cite{Schat:2001xr}\footnote{In a recent study of the $1/N_{c}$ expansion for the negative parity baryons, however, it is found that the mass of the $\Lambda(1405)$ cannot be reproduced~\cite{Matagne:2011fr} when the exact wavefunction is adopted instead of the truncated wavefunction used in Ref.~\cite{Schat:2001xr}.}. In addition, five-quark picture can reproduce the low mass of $\Lambda(1405)$~\cite{Strottman:1979qu,Zhang:2004xt} because the negative parity state can be expressed by putting four quarks and one antiquark in the $l=0$ orbit. The five-quark picture, however, predicts more excited baryons in addition to those observed in experiments.

The meson-baryon picture has been studied in the vector-meson exchange models~\cite{Dalitz:1960du,Dalitz:1967fp}. More quantitative discussion was made in the meson-exchange potential models as in Refs.~\cite{MuellerGroeling:1990cw,Shinmura:2010zz}. In these studies, $\Lambda(1405)$ can be dynamically generated from the meson-baryon interactions, without introducing the bare field. It is also worth mentioning that $\Lambda(1405)$ can be described by the kaon bound state approach in the Skyrme model~\cite{Callan:1985hy}. In this approach, strange baryons are described as bound states of a kaon in the background field of the two-flavor Skyrmion. From the quantitative point of view, the binding energy is generally overestimated with typical parameters of the Skyrme model, while the splitting with $\Lambda(1520)$ is rather reasonable. Electromagnetic structures of $\Lambda(1405)$ in this approach were studied in Ref.~\cite{Schat:1994gm}. More quantitative studies have been performed using the three-flavor cloudy bag model~\cite{Veit:1984an,Veit:1985jr,Jennings:1986yg} which is an extension of the quark bag model to incorporate the meson cloud. The model contains both the bare $\Lambda(1405)$ state and the meson-baryon state whose interactions are constrained by chiral symmetry, and the dominance of the meson-baryon component was pointed out. One of the interesting consequences of the cloudy bag model lies in the pole structure in the complex energy plane. In Ref.~\cite{Fink:1990uk}, the meson-baryon scattering amplitude was analyzed in the complex energy plane using several potential models and the cloudy bag model of Ref.~\cite{Veit:1985jr}. By fitting the scattering data, the potential models generate one pole for $\Lambda(1405)$, while in the cloudy bag model, \textit{two poles} were found in the energy region of $\Lambda(1405)$.\footnote{Each pole of the scattering amplitude corresponds to one resonance state, as we will discuss in Section~\ref{subsec:origin}.} This is the first mentioning of the double-pole structure of $\Lambda(1405)$ in literature, which will be discussed in detail in Section~\ref{subsec:doublepole}.

Lattice QCD is the most promising approach to the nonperturbative regime of strong interaction. In lattice QCD, the mass of a hadron is usually extracted from the two-point correlation function using some operator which creates and annihilates the hadron state of interest. The lattice investigation of $\Lambda(1405)$, together with negative parity excited baryons, was started by quenched simulations with three-quark operators~\cite{Melnitchouk:2002eg,Nemoto:2003ft,Lee:2005mr,Burch:2006cc}. It was found that the simple three-quark picture did not reproduce the mass of $\Lambda(1405)$ and some multiquark and/or meson-baryon components would be necessary. In response to these results, the simulation with five-quark operators~\cite{Ishii:2007ym} was performed, while the lowest energy state was found in heavier region than $\Lambda(1405)$. The full QCD simulation with three-quark operator was done in Ref.~\cite{Takahashi:2009bu}, including the mixing of the flavor singlet and octet states. In this case, again, the low mass of 1405 MeV was not reproduced. Thus, at present, the description of $\Lambda(1405)$ is a difficult but challenging issue in lattice QCD. One should also note that $\Lambda(1405)$ is a resonance at physical point, although it is a bound state below the threshold in the heavy pion mass region. The recent development of lattice QCD enables one to describe excited hadrons as resonances in the hadron-hadron scattering~\cite{Aoki:2007rd}. Such an approach may be more appropriate to describe the $\Lambda(1405)$ resonance in the $\bar{K}N$-$\pi\Sigma$ scattering system, rather than the extraction from the two-point correlation functions.

\section{Chiral unitary approach}
\label{sec:chiralunitary}

The chiral unitary approach is a powerful theoretical tool to describe the hadron scattering amplitude including resonances. This approach is based on two guiding principles; chiral symmetry for low energy hadron dynamics, and general properties of the scattering amplitude such as unitarity and analyticity. The $\Lambda(1405)$ resonance in the $S=-1$ meson-baryon scattering is well described in this approach. In this section, we formulate the chiral unitary approach for meson-baryon scattering.

After a brief overview of the framework, we construct effective Lagrangian of chiral perturbation theory for the meson-baryon system, starting from the chiral symmetry of QCD in Section~\ref{subsec:ChPT}. The low energy meson-baryon interaction is derived in Section~\ref{subsec:interaction} up to $\mathcal{O}(p^2)$ order in chiral perturbation theory. The nonperturbative unitarized amplitude is derived in two different ways. One is to use the scattering equation with the on-shell factorization method (Section~\ref{subsec:LSeq}), and the other is based on the N/D method in scattering theory (Section~\ref{subsec:N/D}). We show that both the methods lead to the same scattering amplitude. In Section~\ref{subsec:origin}, we discuss how the resonances are described in the scattering amplitude. The origin of resonances is investigated from the viewpoint of renormalization schemes.

\subsection{Overview of chiral unitary approach}

The low energy QCD exhibits rich spectra and complicated dynamics of hadrons due to color confinement and chiral symmetry breaking. The spontaneous chiral symmetry breaking causes the appearance of the Nambu-Goldstone (NG) bosons as the light pseudoscalar mesons and the dynamical mass generation of hadrons~\cite{Weinberg:1996kr,Hosaka:2001ux}. In addition to these static properties, the dynamics of the NG bosons is constrained by chiral symmetry through the low energy theorems. For instance, the scattering length of the NG boson with a target hadron is determined up to its sign and strength through the Weinberg-Tomozawa theorem~\cite{Weinberg:1966kf,Tomozawa:1966jm}, which works quite well for the $\pi\pi$ and $\pi N$ channels. The dynamics of hadrons and NG bosons are concisely summarized in chiral perturbation theory with the systematic power counting scheme~\cite{Weinberg:1979kz,Gasser:1983yg}, in which the low energy theorems are correctly encoded as the leading order result of the perturbative expansion. Thus, chiral perturbation theory describes the low energy limit of the meson-baryon scattering amplitude.

On the other hand, physical scattering amplitude should satisfy the unitarity condition of the S-matrix which follows from the probability conservation during the time evolution of the system. Although a solution of the nonperturbative scattering equation satisfies the unitarity condition, perturbative calculation of the amplitude can spoil the unitarity. For instance, since the amplitude in chiral perturbation theory is expanded in powers of the energy of the NG boson, as we go to higher energy, the amplitude monotonically increases and eventually violates the unitarity bound at certain kinematical scale. This does not cause a serious problem for the description of the low energy behavior of the amplitude, but resonances cannot be treated in perturbative calculation unless they are introduced as explicit degrees of freedom. Thus, the naive extrapolation of the amplitude to the higher energy region is not justified. The importance of the unitarity and nonperturbative dynamics has been pointed out for the $\bar{K}N$-$\pi\Sigma$ system~\cite{Dalitz:1959dn} where the inter-hadron interaction is considered to be strong. In order to extend the low energy chiral interaction to the resonance energy region, it is mandatory to combine the interaction with some kinds of unitarization technique. 

In this way, chiral symmetry for the low energy NG boson dynamics and the unitarity of the scattering amplitude are important guiding principles to construct hadron scattering amplitude. Attempts to combine the unitarity condition with chiral perturbation theory have been started using the inverse amplitude method in meson scattering sector~\cite{Truong:1988zp,Dobado:1990qm,Dobado:1993ha}, and using the Lippmann-Schwinger equation in baryon sector~\cite{Kaiser:1995eg,Oset:1998it}. The recent development of the analysis of baryon resonances owes much to the framework based on the N/D method~\cite{Oller:2000fj}. It has been shown that the chiral unitary approach reproduces hadron scattering amplitude successfully. Thanks to the universality of the chiral interaction for the NG boson dynamics, the approach has been applied to various hadron scatterings with the NG bosons as shown in Section~\ref{subsec:application}. In this section we illustrate the theoretical framework of the chiral unitary approach for $s$-wave meson-baryon scattering amplitude in detail. See also the review article of the chiral unitary approach~\cite{Oller:2000ma}.

\subsection{Chiral symmetry in QCD and chiral perturbation theory}
\label{subsec:ChPT}

The strong interaction is governed by quantum chromodynamics (QCD) which is a color SU(3) gauge theory of quarks as fundamental fields and gluons as gauge fields. The three-flavor ($u$, $d$ and $s$) massless QCD Lagrangian is given by
\begin{align}
    \mathcal{L}_{\text{QCD}}^{0}
    &=-\frac{1}{2}\tr [G_{\mu\nu}G^{\mu\nu}]
    +\bar{q}i\gamma^{\mu}D_{\mu}q
    \label{eq:QCD0}, \\
    &\quad 
    G_{\mu\nu}=\partial_{\mu}A_{\mu}-\partial_{\nu}A_{\mu}
    -ig[A_{\mu},A_{\nu}],
    \quad 
    D_{\mu}=\partial_{\mu}-igA_{\mu},
    \quad
    A_{\mu}=\sum_{a}T^{a}A_{\mu}^{a}
    \nonumber ,
\end{align}
where $q$ is the quark field, $A^{a}_{\mu} (a=1\sim 8)$ are the gluon fields, $T^a=\lambda^a/2$ are the generators of the color SU(3) group with Gell-Mann matrices $\lambda^a$, and $g$ is the gauge coupling constant. The quark field is represented as a three-component column vector in color space, with three components in flavor space. 

To appreciate chiral symmetry in the QCD Lagrangian~\eqref{eq:QCD0} it is useful to define the left- and right-handed quarks as
\begin{equation}
	q_L=P_Lq , \quad
	q_R=P_Rq  , \quad
    \nonumber
\end{equation}
with projection operators $P_{L,R}=(1\mp \gamma_5)/2$. In terms of the left- and right-handed quarks, the Lagrangian~\eqref{eq:QCD0} can be 
expressed as
\begin{equation}
    \mathcal{L}_{\text{QCD}}^0
    =-\frac{1}{2}\tr [G_{\mu\nu}G^{\mu\nu}]
    +\bar{q}_{L}i\gamma^{\mu}D_{\mu}q_{L}
    +\bar{q}_{R}i\gamma^{\mu}D_{\mu}q_{R} .
    \nonumber
\end{equation}
Here $q_L$ and $q_R$ are separated from each other, and the Lagrangian is invariant under independent unitary transformations of left- and right-handed quark fields. Hence the theory has a global symmetry U$(3)_R\,\times\, $U$(3)_L$. Among this symmetry, U$(1)_A$ is broken by axial anomaly by quantum correction~\cite{Hooft:1986nc}, while U$(1)_{V}$ trivially holds as the quark number conservation in strong interaction. Removing these U$(1)$ subgroups, we are left with \chiral symmetry, that we refer to as chiral symmetry of QCD. An element of chiral symmetry group $g=(R,L)\in $ \chiral transforms the quark fields as 
\begin{equation}
    \begin{split}
	q_{L} &\to Lq_L,\quad
	L=e^{i\theta^{a}_LT^{a}}\in SU(3)_{L}  \\
	q_{R} &\to Rq_{R},\quad
	R=e^{i\theta^{a}_RT^{a}}\in SU(3)_{R} 
    \end{split}
    \quad (a=1\sim 8)  ,
   \nonumber
\end{equation}
where $\theta^{a}_{L,R}$ are real transformation parameters. 

The quark condensate is the vacuum expectation value of the $\bar{q}q$ operator. This is not invariant under chiral symmetry, since it mixes the left- and right-handed quarks:
\begin{equation}
    \bra{0} \bar{q}q \ket{0}
    =
    \bra{0} \bar{q}_Rq_L+\bar{q}_Lq_R \ket{0}
    .
    \nonumber
\end{equation}
On the other hand, it is invariant under the vector subgroup SU$(3)_V=\{(e^{i\theta_V^{a}T^{a}}, e^{i\theta_V^{a}T^{a}})\}$ which rotates the left- and right-handed fields simultaneously in the same direction. In this way, a finite quark condensate leads to the breakdown of \chiral symmetry to the subgroup SU$(3)_V$. This is called spontaneous breaking of chiral symmetry, where the vacuum expectation value breaks the symmetry of the Lagrangian. When the symmetry is spontaneously broken, Nambu-Goldstone theorem~\cite{Nambu:1961tp,Nambu:1961fr,Goldstone:1961eq} ensures that the spectrum of physical particles must contain one bosonic massless particle for each broken symmetry. These bosons are called the Nambu-Goldstone (NG) bosons. In the case of QCD with three flavors, the lightest pseudoscalar mesons ($\pi$, $K$ and $\eta$) are regarded as the (approximate) NG bosons. 

Because quarks have small but nonzero masses, chiral symmetry is also broken explicitly. The quark mass term can be introduced as a diagonal matrix in flavor space:
\begin{equation}
    \mathcal{L}_{\text{QCD}}
    =\mathcal{L}_{\text{QCD}}^{0}
    -\bar{q}\bm{m}q,
    \quad
    \bm{m}=
    \begin{pmatrix}
	m_u & & \\
	& m_d & \\
	& & m_s 
    \end{pmatrix} .
    \nonumber
\end{equation}
The masses of the $u$ and $d$ quarks are as light as several MeV, while the mass of the $s$ quark is about 150 MeV. Although the strange quark is marginal, we consider three-flavor chiral symmetry is a good starting point for the analysis of hadrons. The explicit chiral symmetry breaking generates small masses of the NG bosons, and flavor symmetry breaking induces the mass difference among hadrons belonging to the same representation. These effects will be incorporated in the chiral Lagrangian as perturbative corrections.

Although the fundamental theory is of simple form, at low energy, the nonperturbative effect of the strong interaction causes color confinement where hadrons take the place of the asymptotic degrees of freedom instead of quarks and gluons. Our strategy is to construct an effective field theory with hadronic degrees of freedom, respecting the symmetries of the underlying theory. Conceptually, the chiral Lagrangian $\mathcal{L}_{\text{eff}}$ with effective degrees of freedom $U$ is related with the QCD Lagrangian as
\begin{equation}
    \exp \{iZ\}
    =\int \mathcal{D}q\mathcal{D}\bar{q}\mathcal{D}A_{\mu}
    \exp
    \left\{
    i\int d^4x \mathcal{L}_{\text{QCD}}
    \right\}
    =\int \mathcal{D}U
    \exp
    \left\{
    i\int d^4x \mathcal{L}_{\text{eff}}
    \right\} ,
    \nonumber
\end{equation}
so that both $\mathcal{L}_{\text{QCD}}$ and $\mathcal{L}_{\text{eff}}$ describe the same partition function $Z$. Notice, however, that this procedure is only \textit{schematic}; we follow the ``theorem" introduced by Weinberg~\cite{Weinberg:1979kz} to construct the effective Lagrangian with symmetries being guiding principles. In the effective Lagrangian $\mathcal{L}_{\text{eff}}$, the dynamics of the original fields in the underlying theory is included in the low energy constants that are not determined by the symmetries.

Among several variants of effective field theories, chiral perturbation theory is the most useful approach by virtue of the systematic power counting rule~\cite{Weinberg:1979kz,Gasser:1983yg,Gasser:1985gg,Gasser:1985ux,Gasser:1985pr} (see also review articles~\cite{Pich:1995bw,Ecker:1995gg,Bernard:1995dp}). It is based on the nonlinear realization of chiral symmetry~\cite{Weinberg:1968de,Coleman:1969sm,Callan:1969sn}, which provides a representation for a system with a global symmetry $G$---in the present case, chiral symmetry \chiral---which breaks spontaneously down to a subgroup $H\subset G$---in the present case, SU$(3)_V$. Thus, the spontaneous symmetry breaking of the system is already incorporated at the Lagrangian level. The lowest elementary excitations in QCD are the octet pseudoscalar mesons which are the NG bosons of chiral symmetry breaking. In chiral perturbation theory, the NG boson fields are collected in the matrix form as
\begin{equation}
    \Phi=\begin{pmatrix}
    \frac{1}{\sqrt{2}}\pi^{0}+\frac{1}{\sqrt{6}}\eta & 
    \pi^{+} & K^{+} \\
    \pi^{-} & -\frac{1}{\sqrt{2}}\pi^{0}
    +\frac{1}{\sqrt{6}}\eta & K^{0} \\
    K^{-} & \bar{K}^{0} & -\frac{2}{\sqrt{6}}\eta 
    \end{pmatrix}  ,
    \nonumber
\end{equation}
and the chiral fields $U$ and $u$ are defined by
\begin{equation}
    U(\Phi)
    =\exp\left\{\frac{i\sqrt{2}\Phi}{f}\right\},\quad
    u(\Phi)
    =\exp\left\{\frac{i\Phi}{\sqrt{2}f}\right\},\quad
    U(\Phi)=u^{2}(\Phi)
    \nonumber ,
\end{equation}
where $f$ is a normalization constant of the NG boson field and corresponds to the meson decay constant in the chiral limit at tree level. The transformation lows of these fields under $g=(R,L) \in$ \chiral are given by
\begin{align}
    U&\stackrel{g}{\to}
    RUL^{\dag},\quad
    U^{\dag}\stackrel{g}{\to}
    LU^{\dag}R^{\dag}  ,
    \quad
    u\stackrel{g}{\to}
    Ru h^{\dag}
    =hu L^{\dag}  , \quad
    u^{\dag} \stackrel{g}{\to}
    Lu^{\dag} h^{\dag}
    =hu^{\dag} R^{\dag}  ,
    \nonumber
\end{align}
where $h(g,u)\in $ SU(3)$_V$ is determined according to the group element $g$ and the NG boson field $\Phi$.

We further introduce external fields in QCD Lagrangian~\eqref{eq:QCD0} as
\begin{align}
    \mathcal{L}_{\text{QCD}}^{\text{ext}}
    =&\mathcal{L}_{\text{QCD}}^{0}
    +\bar{q}_L\gamma^{\mu}l_{\mu}q_L
    +\bar{q}_R\gamma^{\mu}r_{\mu}q_R
    -\bar{q}(s-i\gamma_5p)q ,
    \label{eq:QCDext}
\end{align}
where $l_{\mu},r_{\mu},s$ and $p$ are left-handed vector, right-handed vector, scalar and pseudoscalar fields, respectively. The quark mass term can be introduced as the scalar field:
\begin{equation}
    s=\bm{m}
    =\begin{pmatrix}
	m_u & & \\
	& m_d & \\
	& & m_s 
    \end{pmatrix} 
    \label{eq:massterm} ,
\end{equation}
while the $l_{\mu}$ and $r_{\mu}$ fields can be used to introduce the electromagnetic and weak currents. In the effective field theory, it is convenient to define $\chi$ field as
\begin{align}
    \chi = & 2B_0(s+ip)
    , \quad
    \chi^{\dag}=2B_0(s-ip) ,
    \nonumber
\end{align}
with a real constant $B_0$. The extended Lagrangian~\eqref{eq:QCDext} is invariant under local \chiral, if the external fields obey the following transformation rules:
\begin{align}
    \chi&\stackrel{g}{\to} R \chi L^{\dag}  ,
    \quad
    \chi^{\dag}\stackrel{g}{\to} L \chi^{\dag} R^{\dag}  ,
    \quad
    l_{\mu} \stackrel{g}{\to}
	Ll_{\mu}L^{\dag}
	+iL\partial_{\mu}L^{\dag},
	\quad
	r_{\mu} \stackrel{g}{\to}
	Rr_{\mu}R^{\dag}
	+iR\partial_{\mu}R^{\dag}
    \label{eq:transformation} .
\end{align}
Since it is a local transformation, derivatives of the field $U(\Phi)$ should be given by covariant derivatives
\begin{equation}
    \nabla_{\mu}U=\partial_{\mu}U
    -ir_{\mu}U+iUl_{\mu},\quad
    \nabla_{\mu}U\stackrel{g}{\to}
    R\nabla_{\mu}UL^{\dag}.
    \nonumber
\end{equation} 
Note that the specific choice of the scalar field~\eqref{eq:massterm} does not follow the transformation law~\eqref{eq:transformation} and hence breaks chiral symmetry explicitly. We construct the chiral invariant effective Lagrangian with the $\chi$ field, and then introduce the mass term so that the effective Lagrangian breaks chiral symmetry as in the same way with the underlying QCD does.

With these preliminaries, let us construct the effective field theory by organizing the most general Lagrangians which are invariant under chiral transformation. We introduce chiral counting rule by considering the momentum of the meson $p^{\mu}$  as a small quantity in comparison with the chiral symmetry breaking scale of $4\pi f\sim 1$ GeV. Defining $U$ as a quantity of order $\mathcal{O}(1)$, we count terms with $n$ derivatives of the $U$ field as $\mathcal{O}(p^{n})$. The chiral counting for the external fields is then determined as
\begin{equation}
    U,\ u:\mathcal{O}(1)  , \quad
    \nabla_{\mu} U,\  l_{\mu},\ r_{\mu} , : \mathcal{O}(p)  , \quad
    \chi :\mathcal{O}(p^{2})  .
    \nonumber
\end{equation}
The terms of the effective Lagrangian for the NG bosons are sorted out in powers of the chiral orders:
\begin{equation}
    \mathcal{L}_{\text{eff}}(U)
    =\sum_{n=1}^{\infty}\mathcal{L}_{2n}^{M}(U)
    \label{eq:efflag} 
    =\frac{f^{2}}{4}
    \Tr(\nabla_{\mu}U^{\dag}\nabla^{\mu}U
    +U^{\dag}\chi + \chi^{\dag} U)
    +\dots ,
    \nonumber
\end{equation}
where $\mathcal{L}_{2n}^{M}$ represents the terms of $\mathcal{O}(p^{2n})$. To analyze the low energy dynamics of the NG bosons, the terms with lower powers are relevant. The factor $f^2/4$ of the leading order term has been chosen for the normalization of the kinetic term. The higher order terms are constructed so that they are invariant under chiral transformation, with coefficients (low energy constants) which can not be determined by the symmetry argument. Although the chiral Lagrangian contains derivative couplings, renormalizability is ensured order by order, by virtue of the chiral counting rule for the amplitude. Namely, the divergences of the loop diagram can be systematically tamed by the counter terms at a given order. 

Now we turn to the baryons which are introduced as matter fields in the nonlinear realization~\cite{Coleman:1969sm,Callan:1969sn}.  The octet baryon fields are collected as
\begin{equation}
    B
    =\begin{pmatrix}
    \frac{1}{\sqrt{2}}\Sigma^{0}+\frac{1}{\sqrt{6}}\Lambda & 
    \Sigma^{+} & p \\
    \Sigma^{-} & -\frac{1}{\sqrt{2}}\Sigma^{0}
    +\frac{1}{\sqrt{6}}\Lambda & n \\
    \Xi^{-} & \Xi^{0} & -\frac{2}{\sqrt{6}}\Lambda
    \end{pmatrix} ,
    \nonumber
\end{equation}
which transforms under $g\in$ \chiral as
\begin{equation}
	B\stackrel{g}{\to} hBh^{\dag} , \quad
	\bar{B}\stackrel{g}{\to}
	h\bar{B}h^{\dag} 
    \nonumber ,
\end{equation}
with $h(g,u)\in $ SU(3)$_V$. For baryons, the mass term $M_0\Tr(\bar{B}B)$ is allowed even in the chiral limit. The mass term brings the additional scale $M_0$ in the theory, which causes problems in the counting rule of Lagrangian and eventually in the systematic renormalization program. An elegant method to avoid this difficulty is the heavy baryon chiral perturbation theory~\cite{Jenkins:1990jv}, where the baryon fields are treated as heavy static fermions by taking the limit $M_0\to \infty$. Here we follow Refs.~\cite{Krause:1990xc,Oller:2006yh,Frink:2006hx} to construct the relativistic chiral Lagrangian with keeping the common mass of the octet baryons $M_0$ finite.\footnote{In this paper we utilize chiral perturbation theory for the meson-baryon scattering amplitude up to $\mathcal{O}(p^2)$ where no loop diagram appears. At $\mathcal{O}(p^3)$, an appropriate renormalization procedure in the relativistic scheme~\cite{Becher:1999he,Schindler:2003xv} must be introduced.} We define the following quantities 
\begin{align}
	\chi_+ &=
	u\chi^{\dag}u
	+u^{\dag}\chi u^{\dag} , \quad
	\chi_- =
	u\chi^{\dag}u
	-u^{\dag}\chi u^{\dag} , \nonumber
    \\
    u_{\mu}
    &=i\{u^{\dag}(\partial_{\mu} -ir_{\mu})u
    -u(\partial_{\mu}-il_{\mu})u^{\dag}\} , \nonumber \\
    \Gamma_{\mu}
    &=\frac{1}{2}\{u^{\dag}(\partial_{\mu} -ir_{\mu})u
    +u(\partial_{\mu}-il_{\mu})u^{\dag}\} .
    \nonumber
\end{align}
The latter two quantities are related to the vector ($V_{\mu}$) and axial vector ($A_{\mu}$) currents as $A_{\mu}= -u_{\mu}/2$ and 
$V_{\mu}=-i\Gamma_{\mu}$. These quantities are transformed as $O\stackrel{g}{\to} hO h^{\dag}$, except for the chiral connection $\Gamma_{\mu}$, which transforms as
\begin{equation}
    \Gamma_{\mu}\stackrel{g}{\to} h\Gamma_{\mu} h^{\dag} 
    +h\partial_{\mu} h^{\dag} .
    \nonumber
\end{equation}
The covariant derivatives for the octet baryon fields can be defined as
\begin{align}
    D_{\mu}B= \partial_{\mu}B+[\Gamma_{\mu},B] .
    \nonumber
\end{align}
The power counting rule for baryon fields is given by
\begin{align}
    B,\ \bar{B}:\mathcal{O}(1)  
    , \quad
    u_{\mu},\ \Gamma_{\mu},\
    (i\Slash{D}-M_0)B:\mathcal{O}(p) 
    , \quad
    \chi^{\pm}:\mathcal{O}(p^{2})  .   \nonumber 
\end{align}
With these counting rules, we can construct the most general effective Lagrangian for meson-baryon system as 
\begin{equation}
    \mathcal{L}_{\text{eff}}(B,U)
    =
    \sum_{n=1}^{\infty}
    [\mathcal{L}_{2n}^{M}(U)+\mathcal{L}_{n}^{MB}(B,U)] ,
    \nonumber
\end{equation}
where $\mathcal{L}_{n}^{MB}(B,U)$ consists of bilinears of $B$ field with the chiral order $\mathcal{O}(p^n)$. In the lowest order $\mathcal{O}(p)$, we have
\begin{align}
    \mathcal{L}_{1}^{MB}
    &=\Tr\left(\bar{B}(i\Slash{D}-M_{0})B
    +\frac{D}{2}(\bar{B}\gamma^{\mu}\gamma_{5}
    \{u_{\mu},B\}) 
    +\frac{F}{2}(\bar{B}\gamma^{\mu}\gamma_{5}
    [u_{\mu},B])\right) ,
    \label{eq:LOLag}
\end{align}
where $D$ and $F$ are low energy constants related to the axial charge of the nucleon $g_A=D+F$, and $M_0$ denotes the common mass of the octet baryons. Among many terms of the next-to-leading order Lagrangian~\cite{Krause:1990xc,Oller:2006yh,Frink:2006hx}, the relevant terms to the meson-baryon scattering are
\begin{align}
    \mathcal{L}_{2}^{MB}
    =&b_D\Tr \bigl(\bar{B}\{\chi_+,B\}\bigr)
    +b_F\Tr \bigl(\bar{B}[\chi_+,B]\bigr)
    +b_0\Tr(\bar{B}B)\Tr(\chi_+)\nonumber \\
    &+d_1\Tr \bigl(\bar{B}\{u^{\mu},[u_{\mu},B]\}\bigr)
    +d_2\Tr \bigl(\bar{B}[u^{\mu},[u_{\mu},B]]\bigr)
    \nonumber \\
    &+d_3\Tr(\bar{B}u_{\mu})\Tr(u^{\mu}B) 
    +d_4\Tr(\bar{B}B)\Tr(u^{\mu}u_{\mu}) ,
    \label{eq:NLOLag}
\end{align}
where $b_i$ and $d_i$ are the low energy constants. The first three terms are proportional to the $\chi$ field and hence to the quark mass term.  Thus, they are responsible for the mass splitting of baryons. Indeed, Gell-Mann--Okubo mass formula follows from the tree level calculation with isospin symmetric masses $m_u=m_d=\hat{m}\neq m_s$.

\subsection{Low energy meson-baryon interaction}
\label{subsec:interaction}

\begin{figure}[tb]
\begin{center}
  \includegraphics[width=0.75\textwidth,bb=0 0 630 100]{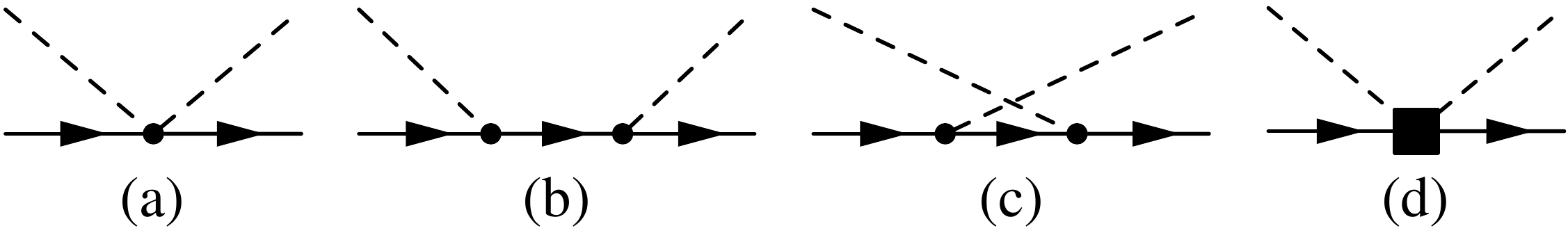}
\begin{minipage}[t]{16.5 cm}
\caption{Feynman diagrams for the meson-baryon interactions in chiral perturbation theory. (a) Weinberg-Tomozawa interaction, (b) s-channel Born term, (c) u-channel Born term, (d) NLO interaction. The dots represent the $\mathcal{O}(p)$ vertices while the square denotes the $\mathcal{O}(p^2)$ vertex. \label{fig:ChPT}}
\end{minipage}
\end{center}
\end{figure}

Here we derive the $s$-wave low energy meson-baryon interaction up to the order $\mathcal{O}(p^2)$ in momentum space. In the three-flavor sector, several meson-baryon channels participate in the scattering, which are labeled by the channel index $i$. The scattering amplitude from channel $i$ to $j$ can be written as $V_{ij}(W,\Omega,\sigma_i,\sigma_j)$ where $W$ is the total energy of the meson-baryon system in the center-of-mass frame, $\Omega$ is the solid angle of the scattering, and  $\sigma_i$ is the spin of the baryon in channel $i$. Since we are dealing with the scattering of the spinless NG boson off the spin 1/2 baryon target, the angular dependence vanishes and the spin-flip amplitude does not contribute after the $s$-wave projection and the spin summation. Thus, the $s$-wave interaction depends only on the total energy $W$ as
\begin{align}
    V_{ij}(W)
    =\frac{1}{8\pi}
    \sum_{\sigma}\int d\Omega \ V_{ij}(W,\Omega,\sigma,\sigma)
    \label{eq:swave} .
\end{align}
In chiral perturbation theory up to $\mathcal{O}(p^2)$, there are four kinds of diagrams as shown in Fig.~\ref{fig:ChPT}. For the $s$-wave amplitude, the most important piece in the leading order terms is the Weinberg-Tomozawa (WT) contact interaction (a). The covariant derivative in Eq.~\eqref{eq:LOLag} generates this term which can also be derived from chiral low energy theorem. At order $\mathcal{O}(p)$, in addition to the WT term, there are s-channel Born term (b) and u-channel term (c) which stem from the axial coupling terms in Eq.~\eqref{eq:LOLag}. Although they are in the same chiral order with the WT term (a), the Born terms mainly contribute to the $p$-wave interaction and the $s$-wave component is in the higher order of the nonrelativistic expansion~\cite{Weinberg:1996kr}. With the terms in the next-to-leading order Lagrangian~\eqref{eq:NLOLag}, the diagram (d) gives the $\mathcal{O}(p^2)$ interaction. In summary, the tree-level meson-baryon amplitude is given by 
\begin{align}
    V_{ij}(W,\Omega,\sigma_i,\sigma_j)
    =V_{ij}^{\text{WT}}(W,\Omega,\sigma_i,\sigma_j)
    +V_{ij}^{\text{s}}(W,\Omega,\sigma_i,\sigma_j)
    +V_{ij}^{\text{u}}(W,\Omega,\sigma_i,\sigma_j)
    +V_{ij}^{\text{NLO}}(W,\Omega,\sigma_i,\sigma_j)
    \label{eq:interaction} ,
\end{align}
where $V_{ij}^{\text{WT}}$, $V_{ij}^{\text{s}}$, $V_{ij}^{\text{u}}$ and $V_{ij}^{\text{NLO}}$ terms correspond to the diagrams (a), (b), (c) and (d) in Fig.~\ref{fig:ChPT}, respectively. In the following we derive the $s$-wave amplitude $V_{ij}(W)$ by calculating these diagrams.

Let us first consider the WT interaction (a). By expanding the covariant derivative term in Eq.~\eqref{eq:LOLag} in powers of the meson field $\Phi$, we obtain the meson-baryon four-point vertex
\begin{align}
    \mathcal{L}^{\text{WT}}
    =\frac{1}{4f^2}\Tr \left(\bar{B}i\gamma^{\mu}[
    \Phi\partial_{\mu}\Phi
    -(\partial_{\mu}\Phi)\Phi
    ,B] \right) .
    \label{eq:WTLag}
\end{align}
The tree-level amplitude by this term is given by
\begin{align}
    V_{i j}^{\text{WT}}(W,\Omega,\sigma_i,\sigma_j)
    =&-
    \frac{C_{i j}}{4 f^2}
    \sqrt{\frac{M_i+E_i}{2M_i}}\sqrt{\frac{M_j+E_j}{2M_j}} 
    \nonumber \\
    &\times (\chi^{\sigma_i})^T
    \Biggl[2W-M_i-M_j+(2W+M_i+M_j) 
    \frac{\bm{q}_i\cdot\bm{q}_j
    +i(\bm{q}_i\times\bm{q}_j)\cdot\bm{\sigma}}
    {(M_i+E_i)(M_j+E_j)}
    \Biggr]\chi^{\sigma_j} ,
    \label{eq:WTterm}
\end{align}
where $\bm{q}_i$, $M_i$ and $E_i$ are the momentum, the mass and the energy of the baryon in channel $i$, and $\chi^{\sigma_i}$ is the two-component Pauli spinor for the baryon in channel $i$. Applying the $s$-wave projection~\eqref{eq:swave}, we obtain the WT interaction
\begin{equation}
    V_{i j}^{\text{WT}}(W) 
    = -  \frac{C_{i j}}{4 f^2}(2W - M_{i}-M_{j}) 
    \sqrt{\frac{M_i+E_i}{2M_i}}\sqrt{\frac{M_j+E_j}{2M_j}} .
    \label{eq:WTtermswave}
\end{equation}
The $C_{ij}$ coefficients express the sign and the strength of the interaction for this channel. With the SU(3) isoscalar factors~\cite{PRSLA268.567,deSwart:1963gc}, it is given by~\cite{Hyodo:2006yk,Hyodo:2006kg}
\begin{align}
    C_{ij}
    =&\sum_{\alpha}[6-C_2(\alpha)]
    \iso{8}{8}{\alpha}
    {I_{\bar{i}} , Y_{\bar{i}}}{I_{i} , Y_{i}}{I , Y} 
    \iso{8}{8}{\alpha}
    {I_{\bar{j}} , Y_{\bar{j}}}{I_{j} , Y_{j}}{I , Y} ,
    \label{eq:CijMB} \\
    &Y = Y_{\bar{i}} +Y_{i} = Y_{\bar{j}} + Y_{j} ,
    \quad
    I = I_{\bar{i}} +I_{i} = I_{\bar{j}} + I_{j} ,
    \nonumber 
\end{align}
where $\alpha$ is the SU(3) representation of the meson-baryon system with $C_2(\alpha)$ being its quadratic Casimir, $I_{i}$ and $Y_{i}$ are the isospin and hypercharge of the particle in channel $i$ ($i$ stands for the baryon and $\bar{i}$ for the meson). Explicit values of $C_{ij}$ for the $S=-1$ meson-baryon scattering can be found in Ref.~\cite{Oset:1998it}. It is remarkable that the sign and the strength of the interaction~\eqref{eq:WTtermswave} are fully determined by the group theoretical factor $C_{i j}$. This is because the low energy constant is absent in the Lagrangian~\eqref{eq:WTLag}, as it is derived from the covariant derivative. In the language of current algebra, this is a consequence of the vector current conservation (Weinberg-Tomozawa theorem)~\cite{Weinberg:1966kf,Tomozawa:1966jm}. Indeed, at the threshold of the $\pi N\to\pi N$ amplitude, Eq.~\eqref{eq:WTtermswave} gives the $\pi N$ scattering length (the relation of the T-matrix with the nonrelativistic scattering amplitude is summarized in Appendix)
\begin{equation}
    a_{\pi N\to \pi N}
    = 
    \begin{cases}
      \vspace{0.3cm}
      \dfrac{M_N}{4\pi (M_N+m_{\pi})}
      \dfrac{m_{\pi}}{f^2} & \text{for } I=1/2 \\
      -\dfrac{M_N}{8\pi (M_N+m_{\pi})}
      \dfrac{m_{\pi}}{ f^2} & \text{for } I=3/2
    \end{cases}  ,
    \label{eq:piNslength}
\end{equation}
in accordance with the low energy theorem. 

It is also remarkable that the phenomenological vector-meson exchange potential~\cite{Dalitz:1967fp} leads to the same channel couplings with $C_{ij}$ when the flavor SU(3) symmetric coupling constants are used. In fact, with the KSRF relation $g_{V}^2=m_{V}^2/2f^2$~\cite{Kawarabayashi:1966kd,Riazuddin:1966sw}, the vector-meson exchange potential reduces to the contact interaction $V\propto C_{ij}/f^2$ in the limit $m_{V}\to \infty$.

Another important feature of the WT interaction~\eqref{eq:WTtermswave} is the dependence on the total energy $W$. This is a consequence of the derivative coupling nature of the NG boson in the nonlinear realization. The energy dependence is an important aspect for the discussion of the $s$-wave resonance state.

Next we turn to the Born diagrams (b) and (c) in Fig.~\ref{fig:ChPT}. By expanding the axial coupling terms in Eq.~\eqref{eq:LOLag} in powers of the meson field $\Phi$, we obtain the meson-baryon Yukawa vertex
\begin{align}
    \mathcal{L}^{\text{Yukawa}}
    =-\frac{1}{\sqrt{2}f}
    \Tr\left(D(\bar{B}\gamma^{\mu}\gamma_{5}
    \{\partial_{\mu}\Phi
    ,B\}) 
    +F(\bar{B}\gamma^{\mu}\gamma_{5}
    [\partial_{\mu}\Phi,B])\right) .
    \nonumber 
\end{align}
Using the three-point vertex, we calculate the s-channel Born diagram [Fig.~\ref{fig:ChPT} (b)] which leads to 
\begin{align*}
    V^{\text{s}}_{ij}(W,\Omega,\sigma_i,\sigma_j)
    =&
    \sum_{k=1}^{8}
    \frac{C_{\bar{i}i,k}^{(c)}C_{\bar{j}j,k}^{(c)}}{12f^2}
    \frac{1}{s-M_k^2} 
    \sqrt{\frac{M_i+E_i}{2M_i}}\sqrt{\frac{M_j+E_j}{2M_j}}
    \\
    &\times (\chi^{\sigma_i})^T
    \Biggl[(W-M_k)(s-W(M_i+M_j)+M_iM_j) \\
    & 
    +(W+M_k)(s+W(M_i+M_j)+M_iM_j)
    \frac{\bm{q}_i\cdot\bm{q}_j
    +i(\bm{q}_i\times\bm{q}_j)\cdot\bm{\sigma}}
    {(M_i+E_i)(M_j+E_j)}\Biggr]\chi^{\sigma_j} ,
\end{align*}
where $s=W^2$ and channel $k$ denotes the intermediate baryon. The coefficients $C_{\bar{i}i,k}^{(c)}$ are given by $D$ and $F$ constants whose explicit forms are shown in Ref.~\cite{Borasoy:2005ie}. In the same way, the u-channel Born diagram leads to
\begin{align*}
    V^{\text{u}}_{ij}(W,\Omega,\sigma_i,\sigma_j)
    =&
    -
    \sum_{k=1}^{8}
    \frac{C_{\bar{j}k,i}^{(c)}C_{\bar{i}k,j}^{(c)}}{12f^2}
    \frac{1}{u-M_k^2} 
    \sqrt{\frac{M_i+E_i}{2M_i}}\sqrt{\frac{M_j+E_j}{2M_j}}
    \\
    &
    \times (\chi^{\sigma_i})^T\Biggl[
    u(W+M_k)
    +W(M_iM_j+M_k(M_i+M_j))\\
    &-M_iM_jM_k -M_i^2(M_j+M_k)-M_j^2(M_i+M_k)\\
    &+
    \bigl\{u(W-M_k)
    +W(M_iM_j+M_k(M_i+M_j)) \\
    &+M_iM_jM_k +M_i^2(M_j+M_k)+M_j^2(M_i+M_k) 
    \bigr\}
    \frac{\bm{q}_i\cdot\bm{q}_j
    +i(\bm{q}_i\times\bm{q}_j)\cdot\bm{\sigma}}
    {(M_i+E_i)(M_j+E_j)}
    \Biggr] \chi^{\sigma_j} ,
\end{align*}
with 
\begin{align*}
    u
    =&
    -s+m_i^2 +m_j^2
    +2E_i E_j-2\bm{q}_i\cdot\bm{q}_j ,
\end{align*}
where $m_i$ is the mass of the meson in channel $i$. After $s$-wave projection, we obtain
\begin{align*}
    V^{\text{s}}_{ij}(W)
    =&
    \sum_{k=1}^{8}
    \frac{C_{\bar{i}i,k}^{(c)}C_{\bar{j}j,k}^{(c)}}{12f^2}
    \frac{1}{W+M_k}
    [s-W(M_i+M_j)+M_iM_j]
    \sqrt{\frac{M_i+E_i}{2M_i}}\sqrt{\frac{M_j+E_j}{2M_j}} , \\
    V^{\text{u}}_{ij}(W)
    =&
    -\sum_{k=1}^{8}
    \frac{C_{\bar{j}k,i}^{(c)}C_{\bar{i}k,j}^{(c)}}{12f^2} \\*
    &
    \times \Biggl[
    W+M_k
    -\frac{(M_i+M_k)(M_j+M_k)}{2(M_i+E_i)(M_j+E_j)}
    (W-M_k+M_i+M_j) \\*
    &+\frac{(M_i+M_k)(M_j+M_k)}{4\bar{q}_i\bar{q}_j}
    \bigl\{
    W+M_k-M_i-M_j \\*
    &-\frac{s+M_k^2-m_i^2-m_j^2-2E_iE_j}{2(M_i+E_i)(M_j+E_j)}
    (W-M_k+M_i+M_j)\bigr\} \\
    &\times
    \ln\frac{s+M_k^2-m_i^2-m_j^2-2E_iE_j-2\bar{q}_i\bar{q}_j}
    {s+M_k^2-m_i^2-m_j^2-2E_iE_j+2\bar{q}_i\bar{q}_j}
    \Biggr] 
    \sqrt{\frac{M_i+E_i}{2M_i}}\sqrt{\frac{M_j+E_j}{2M_j}} ,
\end{align*}
where the three-momentum in the center-of-mass frame is given by
\begin{align}
    \bar{q}_i
    =&
    \frac{\sqrt{[W^2-(M_i-m_i)^2][W^2-(M_i+m_i)^2]}}{2W}
    \label{eq:threemomentum} .
\end{align}

The studies of the chiral unitary approach with the next to leading order terms were performed in Refs.~\cite{Borasoy:2004kk,Borasoy:2005ie,Oller:2005ig,Oller:2006jw,Borasoy:2006sr}. The expansion of the Lagrangian~\eqref{eq:NLOLag} leads to the four-point vertices
\begin{align}
    \mathcal{L}^{\text{NLO}}
    =&-\frac{b_DB_0}{f^2}
    \Tr \bigl(\bar{B}\{
    \bm{m}\Phi^2
    +2\Phi\bm{m}\Phi
    +\Phi^2\bm{m}
    ,B\}\bigr)
    -\frac{b_FB_0}{f^2}\Tr \bigl(\bar{B}[\bm{m}\Phi^2
    +2\Phi\bm{m}\Phi
    +\Phi^2\bm{m},B]\bigr) \nonumber \\
    & -\frac{4b_0B_0}{f^2}
    \Tr(\bar{B}B)\Tr(\bm{m}\Phi^2)\nonumber \\
    &-\frac{2d_1}{f^2}\Tr \bigl(\bar{B}\{\partial^{\mu}\Phi,
    [\partial_{\mu}\Phi,B]\}\bigr)
    -\frac{2d_2}{f^2}\Tr \bigl(\bar{B}[\partial^{\mu}\Phi,
    [\partial_{\mu}\Phi,B]]\bigr)
    \nonumber \\
    &-\frac{2d_3}{f^2}
    \Tr(\bar{B}\partial_{\mu}\Phi)\Tr(\partial^{\mu}\Phi B)
    -\frac{2d_4}{f^2}\Tr(\bar{B}B)
    \Tr(\partial^{\mu}\Phi \partial_{\mu}\Phi) .
    \nonumber
\end{align}
With these Lagrangians, we obtain the scattering amplitude as
\begin{align*}
    V^{\text{NLO}}_{ij}(W,\Omega,\sigma_i,\sigma_j)
    =&
    \sqrt{\frac{M_i+E_i}{2M_i}}\sqrt{\frac{M_j+E_j}{2M_j}}
    \frac{1}{f^2}
    (C_{\bar{i}i,\bar{j}j}^{(b_1)}
    -2(E_iE_j-\bm{q}_i\cdot\bm{q}_j)
    C_{\bar{i}i,\bar{j}j}^{(b_2)}) \\
    &\times (\chi^{\sigma_i})^T
    \left[1-\frac{\bm{q}_i\cdot\bm{q}_j
    +i(\bm{q}_i\times\bm{q}_j)\cdot\bm{\sigma}}
    {(M_i+E_i)(M_j+E_j)}\right] \chi^{\sigma_j}
    .
\end{align*}
The coefficients $C_{\bar{i}i,\bar{j}j}^{(b_1)}$ and $C_{\bar{i}i,\bar{j}j}^{(b_2)}$ include the low energy constants $b_i$ and $d_i$, whose explicit forms are given in Ref.~\cite{Borasoy:2005ie}. The $s$-wave projection provides
\begin{align*}
    V^{\text{NLO}}_{ij}(W)
    =&
    \frac{1}{f^2}
    \left[
    C_{\bar{i}i,\bar{j}j}^{(b_1)}
    -2\left(
    E_iE_j
    +\frac{\bar{q}_i^2\bar{q}_j^2}{3(M_i+E_i)(M_j+E_j)}
    \right)
    C_{\bar{i}i,\bar{j}j}^{(b_2)}
    \right]
    \sqrt{\frac{M_i+E_i}{2M_i}}\sqrt{\frac{M_j+E_j}{2M_j}}
    .
\end{align*}

\subsection{Nonperturbative amplitude in scattering equation}
\label{subsec:LSeq}

In this section we construct the nonperturbative scattering amplitude based on the scattering equation. In quantum mechanics, once the potential is given, the scattering amplitude can be obtained by solving the Lippmann-Schwinger equation. In the chiral unitary approach, we regard the two-body interaction obtained in  chiral perturbation theory as the potential of the scattering equation. This strategy is similar to the study of the nuclear force; the potential is first constructed to reproduce the tree amplitude of the meson-exchange diagram in Born approximation, which is then used in the scattering equation to calculate the observables~\cite{Nagels:1975fb,Nagels:1976xq,Machleidt:1987hj}. To take into account the transitions among the meson-baryon channels with the same quantum numbers, we consider the coupled-channel scattering equation which is a matrix equation in channel space.

A short remark on the particle masses is in order. Since the quark mass term $\chi$ is counted as $\mathcal{O}(p^2)$, at the leading order $\mathcal{O}(p)$ in chiral perturbation theory, all the octet baryons have the common mass $M_0$ and the NG bosons are massless. On the other hand, in the chiral unitary approach, physical hadron masses are adopted in the scattering equation with the leading order interaction kernel. This is because the threshold energy is so important for the physics of resonances that we separately consider the kinematics of the meson-baryon systems from the chiral expansion of the interaction kernel. Conceptually, we may regard that the 1PI self-energy diagrams are already summed up to give the physical masses of the hadrons, and we expand the low energy meson-baryon four-point correlation function in powers of the meson momenta to derive the interaction kernel. For the same reason, we may utilize the physical meson decay constant $f$ to take into account the field renormalization of the NG bosons. We will further discuss the decay constant in relation with the on-shell factorization.

\begin{figure}[tb]
\begin{center}
  \includegraphics[width=0.65\textwidth,bb=0 0 700 80]{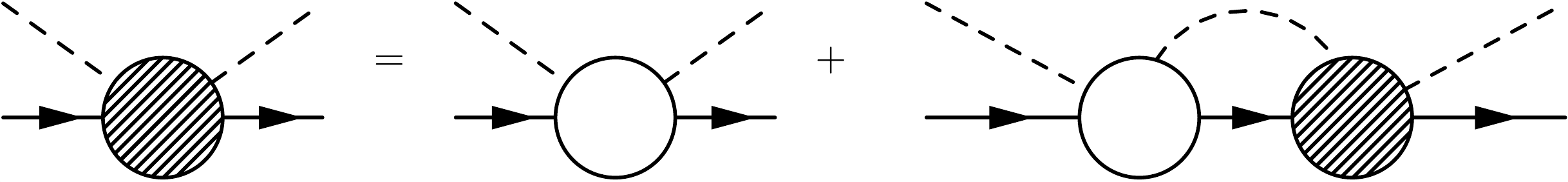}
\begin{minipage}[t]{16.5 cm}
\caption{Schematic illustration of the Lippmann-Schwinger equation~\eqref{eq:LSE1}. The shaded (empty) blob represents the $t$-matrix (potential $v$). Free Green's function $g$ is expressed by the intermediate meson-baryon loop. \label{fig:LSE}}
\end{minipage}
\end{center}
\end{figure}

\begin{figure}[tb]
\begin{center}
  \includegraphics[width=0.75\textwidth,bb=0 0 800 80]{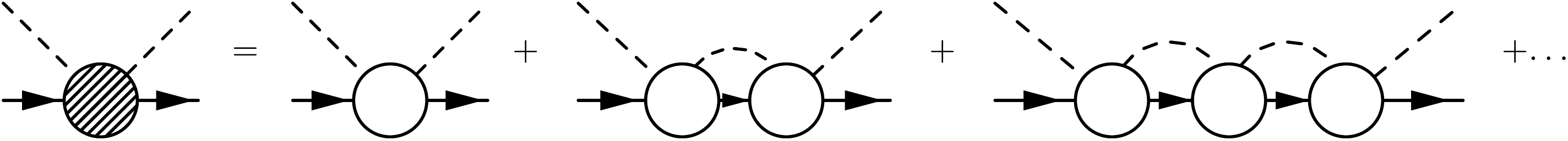}
\begin{minipage}[t]{16.5 cm}
\caption{Schematic illustration of the expanded Lippmann-Schwinger equation~\eqref{eq:LSEeqpand}. The shaded (empty) blob represents the $t$-matrix (potential $v$). Free Green's function $g$ is expressed by the intermediate meson-baryon loop.\label{fig:LSEexpand}}
\end{minipage}
\end{center}
\end{figure}

In the following, we consider the nonrelativistic amplitude and $S$-matrix in $s$ wave $(l=0)$ after partial wave decomposition. The Lippmann-Schwinger equation for the two-body scattering amplitude $t$ with the energy $E$ is given in the operator form as~\cite{Taylor}
\begin{equation}
    t(E) = v + v g(E) t(E) ,
    \label{eq:LSE1}
\end{equation}
where $v$ is the potential and $g$ is the two-body free Green's operator. The equation is schematically displayed in Fig.~\ref{fig:LSE}. Substituting the whole right hand side into the $t$ operator in the right hand side successively, the scattering amplitude can be expressed as the infinite sum of $v$ and $g$ as
\begin{equation}
   t = v + v\ g\ v + v\ g\ v\ g\ v + \cdots ,
   \label{eq:LSEeqpand}
\end{equation}
as illustrated in Fig.~\ref{fig:LSEexpand}. In this way, the solution of the scattering equation~\eqref{eq:LSE1} includes the nonperturbative resummation of the multiple scattering which can generate bound states and resonances dynamically.

We label the scattering state in channel $i$ by its on-shell three-momentum $k_{i}=\sqrt{2\mu_{i}(E-E_{i}^{0})}$ where $\mu_i$ is the reduced mass of the system and $E_{i}^{0}$ is the appropriate threshold energy for channel $i$. Taking the matrix element of Eq.~\eqref{eq:LSE1} by the scattering state $\ket{k_i}$ and inserting the complete set of the intermediate states, we obtain the integral equation for the scattering amplitude 
\begin{align}
   \bra{k_i}t(E)\ket{k_j}
   =& \bra{k_i}v\ket{k_j}
   + \sum_k\int d^3q_k
   \frac{\bra{k_i}v\ket{q_k}\bra{q_k}t(E)\ket{k_j}}
   {E-E_{q_k}+i\epsilon} 
    \nonumber ,
\end{align}
where $E_{q_k}=q_k^2/2\mu_k+E_{k}^{0}$. Note that the intermediate momentum $q_{k}$ can be off shell. Defining the $t$ matrix element $T_{ij}$, the interaction kernel $V_{ij}$ and the energy denominator $\tilde{G}_k$ as
\begin{align}
   T_{ij}(E;k_i,k_j)
   =&\bra{k_i}t(E)\ket{k_j},\quad
   V_{ij}(k_i,k_j)
   =\bra{k_i}v\ket{k_j},\quad
   \tilde{G}_k(E;q_k)
   =\frac{1}{E-E_{q_k}+i\epsilon} ,
   \nonumber
\end{align}
the scattering equation can be written as
\begin{align}
   T_{ij}(E;k_i,k_j)
   =& V_{ij}(k_i,k_j)
   + \sum_k\int d^3q_k
   V_{ik}(k_i,q_k)
   \tilde{G}_k(E;q_k)
   T_{kj}(E;q_k,k_j)
    \label{eq:LSE2} .
\end{align}
This is an integral equation for the amplitude $T_{ij}(E;k_i,k_j)$. However, in certain circumstances, Eq.~\eqref{eq:LSE2} reduces to a tractable algebraic equation. For instance, if the $V_{ij}(k_i,k_j)$ is just a constant and independent of the momenta, then the kernel $V_{ik}$ can be factored out from the $q_k$ integration in Eq.~\eqref{eq:LSE2}. The expanded form~\eqref{eq:LSEeqpand} ensures that the $T_{kj}$ is also independent of the external momenta, so the $q_k$ integration in Eq.~\eqref{eq:LSE2} acts only on the energy denominator. The same factorization is achieved by the use of the separable form factor for the off-shell momentum dependence of the interaction kernel as in Ref.~\cite{Kaiser:1995eg}. Then we can perform the momentum integral separately, and obtain
\begin{align}
   T_{ij}(E)
   =& V_{ij}
   + \sum_k
   V_{ik}
   G_k(E)
   T_{kj}(E),
   \quad
   G_k=\int d^3q_k\tilde{G}_k(E;q_k)
   \nonumber .
\end{align}
The solution of this equation can be given in matrix form as
\begin{equation}
   T = (V^{-1}-G)^{-1}
   \label{eq:Tamp} .
\end{equation}
A similar factorization method has been developed in the chiral unitary approach~\cite{Oller:1997ti,Oset:1998it}, which will be explained below.

To apply the above framework to the meson-baryon scattering, we regard the tree-level chiral amplitude~\eqref{eq:swave} as the interaction kernel $V_{ij}$ which now depends on the total energy. In addition, relativistic kinematics is required to account for the dynamics of the light NG bosons. Thus, denoting the total center-of-mass energy as $W$, we replace 
\begin{align}
   E\to & W,\quad
   V_{ij}(k_i,k_j)
   \to V_{ij}(W;k_i,k_j),\quad
   \nonumber \\
   \tilde{G}_k(E;q_k)
   \to & 
   \tilde{G}_k(W;q_k)=
   \frac{2M_k}{(P-q_k)^{2}-M_k^{2}+i\epsilon}
   \frac{1}{q_k^{2}-m_k^{2}+i\epsilon} ,\quad
   \int d^3q_k
   \to i\int \frac{d^4q_k}{(2\pi)^4} ,
   \nonumber
\end{align}
where $M_k$ and $m_k$ are the masses of the baryon and the meson in channel $k$, and the total momentum is given by $P^{\mu}=(W,\bm{0})$ in the center-of-mass system. Note that the momentum variable $q_{k}^{\mu}$ now represents the four momentum and we neglect the spatial components of the baryon propagator. Then the scattering equation becomes
\begin{align}
   T_{ij}(W;k_i,k_j)
   =& V_{ij}(W;k_i,k_j)
   + \sum_k i\int \frac{d^4q_k}{(2\pi)^4}
   V_{ik}(W;k_i,q_k)
   \tilde{G}_k(W;q_k)
   T_{kj}(W;q_k,k_j)
   . \nonumber 
\end{align}
Here the initial and final momenta $k_{i}$ and $k_{j}$ are on shell, while the intermediate $q_{k}$ is off shell. It is this off-shell momentum that prevents the factorization of $T$ out of the $q_{k}$ integral.

\begin{figure}[tb]
\begin{center}
  \includegraphics[width=0.70\textwidth,bb=0 0 750 100]{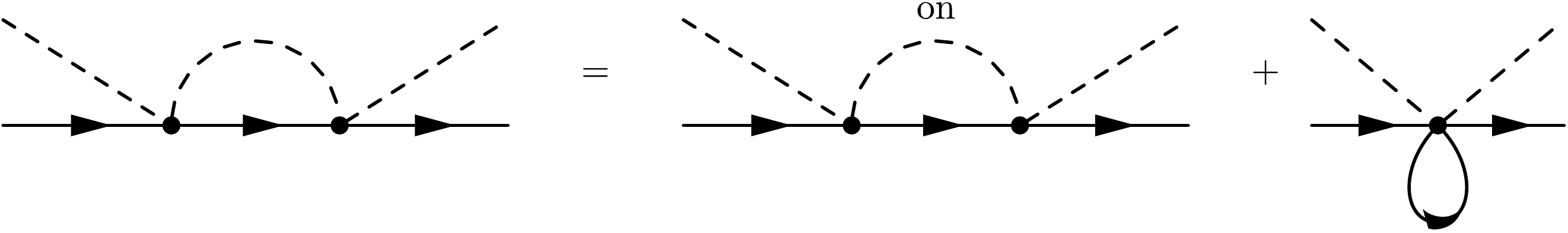}
\begin{minipage}[t]{16.5 cm}
\caption{Diagrammatic illustration of the on-shell factorization of Eq.~\eqref{eq:onshell}.\label{fig:factorization}}
\end{minipage}
\end{center}
\end{figure}

Now let us consider the on-shell factorization, following Refs.~\cite{Oller:1997ti,Oset:1998it}. We first note that the on-shell part of the interaction kernel $V$ for the meson is given with $q_j^2=m_j^2$, so the rest of the interaction kernel is proportional to $(q_j^2-m_j^2)$. Thus, the interaction kernel can be decomposed as
\begin{align}
   V_{ij}(W;k_i,q_j)
   =V_{ij}^{\text{on}}(W;k_i,q_j^2=m_j^2)
   +\beta_{ij} (q_j^2-m_j^2) ,
   \nonumber
\end{align}
with the on-shell part $V^{\text{on}}_{ij}$ and a coefficient $\beta_{ij}$ which controls the off-shell contribution. Substituting this into the momentum integral, we find
\begin{align}
   &i\int \frac{d^4q_k}{(2\pi)^4} 
   V_{ik}(W;k_i,q_k)\frac{2M_k}{(P-q_k)^{2}-M_k^{2}+i\epsilon}
   \frac{1}{q_k^{2}-m_k^{2}+i\epsilon} 
   \nonumber \\
   =&V_{ik}^{\text{on}}(W;k_i,q_k^2=m_k^2)
   i\int \frac{d^4q_k}{(2\pi)^4} 
   \frac{2M_k}{(P-q_k)^{2}-M_k^{2}+i\epsilon}
   \frac{1}{q_k^{2}-m_k^{2}+i\epsilon} 
   \nonumber \\
   &+i\beta_{ik}\int \frac{d^4q_k}{(2\pi)^4} 
   \frac{2M_k}{(P-q_k)^{2}-M_k^{2}+i\epsilon} .
   \label{eq:onshell}
\end{align}
The first term in the right-hand side is the on-shell part of the interaction $V$, while the second term corresponds to a tadpole diagram, as shown diagrammatically in Fig.~\ref{fig:factorization}. Namely, the off-shell part of the meson in the kernel acts as the ``contraction" of the meson propagator. In the same way, the off-shell part of the baryon contracts the baryon propagator and leads to the meson tadpole diagram. The contribution of these tadpole graphs can be absorbed into the renormalization of the meson-baryon vertex in the leading order. In practical calculations, we use a physical value of the meson decay constant as an renormalized value. In this way, we are left with the on-shell part of the interaction and the loop integral can be performed separately. Defining the loop function as
\begin{align}
   G_k(W) =
   i\int\frac{d^{4}q_k}{(2\pi)^{4}}
   \frac{2M_k}{(P-q_k)^{2}-M_k^{2}+i\epsilon}
   \frac{1}{q_k^{2}-m_k^{2}+i\epsilon} 
   \label{eq:loopfunc} ,
\end{align}
the scattering equation finally becomes
\begin{align}
   T_{ij}(W)
   =& V_{ij}(W)
   + \sum_k
   V_{ik}(W)
   G_k(W)
   T_{kj}(W),
   \label{eq:LSEfinal}
\end{align}
since the on-shell momenta $k_{i}$ are the function of the total energy $W$. This is an algebraic equation and its solution can be written as in Eq.~\eqref{eq:Tamp} with the energy dependence of the kernel $V_{ij}(W)$. 

The integral~\eqref{eq:loopfunc} diverges logarithmically, so we need to introduce some regularization scheme. Adopting the dimensional regularization which preserves the fundamental symmetries, the finite part of the loop function is written as
\begin{align}
  G_k(W)\to 
  G_k(W;a_k(\mu)) =
  &\frac{1}{(4\pi)^{2}}
    \Bigl\{a_k(\mu)+\ln\frac{M_k^{2}}{\mu^{2}}
    +\frac{m_k^{2}-M_k^{2}+W^{2}}{2W^{2}}
    \ln\frac{m_k^{2}}{M_k^{2}}
    \nonumber\\
    &+\frac{\bar{q}_k}{W}
    [\ln(W^{2}-(M_k^{2}-m_k^{2})+2W\bar{q}_k)
    +\ln(W^{2}+(M_k^{2}-m_k^{2})+2W\bar{q}_k) 
    \nonumber\\
    &
    -\ln(-W^{2}+(M_k^{2}-m_k^{2})+2W\bar{q}_k)
    -\ln(-W^{2}-(M_k^{2}-m_k^{2})+2W\bar{q}_k)
    ]\Bigr\} ,  \label{eq:loopfuncDR}
\end{align}
with the three-momentum function $\bar{q}_k$ given in Eq.~\eqref{eq:threemomentum}. The loop function $G_k(W;a_k(\mu))$ is the function of the total energy $W$ and the subtraction constant $a_k(\mu)$ which determines the finite part of the loop function at the renormalization scale $\mu$. Note that there is one renormalization parameter for each channel, since the shift of the constant can be absorbed by the change of the scale as $a_k(\mu^{\prime})-a_k(\mu)=2\ln (\mu^{\prime}/\mu)$. Thus, with the WT interaction, $a_k(\mu)$ are the only free parameters in the chiral unitary approach. The subtraction constants have been used to fit experimental data, but the choice of the parameters is closely related to the origin of resonances as we discuss in Section~\ref{subsec:origin}.

\subsection{Unitarity and the N/D method}
\label{subsec:N/D}

We have seen that the Lippmann-Schwinger equation leads to the nonperturbative scattering amplitude as in Eq.~\eqref{eq:Tamp}. Let us derive the same amplitude in the framework of scattering theory, based on Refs.~\cite{Oller:1998zr,Meissner:1999vr,Oller:2000fj}. This method is less intuitive than the one with the scattering equation, but the connection with the unitarity of the S-matrix is apparent. For simplicity, we deal with the single-channel meson-baryon scattering in this section. 

In the scattering theory, the asymptotic completeness leads to the unitarity of the S-matrix operator:
\begin{equation}
    S^{\dag}S=SS^{\dag}= 1 .
    \nonumber
\end{equation}
As shown in Eq.~\eqref{eq:imT} in Appendix, the unitarity condition leads to the optical theorem for the forward scattering with total momentum $P^{\mu}$:
\begin{align}
    \im T(s)
    =&-\frac{1}{2}\sum_n\int d\Pi_{q}^{(n)}(P)
    |T_{qP}|^2 ,\nonumber
\end{align}
where $d\Pi_{q}^{(n)}(P)$ denotes the $n$-body phase space with the total momentum $P^{\mu}$ given in Eq.~\eqref{eq:phasespace} and we write the amplitude as a function of the Mandelstam variable $s=P^{\mu}P_{\mu}=W^2$. This relation shows that the scattering amplitude $T(s)$ has an imaginary part when a intermediate state can go on shell for given $s$. The lowest one in the present case is the two-body meson-baryon state, so the imaginary part of the amplitude appears above the meson-baryon threshold. When the energy variable $s$ is analytically continued to the complex plane, the imaginary part on the real axis causes a branch cut which is called unitarity (right-hand) cut. Restricting the model space to the meson-baryon two-body intermediate states, we can write the imaginary part of the inverse amplitude as
\begin{equation}
    \im T^{-1}(s) = \frac{\rho(s)}{2}
    \quad
    \text{for } s\geq s_+
    \label{eq:Tinv} ,
\end{equation}
where the two-body phase space factor is given by 
\begin{align}
    \rho(s) 
    = \frac{M\sqrt{(s-s_-)(s-s_+)}}{4\pi s}
    \label{eq:phasespace2} ,
\end{align}
with $s_{\pm}=(M\pm m)^2$.

Now we follow the N/D method~\cite{Chew:1960iv,Bjorken:1960zz} to construct the general form of the scattering amplitude. We first consider the kinematical singularities of the amplitude. In addition to the unitarity cut discussed above, in the relativistic kinematics, the unphysical (left-hand and circular) cuts appear due to the crossed diagrams. In the N/D method, we decompose the amplitude $T$ into the $N$ and $D$ functions, and let $D$ ($N$) be exclusively responsible for the unitarity cut (unphysical cuts):
\begin{align}
    T(s)
    =&\frac{N(s)}{D(s)} \nonumber \\
    \im N(s) 
    =& 0\quad \text{for}\quad s\geq s_{+} \nonumber\\
    \im D(s) 
    =& 0\quad \text{for unphysical cuts} . \nonumber
\end{align}
Then the imaginary parts of the $N$ and $D$ functions are given by
\begin{align}
    \im D(s) 
    =& \frac{\rho(s)}{2}N(s)\quad \text{for}\quad s\geq s_{+} 
    \nonumber\\
    \im N(s) 
    =& \im [T(s)] \times D(s)\quad \text{for unphysical cuts} ,
    \nonumber
\end{align}
where we have used Eq.~\eqref{eq:Tinv}. Apart from these branch cuts, the scattering amplitude is meromorphic (analytic except for possible poles) as a consequence of causality~\cite{Taylor,Bohm:2001}. The possible poles and zeros of the amplitude are called the Castillejo-Dalitz-Dyson (CDD) poles~\cite{Castillejo:1956ed} which are interpreted as the effect of independent particles participating in the scattering~\cite{PR124.264}. Thus, using the dispersion relation, we can write the $N$ and $D$ functions as
\begin{align}
    D(s)
    =& \frac{1}{2\pi}
    \int_{s_{+}}^{\infty}
    ds^{\prime}
    \frac{\rho(s^{\prime})N(s^{\prime})}{s^{\prime}-s}
    +(\text{subtractions})+(\text{pole terms}) ,
    \label{eq:right} \\
    N(s)
    =& \frac{1}{\pi}
    \int_{\text{unphysical}}
    ds^{\prime}
    \frac{\im [T(s^{\prime})]D(s^{\prime})}
    {s^{\prime}-s}
    +(\text{subtractions})+(\text{pole terms})
    \label{eq:left} ,
\end{align}
where subtraction terms may appear depending on the properties of the integrand, and possible CDD pole terms are also included. 

Equations.~\eqref{eq:right} and \eqref{eq:left} are the coupled integral equations to determine the $N$ and $D$ functions, which can be solved by knowing $\im T$ on the unphysical cuts. In the present study, we are interested in the physical scattering amplitude and resonances above the threshold, where the effect from the unphysical cuts is considered to be small. Thus we set $N=1$ and neglect the dispersion integral from the unphysical cuts. Then the $D$ function, which is now the inverse amplitude, is given by the dispersion integral of $\rho(s)$ as\footnote{Here we change the energy variable from $s$ to $W=\sqrt{s}$. It is shown that the dispersion integral is of the same form in complex $W$ plane~\cite{Meissner:1999vr}.}
\begin{equation}
    T^{-1}(W) = \sum_{i}\frac{R_i}{W-W_i}
    +\tilde{a}(s_0)
    +\frac{s-s_0}{2\pi}\int_{s_+}^{\infty}ds^{\prime}
    \frac{\rho(s^{\prime})}
    {(s^{\prime}-s)(s^{\prime}-s_0)}
    \label{eq:TampNoD} .
\end{equation}
The first term corresponds to the CDD pole contributions, whose position $W_i$ and residue $R_i$ cannot be determined within the scattering theory. Single subtraction is performed at the subtraction point $s_0$ with the subtraction constant $\tilde{a}(s_0)$ to tame the divergence. From the explicit form of the two-body phase space~\eqref{eq:phasespace2}, it is understood that the single subtraction is sufficient to obtain the finite dispersion integral. Equation~\eqref{eq:TampNoD} is the general form of the amplitude which is compatible with the unitarity condition~\eqref{eq:Tinv}.

Next we match the amplitude with that in chiral perturbation theory at low energy~\cite{Oller:2000fj}. We first calculate the imaginary part of the loop function $G(W)$ in Eq.~\eqref{eq:loopfunc} by Cutkosky rule: 
\begin{align}
   \im G(W) 
   =&
   -\frac{1}{2}\int\frac{d^{3}p_1}{(2\pi)^{3}}
   \frac{2M}{2E}
   \int\frac{d^{3}p_2}{(2\pi)^{3}}
   \frac{1}{2\omega}
   (2\pi^4)\delta^4(P-p_1-p_2) 
   =
   -\frac{\rho(s)}{2} ,
   \nonumber 
\end{align}
where we denote $p_1=P-q$ and $p_2=q$. Thus, the imaginary part of the loop function $G(W)$ is also given by the phase space function~\eqref{eq:phasespace2}. Because the dimensional regularization preserves the analyticity, the dispersion integral plus the subtraction constant in Eq.~\eqref{eq:TampNoD} can be identified as the finite part of the loop integral in Eq.~\eqref{eq:loopfuncDR}. In this way, denoting the CDD pole contributions as $\mathcal{T}^{-1}$, we can write the general amplitude as
\begin{align}
    T(W) 
    =& 
    [\mathcal{T}^{-1}(W)
    -G(W)]^{-1} \nonumber \\
    =& 
    \mathcal{T}(W)
    +\mathcal{T}(W)G(W)\mathcal{T}(W)+
    \dots \label{eq:TampNoDexpand} .
\end{align}
We determine $\mathcal{T}(W)$ by the systematic matching of this amplitude with the low energy expansion~\cite{Oller:2000fj}. For instance, at $\mathcal{O}(p)$, the right hand side of Eq.~\eqref{eq:TampNoDexpand} should coincide with the leading order amplitude in chiral perturbation theory as
\begin{align}
    \mathcal{T}_1
    =& 
    V_1(W)  \nonumber ,
\end{align}
which is given by the WT term and the Born terms. In the same way, we can determine the higher order amplitude of $\mathcal{T}(W)$ as 
\begin{align}
    \mathcal{T}_2
    =& 
    V_2(W)  \nonumber \\
    \mathcal{T}_3
    =& 
    V_3(W) -V_1 G(W)V_1 \nonumber ,
\end{align}
where the one-loop contributions should be subtracted at order $\mathcal{O}(p^3)$ to avoid the double counting. In this way, with the interaction kernel up to $\mathcal{O}(p^2)$ in Eq.~\eqref{eq:interaction}, the final form of the scattering amplitude is given by
\begin{align}
    T(W) 
    =& 
    [V^{-1}(W)
    -G(W)]^{-1} ,
    \nonumber
\end{align}
in accordance with Eq.~\eqref{eq:Tamp}.

Let us here summarize the difference of the chiral unitary approach from the standard calculation in chiral perturbation theory. By virtue of the power counting rule, chiral perturbation theory is renormalizable and crossing symmetric at given order. The unitarity of the amplitude is however satisfied only perturbatively, and resonances cannot be treated dynamically. In the chiral unitary approach, on the other hand, the infinite resummation of diagrams recovers the unitarity of the amplitude. At the same time, the resummation spoils the systematic power counting, so the amplitude $T$ in general depends on the cutoff parameter, the subtraction constant.\footnote{It is possible to eliminate the cutoff dependence by adopting the renormalized interaction kernel $V$, as was done in the meson sector~\cite{GomezNicola:2001as}. In the baryon sector, this requires the computation of the interaction kernel up to $\mathcal{O}(p^3)$.} Since the contributions from the unphysical cuts are neglected, the crossing symmetry is lost in the above formulation.\footnote{In principle, crossing symmetry can be maintained by the use of the Roy equation~\cite{Kaiser:1998fi}.} Thus, it is important to choose the appropriate scheme for the analysis of the system in question. For instance, the leading order chiral interaction for $\pi N$ channel~\eqref{eq:piNslength} is not strong enough to generate bound states and resonances, which is in accordance with the absence of $s$-wave resonances around the $\pi N$ threshold. Therefore, the $s$-wave $\pi N$ system around the threshold can be well described in chiral perturbation theory. On the other hand, in the $\bar{K}N$-$\pi\Sigma$ system, chiral interaction is found to be strong and it is experimentally known that the $\Lambda(1405)$ resonance exist below the $\bar{K}N$ threshold. In the latter case, the unitarization is mandatory to deal with the strong dynamics of the hadronic interaction.

\subsection{Resonances in the amplitude and their origin}
\label{subsec:origin}

We have discussed the construction of the meson-baryon scattering amplitude. If the attraction of the potential is strong enough, the system develops bound states and/or resonances which are expressed as pole singularities in the scattering amplitude. In this section, we show the basic properties of the resonance pole and study the origin of resonances through the renormalization procedure.

The resonance state is expressed in the scattering amplitude as a pole singularity in the second Riemann sheet of the complex energy plane\footnote{In the complex energy plane, the scattering amplitude is defined on a two-sheeted Riemann surface, because the energy is a  function of momentum square $p^{2}$ and the two momenta are mapped onto the same energy. In the first Riemann sheet, only bound state poles can appear as singularities, while resonance poles lie in the second Riemann sheet. }.  The real and imaginary parts of the pole position correspond to the mass $M_R$ and the half width $\Gamma_R/2$ of the resonance state, respectively, and the residue of the pole is the product of the coupling strength $g_i$ of the resonance to the channel $i$. On the real axis, the scattering amplitude can be written as the sum of the Breit-Wigner pole term and the nonresonant background contribution $T_{ij}^{\text{BG}}(W)$ which is assumed to be a slowly varying function of $W$: 
\begin{equation}
    T_{ij}(W) = \frac{g_ig_j}{W-M_R+i\Gamma_R/2} 
    + T_{ij}^{\text{BG}}(W) .
    \label{eq:pole}
\end{equation}

Since the amplitude of the chiral unitary approach is given in analytic form, it is easy to perform the analytic continuation to the complex energy plane. In the chiral unitary approach, the Riemann sheet of the amplitude is fully specified by the loop function which contains the information of the relative momentum of the two-body system. In practice, the amplitude in the second Riemann sheet can be obtained by using the following loop function:
\begin{equation}
    G^{II}_i(W) = G_i(W)
    +i
    \frac{M_i\bar{q}_i(W)}{2\pi W} .
    \nonumber
\end{equation} 
From the expression of the amplitude~\eqref{eq:pole}, it is possible to extract the coupling strengths $g_i$ to the meson-baryon channel $i$ by calculating the residues of the pole:
\begin{equation}
    g_ig_j
    = \lim_{W\to z_R} (W-z_R)T_{ij}(W) 
    \nonumber ,
\end{equation}
with $z_{R}=M_R-i\Gamma_R/2$. In this way, the properties of the resonance can be extracted from the pole of the amplitude.

In general, the origin of resonances in the two-body scattering amplitude can be classified into two categories. One is the preexisting elementary state in the free Hamiltonian. If there are open channels at the energy of this state, it acquires a finite width through the couplings to the channels. The other is the state which is dynamically generated by two-body attraction and does not exist in the free Hamiltonian. In the terminology of scattering theory, the former state is introduced additionally to the scattering equation as a pole term called the CDD pole~\cite{Castillejo:1956ed,PR124.264}. Taking only the hadronic degrees of freedom into account, we regard the former (latter) state as the elementary (composite) particle.

The investigation of the compositeness/elementarity of a particle is a long-standing problem~\cite{Salam:1962ap,Weinberg:1962hj,PTP29.877,PR136.B816,Weinberg:1965zz}. For instance, the property of the deuteron is studied using the relation between the compositeness and the binding energy~\cite{Weinberg:1965zz}. In the chiral unitary approach, the nonperturbative resummation can generate composite states dynamically, while the contribution from elementary particles can be included in the interaction kernel $V$. Therefore, it has been considered that the origin of the resonance is identified from the structure of the interaction kernel. For instance, the resonances generated by the WT interaction were considered as the hadronic molecular resonances, since the interaction kernel has no pole term contribution. However, it is found that the CDD pole contribution can also be hidden in the cutoff parameter in the loop function~\cite{Hyodo:2008xr}. 

Let us discuss the origin of resonances in the amplitude in the chiral unitary approach. For simplicity, we consider the single-channel case. The extension to the multi-channel problem is discussed in Ref.~\cite{Hyodo:2008xr}. As explained in Section~\ref{subsec:LSeq}, the subtraction constant has to be introduced to determine the finite part of the loop function. In the conventional approach, the subtraction constant is regarded as a cutoff parameter and is used to fit experimental data. Although the choice of the cutoff is not constrained \textit{a priori}, certain value of the subtraction constant can bring in a seed of the resonance. To demonstrate this, let us examine the loop function $G(W)$ in detail. Discretizing the energy of the intermediate state as $W_n$, the loop function in Eq.~\eqref{eq:loopfunc} can be schematically represented by
\begin{equation}
   G(W) \simeq \frac{1}{2\pi} 
   \sum_{n}\frac{\rho(W_{n})}{W-W_{n}+i\epsilon} ,
   \nonumber
\end{equation}
where $\rho(W_n)$ is a real non-negative function representing the spectral density of the state $n$. If the model space consists exclusively of the meson-baryon scattering states, the energy of the lowest-lying state $W_0$ is given by the threshold, and $W_n\geq W_0$ for all $n$. Noting that the spectral function is positive definite, the loop function $G(W)$ should be negative for the energy below the threshold, $W<W_{0}$. However, if we choose the subtraction constant $a$ freely, the loop function below the threshold can become positive. 

The meaning of the violation of negativity of the loop function is understood in the following way. In the first place, the interaction $V$ and the loop function $G$ are entangled and cannot be determined separately. This is schematically written by introducing $a$ dependence in $V$ as
\begin{equation}
   T(W) = \frac{1}{V(W;a)^{-1}-G(W;a)}.
   \nonumber
\end{equation}
There are infinitely many combinations of $V(W;a)$ and $G(W;a)$ which reproduce the same amplitude $T(W)$. The formulation in the previous sections was based on the phenomenological renormalization scheme, in which the interaction $V$ is determined by chiral perturbation theory first and the subtraction constant in the loop function $G$ is fixed by experimental data. With the Weinberg-Tomozawa interaction, the scattering amplitude is given by
\begin{equation}
   T(W) = \frac{1}
   {V^{\text{WT}}(W)^{-1}-G(W;a_{\text{pheno}})} ,
   \label{eq:pheno}
\end{equation}
where $a_{\text{pheno}}$ denotes the subtraction constant chosen to reproduce experimental data. In this scheme, the subtraction constant represents the effects which are not included in the interaction kernel. Although the low energy constants in the higher order terms contain the information of the contracted resonance propagator~\cite{Ecker:1988te,Bernard:1995dp}, it is clear from Eq.~\eqref{eq:WTtermswave} that there is no s-channel resonance contribution in the WT interaction. This means that, in the framework of Eq.~\eqref{eq:pheno}, the loop function $G$ should be responsible for the contribution from independent particles, if they exist. This effect is considered as the CDD pole contribution hidden in the loop function. In this case, the origin of the resonance in the amplitude $T$ is not exclusively related to the structure of the interaction kernel $V$.

To discuss the origin of the resonance in a transparent manner, it is desirable to construct the renormalization condition such that the loop function is free from the CDD pole contribution and all the nontrivial structures are embedded in the interaction kernel. To this end, the \textit{natural renormalization scheme} has been proposed in Ref.~\cite{Hyodo:2008xr}. As discussed above, when the model space of the theory contains only the two-body meson-baryon scattering states, the loop function necessarily has negative values below the threshold: $G(W;a)\le 0$ for $W\le M+m$. In addition, since the interaction kernel $V$ is derived in chiral perturbation theory, we require that the amplitude $T(W)$ should reduce to the tree level $V(W)$ at some matching scale $W=\mu_m$. This is achieved by $G(\mu_{m};a)=0$ where the scale $\mu_{m}$ should be chosen in the low energy region $M\leq \mu_m\leq M+m$. Noting that the loop function is a monotonically decreasing function of $W$ in the region of $M \le W \le M+m$, to satisfy both the requirements, we obtain $\mu_{m}=M$ and the renormalization condition is determined uniquely as 
\begin{equation}
   G(W=M;a_{\text{natural}}) = 0.
   \label{eq:anatural}
\end{equation}
The subtraction constant $a_{\text{natural}}$ obtained in this equation is called the natural subtraction constant. In this scheme, since the loop function is determined by the theoretical argument, the interaction kernel $V$ should be adjusted to reproduce experimental data. Thus, the scattering amplitude is expressed by the loop function with the natural subtraction constant and the effective interaction in this scheme $V^{\text{natural}}$ as 
\begin{equation}
   T(W) = \frac{1}
   {V^{\text{natural}}(W;a_{\text{natural}})^{-1}
   -G(W;a_{\text{natural}})} .
   \label{eq:natural}
\end{equation}
By the definition of the natural subtraction constant, the loop function does not contain any seed of resonance, so the nontrivial origin of resonances should be attributed to the interaction kernel $V$. If Eq.~\eqref{eq:natural} describes well the observed scattering amplitude without introducing explicit pole terms in $V$, resonances appearing in the scattering amplitude can be concluded to be dynamically generated resonances and their structures are dominated by the hadronic molecular type components.

From the viewpoint of the renormalization theory, Eqs.~\eqref{eq:pheno} and \eqref{eq:natural} are the different expressions of the same physical amplitude $T(W)$. So, by equating two amplitudes, we obtain
\begin{align}
   V^{\text{WT}}(W)^{-1}&-G(W;a_{\rm pheno})
   =V^{\text{natural}}(W;a_{\rm natural})^{-1}
   -G(W;a_{\rm natural}) .
   \nonumber
\end{align}
The left hand side is completely determined by chiral perturbation theory and data fitting, while the natural subtraction constant in the right hand side is fixed by Eq.~\eqref{eq:anatural} theoretically. Thus, solving this equation for $V^{\text{natural}}$, we obtain the effective interaction in the natural scheme. This can be expressed as the sum of the original WT interaction and a pole term:
\begin{equation}
   V^{\text{natural}}(W;a_{\rm natural})
   = V^{\text{WT}}(W) + \frac{C}{2f^2}
   \frac{(W-M)^2}{W-M_{\text{eff}}} ,
   \label{eq:Vnatural}
\end{equation}
where we have used the explicit forms of $V^{\rm WT}(W)$ in Eq.~\eqref{eq:WTtermswave} and $G(W;a)$ in Eq.~\eqref{eq:loopfuncDR} for the single-channel scattering and ignored the small energy dependence of the kinematical factor in the interaction $(M+E)/2M\to 1$. The mass of the pole term is given by the difference of the subtraction constants $\Delta a \equiv a_{\rm natural}-a_{\rm pheno}$ as
\begin{equation}
   M_{\text{eff}} \equiv M - \frac{16\pi^2f^2}{C M\Delta a} .
   \label{eq:effectivemass}
\end{equation}
It is important to note that the numerator of the pole term in Eq.~\eqref{eq:Vnatural} is quadratic in the meson energy $\omega\sim W-M$, so this term is in higher order than the WT interaction. Namely, the low energy theorem is maintained in the effective interaction~\eqref{eq:Vnatural}.

Equation~\eqref{eq:Vnatural} shows that any slight difference of the phenomenologically determined subtraction constant from the natural value introduces a pole term in the interaction kernel in the natural renormalization scheme. For a molecular type resonance, the phenomenological subtraction constant is close to the natural value and $\Delta a $ becomes small. Therefore, the effective mass~\eqref{eq:effectivemass} takes a large value and the pole term goes away from the relevant energy region of resonances. On the other hand, as the deviation of the phenomenological constant from the natural value increases, the pole mass approaches the baryon mass $M$ and the pole term will substantially affect the amplitude around the resonance. In this case, the origin of the resonance can be attributed to this pole term, rather than the dynamical meson-baryon molecule. It is important that the contribution from the independent particle can be hidden in the loop function through the renormalization procedure. The natural renormalization scheme is used to analyze the origin of the physical baryon resonances in Section~\ref{subsec:Lambda1405MBstate}.

\section{The structure of $\Lambda(1405)$ in chiral dynamics}
\label{sec:structureL1405}

In this section, based on chiral unitary approach introduced in the previous sections, we discuss the structure of the $\Lambda(1405)$ resonance located between the $\pi\Sigma$ and $\bar KN$ thresholds with isospin $I=0$ and strangeness $S=-1$. As we see below, the $s$-wave scattering amplitude in chiral unitary approach with constraints of experimental data describes well both the $K^{-}p$ scattering cross sections and the $\Lambda(1405)$ mass spectrum below the $\bar KN$ threshold. This  implies that we have a microscopical description of $\Lambda(1405)$  in terms of hadronic degrees of freedom in hand, which enables us to discuss the properties and structure of the $\Lambda(1405)$ resonance in various aspects. 

After giving brief history of the study of $\Lambda(1405)$ in chiral unitary approach and some details of the model used in this section, we first find in Section~\ref{subsec:doublepole} that $\Lambda(1405)$ is composed of two states which have different coupling nature to the $\pi\Sigma$ and $\bar KN$ channels. As a consequence of this property, the $\Lambda(1405)$ mass spectrum can be dependent on the channel by which it is initiated. In Section~\ref{subsec:spectrum}, we discuss experimental observations of the $\Lambda(1405)$ spectrum to discriminate its pole structure. Production experiments are required to investigate the spectrum below the $\bar KN$ threshold.

Next we turn to the internal structure of $\Lambda(1405)$ in Sections~\ref{subsec:Lambda1405MBstate}, \ref{subsec:Nc}, and \ref{subsec:elemag}. The meson-baryon nature of $\Lambda(1405)$ is usually taken for granted in the phenomenological applications to the $\bar{K}N$ interaction. Although it is a plausible scenario, the picture should be examined in a reasonable way. We present three different analyses on this issue, showing that the meson-baryon molecular structure of $\Lambda(1405)$ is indeed realized in the amplitude of the chiral unitary approach. The techniques discussed in these sections can be applied generally to dynamically generated resonances. 

Finally we review the applications of chiral unitary approach to other scattering systems. Universal feature of the leading order chiral interaction leads to the generation of various resonances in the similar manner with $\Lambda(1405)$. The systematics of the generated resonances is shown to be closely related to the group theoretical property of the WT interaction.

\subsection{Application of chiral unitary approach to $\Lambda(1405)$}

To study the structure of the $\Lambda(1405)$ resonance, we apply the chiral unitary approach to the strangeness $S=-1$ and charge $Q=0$ meson-baryon scattering. In the particle basis, this sector contains ten coupled channels, $K^- p$, $\bar{K}^0 n$, $\pi^0 \Lambda$, $\pi^0 \Sigma^0$, $\pi^+ \Sigma^-$, $\pi^- \Sigma^+$, $\eta \Lambda$, $\eta\Sigma^0$, $K^0\Xi^0$ and $K^+\Xi^-$. In the isospin basis, the channels are classified as
\begin{align}
   \begin{cases}
   \bar{K}N, \pi\Sigma, \eta\Lambda, K\Xi
   & I=0 \\
   \bar{K}N, \pi\Sigma, \pi\Lambda, \eta\Sigma,  K\Xi
   & I=1 \\
   \pi\Sigma
   & I=2
   \end{cases}
   ,
\end{align}
which are related to the particle basis through the SU(2) Clebsch-Gordan coefficients with the phase convention summarized in Appendix. 

In the framework of the chiral unitary approach, the $\Lambda(1405)$ resonance was first discussed in Ref.~\cite{Kaiser:1995eg} using Lippmann-Schwinger equation with the interaction kernel up to NLO terms in the heavy baryon formalism. The potentials with local and separable form factors were examined. Using the total cross sections of the $K^-p$ scattering amplitude and the threshold branching ratios as constraints, the $\Lambda(1405)$ resonance was well described in the $\pi\Sigma$ spectrum. In Ref.~\cite{Oset:1998it}, full SU(3) channels were included in the coupled-channel equation, and the three-momentum cutoff scheme was adopted to tame the divergence of the loop function. The on-shell factorization scheme for meson-baryon scattering was developed. In Ref.~\cite{Oller:2000fj}, the formulation based on the $N/D$ method was introduced to the meson-baryon scattering, which becomes the foundation of the recent investigations in this field. A comprehensive study up to NNLO interaction was found in Ref.~\cite{Lutz:2001yb}.

After having obtained the good formulation of $\Lambda(1405)$, the properties of $\Lambda(1405)$ were investigated. It was found that there are two poles around $\Lambda(1405)$~\cite{Oller:2000fj} and their physical significance was discussed in Ref.~\cite{Jido:2003cb}. In higher energies than the $\Lambda(1405)$ region, the $\Lambda(1670)$ resonance (the $\Sigma(1620)$ resonance) exists in the isospin $I=0$ ($I=1$) channel of the $S=-1$ scattering. These resonances were successfully reproduced in the chiral unitary approach~\cite{Oset:2001cn}, without altering the description of $\Lambda(1405)$. In Ref.~\cite{Garcia-Recio:2002td}, the $S=-1$ and $I=0$ sector was studied in the fully relativistic treatment of the Bethe-Salpeter (BS) equation. The cut-off and form factor dependences were discussed in Refs.~\cite{Krippa:1998us,Nam:2003ch}, respectively. In order to clarify the SU(3) content of these resonances, the extrapolation of the amplitude to the SU(3) symmetric limit was performed in Ref.~\cite{Jido:2003cb}. As we will see below, this analysis clears up the origin of resonances in terms of SU(3) multiplets. 

The DEAR experiment of the kaonic hydrogen~\cite{Beer:2005qi} gave a large impact on theoretical studies of $S=-1$ meson-baryon amplitude. Although globally successful description of the scattering data had been achieved in previous studies, the total cross sections and the mass spectrum suffer from the experimental errors and uncertainties. The $\bar{K}N$ scattering length is given exactly at the threshold with definite absolute value, so the precise determination is very important to constrain theoretical models. In response to the accurate data, the model for the $S=-1$ meson-baryon scattering was updated by performing systematic $\chi^2$ analyses with the NLO interaction terms~\cite{Borasoy:2004kk,Borasoy:2005ie,Oller:2005ig,Oller:2006jw,Bouzas:2010yb}. The NLO calculation with the separable form factor was also performed in Ref.~\cite{Cieply:2009ea}. A different strategy was taken in Ref.~\cite{Borasoy:2006sr} where the model was fit to low energy scattering data in order to predict the $\bar{K}N$ scattering length. In this way, quantitative refinement of the theoretical model is possible using higher order terms. It is also clear that the subthreshold extrapolation of the $\bar{K}N$ amplitude and the structure of $\Lambda(1405)$ is strongly affected by the $\bar{K}N$ threshold data. Very recently, new measurement by SIDDHARTA experiment~\cite{Bazzi:2011zj} was found to be in line with the KEK measurement~\cite{Iwasaki:1997wf}, rather than the DEAR data~\cite{Beer:2005qi}. Detailed studies with the SIDDHARTA data is called for.

\begin{figure}[tb]
\begin{center}
  \includegraphics[width=0.7\textwidth,bb=0 0 550 350]{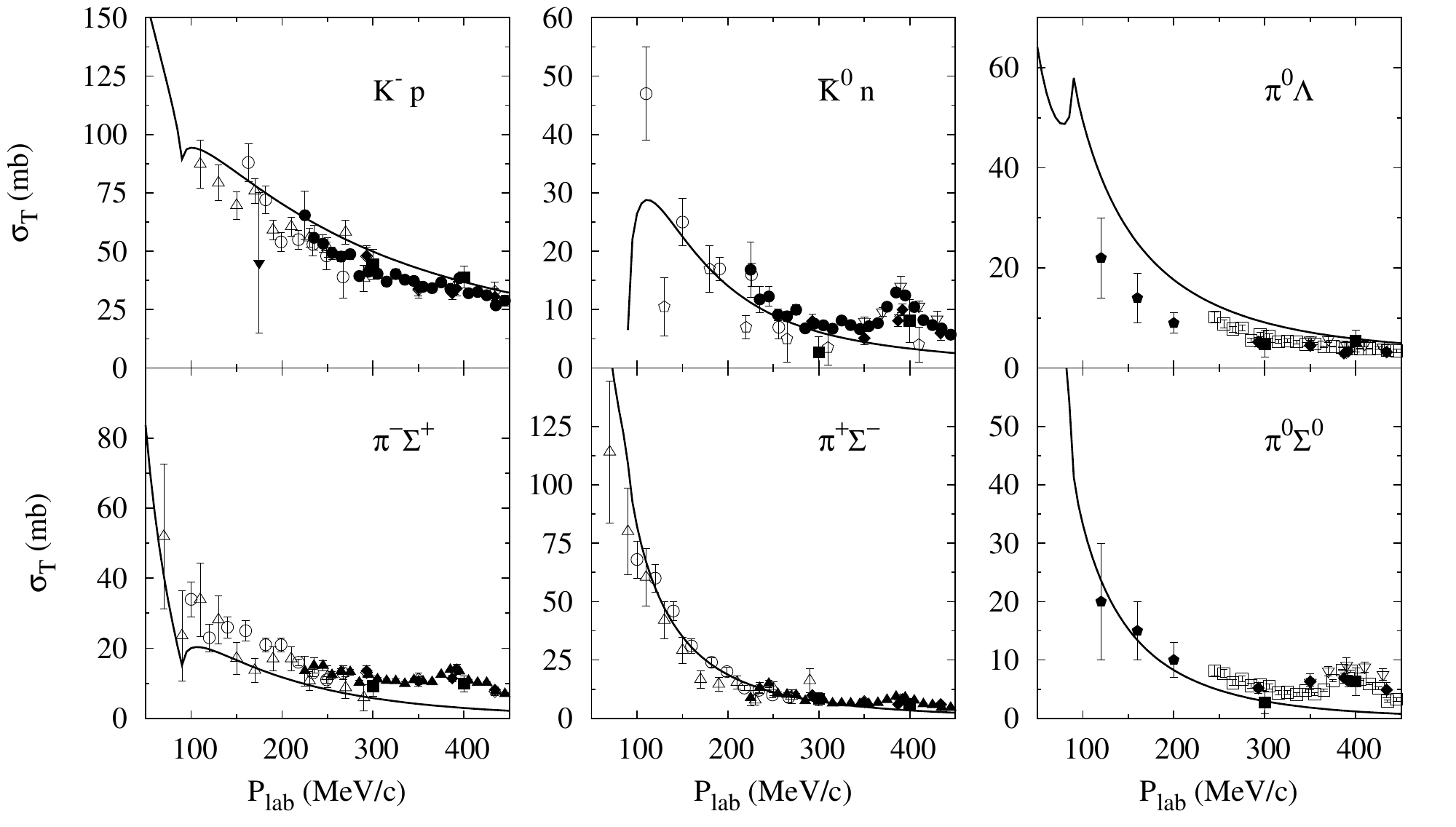}
\begin{minipage}[t]{16.5 cm}
\caption{Total cross sections of $K^{-}p$ to $K^{-}p$, $\bar K^{0} n$, $\pi^{0}\Lambda$, $\pi^{-}\Sigma^{+}$, $\pi^{+}\Sigma^{-}$ and $\pi^{0}\Sigma^{0}$. The solid lines denote the results of the chiral unitary model~\cite{Oset:2001cn,Jido:2002zk}.  The symbols stand for data points taken from 
   Refs.~\cite{Sakitt:1965kh,Ciborowski:1982et,Bangerter:1980px,Mast:1975pv,Mast:1974sx,Nordin:1961zz,Berley:1996zh,FerroLuzzi:1962zz,Watson:1963zz,Eberhard:1959zz,kim66}. 
 \label{fig:KpTotal}}
\end{minipage}
\end{center}
\end{figure}

In the following sections, we adopt the model of Ref.~\cite{Oset:2001cn} to study the structure of the $\Lambda(1405)$ resonance. The interaction kernel is chosen to be the WT term, so the free parameters of the model are the subtraction constants in the loop function. These are determined so as to reproduce the threshold branching ratios~\eqref{eq:branchingratio} as
\begin{align}
& a_{\bar{K} N} = -1.84, \phantom{{}_{\eta \Lambda}} \! \! 
a_{\pi \Sigma} = -2.00, \phantom{{}_{\eta \Sigma}} \! \! 
a_{\pi \Lambda} = -1.83, \nonumber \\ &
a_{\eta \Lambda} = -2.25, \phantom{{}_{\bar{K} N}} \! \! 
a_{\eta \Sigma} = -2.38, \phantom{{}_{\pi \Lambda}} \! \! 
a_{K \Xi} = -2.67 ,
\nonumber
\end{align}
with the renormalization scale $\mu=630$ MeV. With these subtraction constants, the branching ratios are obtained as~\cite{Jido:2002zk}
\begin{align}
	\gamma &=  2.30 ,\quad
	R_c= 0.618 ,\quad
	R_n= 0.257 .
	\nonumber
\end{align}
The low energy total cross sections of $K^-p$ scattering to various final states are well reproduced as shown in Fig.~\ref{fig:KpTotal}. The experimental data shows that the cross section of the double-charge exchange process $K^{-}p\to \pi^{+}\Sigma^{-}$ is larger than the cross sections to the other $\pi\Sigma$ states, which can also be observed from the ratio $\gamma>1$. This feature is well reproduced by the resummation of the chiral unitary approach, since the tree-level WT term for the double-charge exchange process $K^{-}p\to \pi^{+}\Sigma^{-}$ is zero. In addition, the imaginary part of the amplitude in the $\pi\Sigma$ diagonal channel shows a peak structure below the $\bar{K}N$ threshold, which can be interpreted as the $\Lambda(1405)$ resonance.

In this way, we briefly demonstrate that the simple model~\cite{Oset:2001cn} can reproduce the essential feature of the meson-baryon scattering amplitude. It should be nevertheless pointed out that the systematic studies including higher order terms are necessary to obtain more quantitative description of the amplitude~\cite{Borasoy:2004kk,Borasoy:2005ie,Oller:2005ig,Oller:2006jw,Bouzas:2010yb}.

\subsection{Pole structure of $\Lambda(1405)$}
\label{subsec:doublepole}

The resonance state is expressed in the scattering amplitude as a pole singularity in the second Riemann sheet of the complex energy plane. The properties of the resonance state can be extracted from the character of the pole term. In the chiral unitary approach, the scattering amplitude is obtained in an analytic form. Thus, it is easy to make analytic continuation of the scattering amplitude to the second Riemann sheet and to search for the poles there. The real and imaginary parts of the pole position represent the mass and half width of the resonance, respectively, $z_R=M_R-i\Gamma_R/2$, and the residues of the pole of the scattering matrix express the coupling nature of the resonance to the external channels. 

\begin{figure}
\begin{center}
  \includegraphics[width=0.5\textwidth,bb=0 0 550 450]{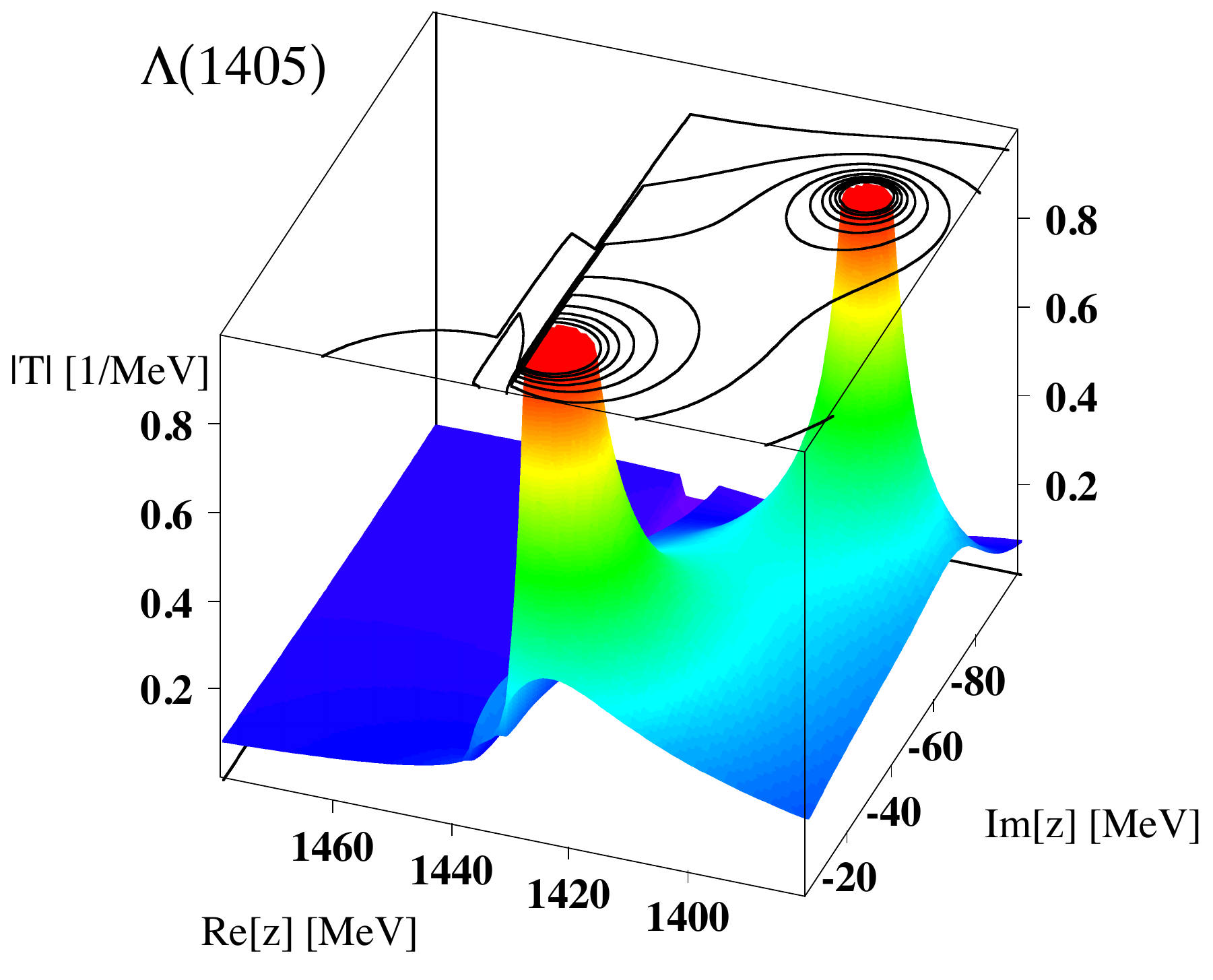} 
\begin{minipage}[t]{16.5 cm}
\caption{Absolute value of the scattering amplitude $|T|$ of the $\bar{K}N$ elastic channel in the second Riemann sheet of the complex energy $z$ plane. \label{fig:pole1405}}
\end{minipage}
\end{center}
\end{figure}

There are two poles in the scattering amplitude with $S=-1$ and $I=0$ around the $\Lambda(1405)$ energies~\cite{Oller:2000fj} and one pole around $\Lambda(1670)$~\cite{Oset:2001cn}. The positions of these poles are shown in Table~\ref{tab:poles}.  For $\Lambda(1405)$, one pole is located at higher energy around 1426 MeV with a narrower width 32 MeV, while the other is sitting at lower energy around 1390 MeV with a larger width 132 MeV. Both poles are found below the $\bar KN$ threshold, so the $\pi\Sigma$ state with $I=0$ is only the open channel for these poles. The pole positions of $\Lambda(1405)$ and their effect on the amplitude on the real axis are illustrated by plotting the absolute value of the scattering amplitude in the complex energy plane in Fig.~\ref{fig:pole1405}. The two poles are located definitely at different positions around the $\Lambda(1405)$ energy in the complex energy plane, while there is only one bump structure of the scattering amplitude on the real axis. Because the real parts of the two poles are close to each other, the contributions of these poles interfere in the amplitude on the real axis. As a consequence, what one can observe experimentally on the real axis is only a single resonance peak. Since the pole of the scattering amplitude can be interpreted as one resonance state, this finding indicates that the nominal $\Lambda(1405)$ is not a single resonance but a superposition of these two independent states with the same quantum numbers~\cite{Jido:2003cb}.

\begin{table}[b]
\begin{center}
\begin{minipage}[t]{16.5 cm}
\caption{The pole positions $z_R$ and the absolute values of the coupling strengths $|g_i|$ in the $S=-1$ and $I=0$ amplitude taken from Ref.~\cite{Jido:2003cb}.}
\label{tab:poles}
\end{minipage}
\begin{tabular}{l|c|c|c}
\hline
  & \multicolumn{2}{c|}{$\Lambda(1405)$} & $\Lambda(1670)$  \\
  \hline
 $z_{R}$ [MeV] & $1390 - 66i$ & $1426 - 16i$ & $1680 - 20i$  \\
 \hline
 $|g_i| (\pi \Sigma)$ & 2.9 & 1.5 & 0.27 \\
 $|g_i| ({\bar K} N)$ & 2.1 & 2.7 & 0.77 \\
 $|g_i| (\eta\Lambda)$ & 0.77 & 1.4 & 1.1 \\
 $|g_i| (K\Xi)$ & 0.61 & 0.35 & 3.5 \\
 \hline
 \end{tabular}
\end{center}
\end{table}%

The presence of the two poles around $\Lambda(1405)$ is more significant for experimental observations due to the coupling nature of these resonance states. The coupling constants of the resonance to the external channels can be extracted from the residues of the scattering matrix at the pole position as seen in Eq.~\eqref{eq:pole}. We show in Table~\ref{tab:poles} the coupling constants of these resonances to the meson-baryon channels obtained with the chiral unitary approach. From this table, it is found that these two poles have clearly different coupling nature to the meson-baryon channels; the higher energy pole dominantly couples to the $\bar{K}N$ channel, while the lower energy pole strongly couples to the $\pi\Sigma$ channel. The larger (smaller) imaginary part of the lower (higher) pole is the consequence of the stronger (weaker) $\pi\Sigma$ coupling. 

\begin{figure}[tb]
\begin{center}
  \includegraphics[width=0.7\textwidth,bb=0 0 450 130]{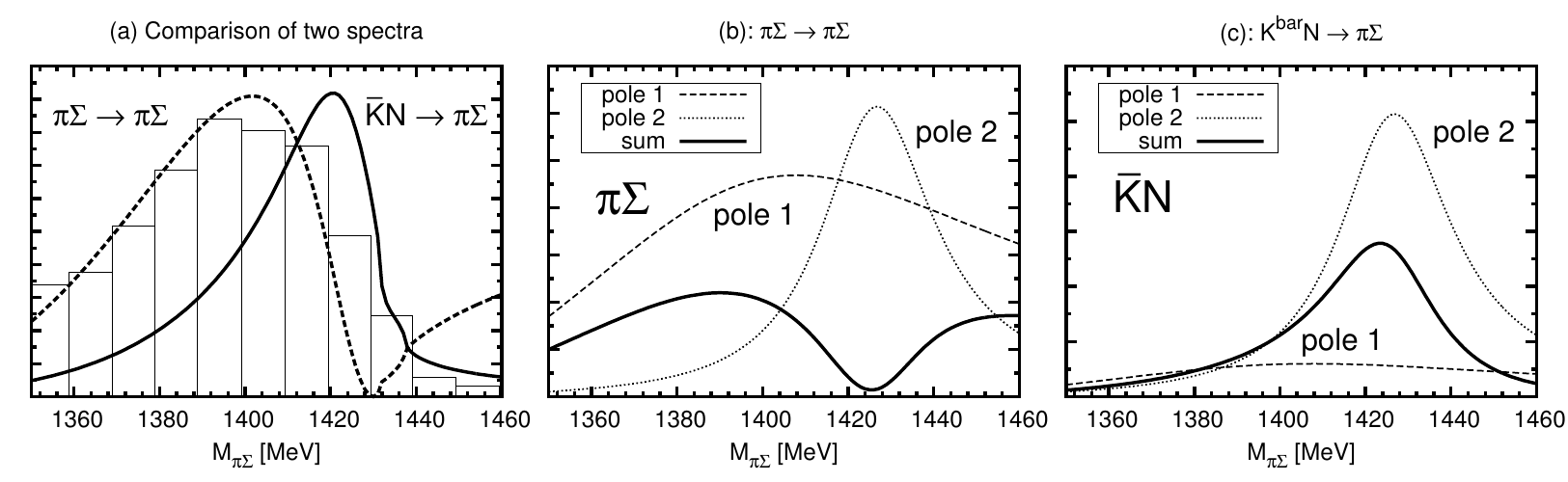}
\begin{minipage}[t]{16.5 cm}
\caption{$\pi\Sigma$ invariant mass spectra with $I=0$ in arbitrary units taken from Ref.~\cite{Jido:2010ag}. See also Ref.~\cite{Jido:2003cb}. 
(a): chiral unitary model calculation of the $\pi \Sigma$ invariant mass spectra of $\bar KN \rightarrow \pi\Sigma$ (solid line) and $\pi\Sigma \rightarrow \pi \Sigma$ (dashed line). The heights are adjusted. The histogram denotes an experimental spectrum of $\pi^{-}\Sigma^{+}$ channel given in Ref.~\cite{Hemingway:1985pz}.
(b) and (c): model calculations of the $\pi\Sigma$ spectra by two Breit-Wigner terms whose parameters are determined by the chiral unitary model. The dashed (dotted) line denotes the spectrum only with pole 1 (pole 2) term, while the solid line shows the spectrum calculated by coherent sum of pole 1 and 2. 
\label{fig:MSI0}}
\end{minipage}
\end{center}
\end{figure}

Due to the different coupling nature of these resonances, the shape of the $\Lambda(1405)$ spectrum can be different depending on the initial and final channels~\cite{Jido:2003cb}. In the $\bar KN \to \pi\Sigma$ amplitude, the initial $\bar KN$ channel gets more contribution from the higher pole with a larger weight. Consequently, the spectrum shape has a peak around 1420 MeV coming from the higher pole, as seen in Fig~\ref{fig:MSI0}(a). This is obviously different from the $\pi\Sigma \to \pi\Sigma$ spectrum which is largely affected by the lower pole. To examine the relevance of two poles, $\pi\Sigma$ invariant mass spectra are calculated in a simple model with two Breit-Wigner pole terms [first term of Eq.~\eqref{eq:pole}]. The pole parameters are determined by the chiral unitary model as in Table~\ref{tab:poles}. The spectra initiated by $\pi\Sigma$ and $\bar KN$ are shown in Fig.~\ref{fig:MSI0}(b) and (c), respectively. The dashed (dotted) line denotes the spectrum only with pole 1 (pole 2) term, while the solid line shows the spectrum calculated by coherent sum of pole 1 and 2. We find that the spectra by only the pole terms well reproduce the results of the full amplitude shown in Fig.~\ref{fig:MSI0}(a). Because the $\bar{K}N$ and $\pi\Sigma$ channels put different weights on the pole 1 and pole 2, the spectrum varies from one to the other. From this analysis, one finds that the $\bar{K}N \to \pi\Sigma$ spectrum in Fig.~\ref{fig:MSI0}(a) is affected mainly by the higher pole which strongly couples to the $\bar{K}N$ state, leading to higher peak position with narrower width, while the $\pi\Sigma \to \pi\Sigma$ spectrum puts more emphasis on the lower energy pole and makes the spectrum with the peak position around 1405 MeV with a broader width.

\begin{figure}[tb]
\begin{center}
  \includegraphics[width=0.8\textwidth,bb=0 0 700 400]{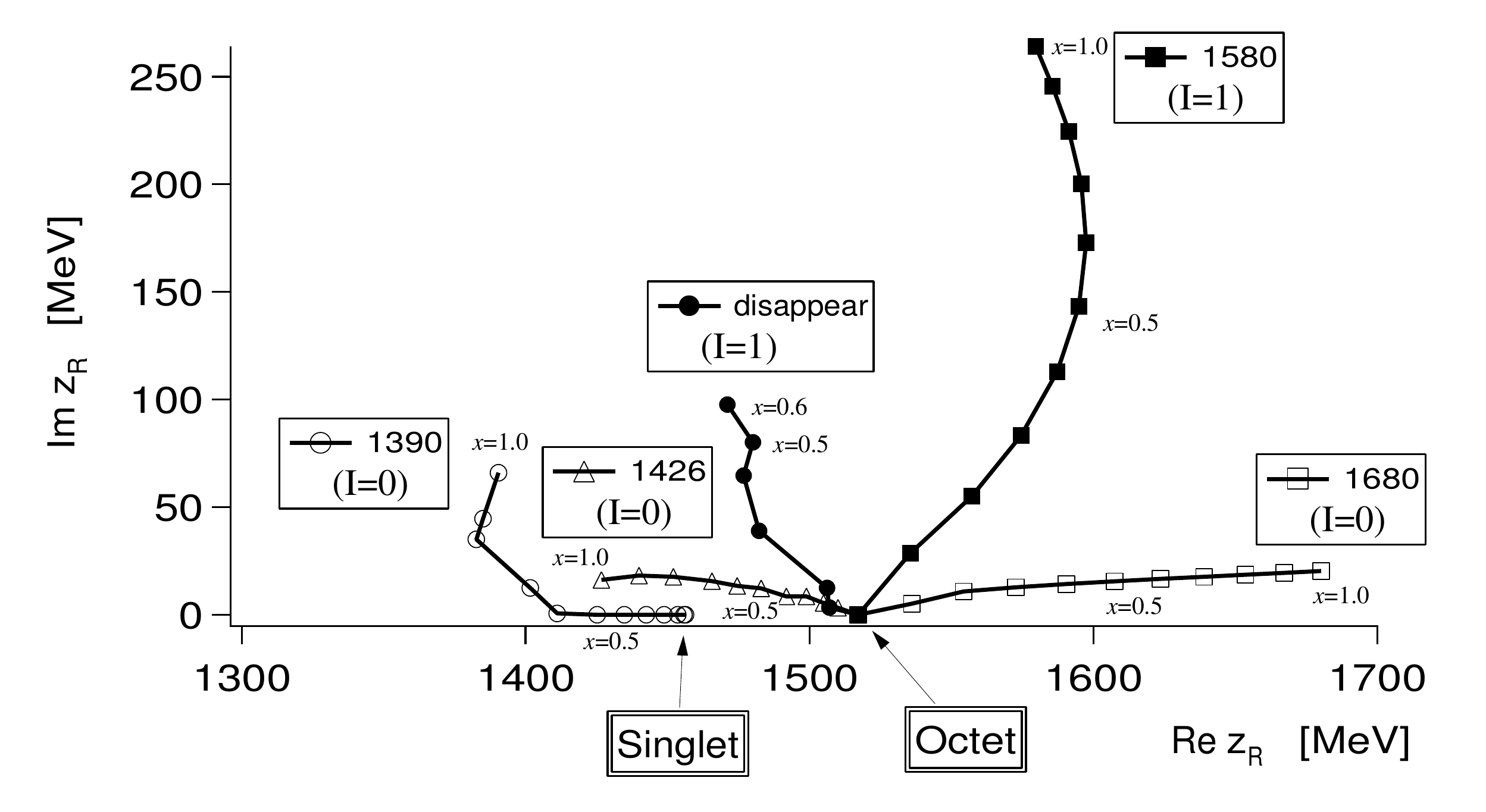}
\begin{minipage}[t]{16.5 cm}
\caption{Trajectories of the poles in the scattering amplitudes obtained by changing the SU(3) breaking parameter $x$ gradually. In the SU(3) symmetric limit ($x=0$), only two poles appear, one is for the singlet and the other for the octet. The symbols correspond to the step size $\delta x =0.1$. Figure is taken from Ref.~\cite{Jido:2003cb}. \label{fig:tracepole}}
\end{minipage}
\end{center}
\end{figure}

The reason of the presence of the two poles originates in two attractive components of the WT interaction~\cite{Jido:2003cb,Hyodo:2007jq}. In the SU(3) basis, the coupling strengths for the $S=-1$ and $I=0$ channels are obtained by Eq.~\eqref{eq:CijMB} with removing the Clebsch-Gordan coefficients as
\begin{equation}
    C_{ij}^{\text{SU(3)}} = 
    \begin{pmatrix}
    6 & & & \\
    & 3 & & \\
    & & 3 & \\
    & & & -2\\
    \end{pmatrix}
    \quad
    \begin{matrix}
    \bm{1} \\ 
    \bm{8} \\
    \bm{8}^{\prime} \\
    \bm{27}
    \end{matrix}
    , 
    \label{eq:CijSU3}
\end{equation}
which is a diagonal matrix because the WT interaction is SU(3) symmetric. In our convention, positive values of the coefficient $C$ represent attractive interaction. Thus, we observe that the meson-baryon interaction is attractive in the singlet and octet channels and altogether three bound states are produced in the SU(3) symmetric limit. Introducing the SU(3) breaking in particle masses, one of the octet channels develops to the $\Lambda(1670)$ resonance, while the singlet and the other octet states evolve to the two poles of $\Lambda(1405)$~\cite{Jido:2003cb}. The pole trajectories along with the SU(3) breaking effect are shown in Fig.~\ref{fig:tracepole}.

\begin{figure}[tb]
\begin{center}
  \includegraphics[width=0.5\textwidth,bb=0 0 300 230]{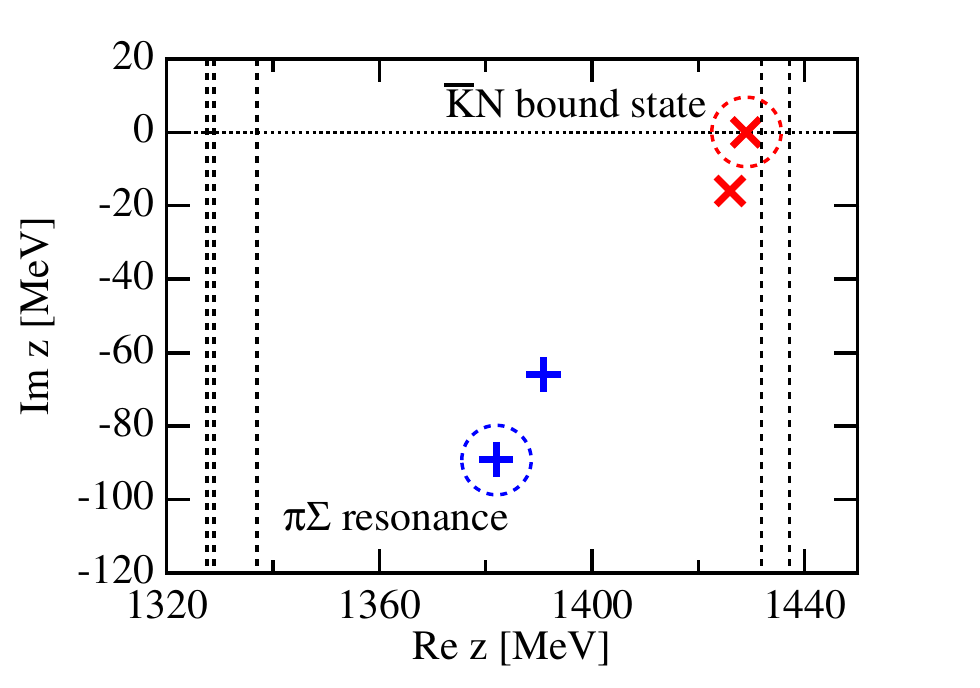}
\begin{minipage}[t]{16.5 cm}
\caption{Pole positions of the scattering amplitude for $\Lambda(1405)$ in the complex energy plane with the parameters in Ref.~\cite{Oset:2001cn}. Crosses enclosed by circles represent poles obtained by switching off the transition couplings. Vertical dashed lines indicate the meson-baryon threshold energies. \label{fig:polesingle}}
\end{minipage}
\end{center}
\end{figure}

Multiplying the SU(3) isoscalar factors as in Eq.~\eqref{eq:CijMB}, we obtain the coupling strengths to the $I=0$ channels in the isospin basis:
\begin{equation}
    C_{ij}^{\text{isospin}}
    = 
    \begin{pmatrix}
    3 & -\sqrt{\frac{3}{2}} & \frac{3}{\sqrt{2}} & 0  \\
      & 4 & 0 & \sqrt{\frac{3}{2}} \\
      &   & 0 & -\frac{3}{\sqrt{2}} \\
      &   &   & 3 \\
    \end{pmatrix}
    \hspace{-1cm}
    \begin{matrix}
    \phantom{\sqrt{\frac{3}{2}}}& \bar{K}N \\ 
    \phantom{\sqrt{\frac{3}{2}}}& \pi\Sigma \\
    \phantom{\frac{3}{\sqrt{2}}}& \eta\Lambda \\
    \phantom{3} & K\Xi
    \end{matrix}
    , 
    \label{eq:Cijisospin}
\end{equation}
where $C_{ji}=C_{ij}$. The diagonal components show that the $\bar{K}N$, $\pi\Sigma$, and $K\Xi$ channels are attractive. Since $\Lambda(1405)$ lies between the $\bar{K}N$ and $\pi\Sigma$ thresholds, it is natural to expect that the two poles originate in the attractive interactions in these channels~\cite{Hyodo:2007jq}. The attraction in the $K\Xi$ channel is responsible for the $\Lambda(1670)$ resonance. This picture can be verified by switching off the transition couplings in Eq.~\eqref{eq:Cijisospin}. In this case, the $\bar{K}N$ channel supports a bound state while a resonance is generated in the $\pi\Sigma$ channel as shown in Fig.~\ref{fig:polesingle}. Ever since it was predicted~\cite{Dalitz:1959dn,Dalitz:1960du}, $\Lambda(1405)$ is interpreted as a kind of Feshbach resonance: the $\bar{K}N$ quasi-bound state embedded in the $\pi\Sigma$ continuum. In the chiral unitary approach, the primary origin of $\Lambda(1405)$ is the $\bar{K}N$ attraction, while the $\pi\Sigma$ continuum is not a simple scattering state but has a strong correlation. As we will see in the later discussion, this structure is also important to determine the subthreshold extrapolation of the $\bar{K}N$ interaction.

It is interesting to note that the higher energy $\bar{K}N$ channel has stronger attraction to support a bound state, and the lower energy $\pi\Sigma$ channel shows a relatively weaker attraction, which is nevertheless strong enough to generate a resonance. Thus, the appearance of the two poles in this energy region is realized by the subtle balance of the two attractive forces. The combination of the two poles, a narrow quasi-bound state at higher energy and a broad resonance at lower energy, is also observed in the meson-meson scattering sector, as $\sigma$ and $f_0(980)$ resonances in $\pi\pi$-$\bar{K}K$ system. Since this system can be obtained by replacing $\Sigma$ by $\pi$ (both are $I=1$) and $N$ by $K$ (both are $I=1/2$) in $\pi\Sigma$-$\bar{K}N$, the universality of the leading order chiral interaction causes the similar pole structure. In this sense, the physics of the lower energy pole of $\Lambda(1405)$ is related to the $\sigma$ meson through chiral symmetry~\cite{Hyodo:2010jp}. In both cases the position of the lower pole is relatively far from the real axis and the determination of the precise location is difficult. Indeed, the studies with the NLO interaction~\cite{Borasoy:2004kk,Borasoy:2005ie,Oller:2005ig,Oller:2006jw,Borasoy:2006sr} reported the large uncertainty in the position of the lower pole of $\Lambda(1405)$ from the constraint of experimental data around the $\bar{K}N$ threshold. This is natural because the pole stems from the $\pi\Sigma$ attraction and thus is not sensitive to the observable around the $\bar{K}N$ threshold. The $\pi\Sigma$ threshold quantities (scattering length and effective range) were suggested for alternative constraints on the position of the lower pole~\cite{Ikeda:2011dx,Lambdacdecay,IkedaLattice}.

At this point, we mention the related works on the two-pole structure of $\Lambda(1405)$. The presence of the two poles in the $\bar{K}N$-$\pi\Sigma$ amplitude was first pointed out in Ref.~\cite{Fink:1990uk} which analyzed the low energy $S=-1$ meson-baryon scattering amplitude in the complex energy plane with several models. The potential models developed one pole for $\Lambda(1405)$, while the cloudy bag model of Ref.~\cite{Veit:1985jr} generated two poles. The mechanism is essentially the same with the chiral unitary approach, since the cloudy bag model respects chiral symmetry for the meson coupling to baryons. Schematically, the interaction in the cloudy bag model~\cite{Veit:1985jr} is given by
\begin{equation}
    V_{ij}^{\text{CBM}}
    =
    \bra{i}H_{c}\ket{j}
    +\sum_{B_0}
    \bra{i}H_{s}\ket{B_0}\frac{1}{E-M_0(B_0)}
    \bra{B_0}H_{s}\ket{j} ,
    \nonumber
\end{equation}
where the first term is the contact meson-baryon four point interaction of $H_c$ and the second term represents the bare state $B_0$ contribution with Yukawa coupling of $H_s$. Apart from the off-shell dependence, the contact term $\bra{i}H_{c}\ket{j}$ has the same structure with the $C_{ij}$ coefficient in the WT interaction, and the bare state which locates at higher energy~\cite{Fink:1990uk}. Therefore, around $\bar{K}N$ and $\pi\Sigma$ energy region, the amplitude is driven by the similar interaction kernel and hence results in the two-pole structure. Recently, in the framework of J\"ulich meson-exchange model, the pole structure of $\Lambda(1405)$ is studied~\cite{Haidenbauer:2010ch}. Using the parameters already determined in Ref.~\cite{MuellerGroeling:1990cw}, it is shown that  $\Lambda(1405)$ is accompanied by two poles as in the chiral unitary model~\cite{Haidenbauer:2010ch}. Although the interaction in the J\"ulich model has many ingredients, the vector-meson exchange part provides the similar structure with the WT interaction, which leads to the generation of two poles in the amplitude.

\subsection{Observation of the $\Lambda(1405)$ spectrum}
\label{subsec:spectrum}

Given the importance of the two-pole structure in various aspects, experimental verification of the structure is an urgent issue. The phenomenological consequence of the two-pole structure is reflected in the $\pi\Sigma$ spectrum to which $\Lambda(1405)$ decays exclusively. Since the two poles couple to $\bar{K}N$ and $\pi\Sigma$ with different weights, the shape of the mass spectrum depends on the initial channel as discussed in the previous section. Ideally, the double pole structure could be studied by comparing the $\bar{K}N \to\pi\Sigma$ and $\pi\Sigma\to \pi\Sigma$ amplitudes as in Fig.~\ref{fig:MSI0}. In practice, however, to access the relevant $\bar{K}N \to \pi\Sigma$ amplitude, the initial $\bar KN$ state must be off the mass shell since $\Lambda(1405)$ is located below the $\bar{K}N$ threshold. It is therefore necessary to utilize $\Lambda(1405)$ production experiments whose final state contains $\pi\Sigma$ and some additional particle(s) $X$ and to reconstruct the $\pi\Sigma$ invariant mass spectrum around 1405 MeV. Schematically the amplitude for the process $I \to X \pi \Sigma$ can be given by inserting intermediate states labeled by $i$ as
\begin{align}
    T_{I\to X\pi\Sigma}
    &=
    \sum_i T_{I\to Xi}\times T_{i\to \pi\Sigma} 
    \nonumber \\
    &\approx
    T_{I\to X\bar{K}N}\times T_{\bar{K}N\to \pi\Sigma}
    +T_{I\to X\pi\Sigma}\times T_{\pi\Sigma\to \pi\Sigma}  ,
    \label{eq:factorization}
\end{align}
where $T_{I\to Xi}$ represents the transition amplitude from the initial state $I$ to the intermediate state $i$ plus $X$, and $T_{i\to \pi\Sigma}$ stands for the amplitude involving $\Lambda(1405)$ created by the intermediate state $i$ and decaying to the final $\pi \Sigma$ state. The second line is obtained by assuming that the intermediate states are dominated by the $\bar KN$ and $\pi\Sigma$ channels, which are most relevant channels for the $\Lambda(1405)$ energies. For a normal resonance represented by a single pole dominating over nonresonant backgrounds, both $T_{\bar KN \to \pi\Sigma}$  and $T_{\pi\Sigma\to \pi\Sigma}$ would have the resonance peak at the same position and the shape of the $\pi\Sigma$ invariant mass should not depend on the choice of $I$ and $X$. On the other hand, if $\Lambda(1405)$ is represented by two states with different weights to the $\bar{K}N$ and $\pi\Sigma$ channels, then the $\pi\Sigma$ spectrum depends on the ratio $T_{I\to X\bar{K}N}/T_{I\to X\pi\Sigma}$, namely the reaction to produce $\Lambda(1405)$. In this way, the comparison of the $\pi\Sigma$ spectra in different reactions will give a hint of the pole structure of $\Lambda(1405)$. 

There have been several studies of the production processes of $\Lambda(1405)$ using the amplitude of the chiral unitary model for the final state interaction in Eq.~\eqref{eq:factorization}. Photoproduction of $\Lambda(1405)$ [$\gamma p \to K^+\Lambda(1405)\to K^+\pi\Sigma$]~\cite{Nacher:1998mi} and the radiative production with a kaon beam [$K^-p\to \gamma\Lambda(1405)\to \gamma\pi\Sigma$]~\cite{Nacher:1999ni} were studied with the amplitude obtained in Ref.~\cite{Oset:1998it}. The isospin interference shown in Eqs.~\eqref{eq:pipSigmam}-\eqref{eq:pi0Sigma0} was first pointed out in Ref.~\cite{Nacher:1998mi}. The result of the photoproduction experiment was reported in Ref.~\cite{Ahn:2003mv} which shows similar spectra of charged $\pi\Sigma$ states with those predicted in Ref.~\cite{Nacher:1998mi}. To understand the new data of the photoproduction of $\Lambda(1405)$ recently obtained in Refs.~\cite{Niiyama:2008rt,Moriya:2009mx}, more detailed study of each reaction process will be highly desired. The reaction study in connection with the double-pole structure has been started in Ref.~\cite{Hyodo:2003jw} where the $\pi^-$ induced reaction $\pi^-p\to K^0\Lambda(1405)\to K^0\pi\Sigma$ was investigated in relation with the experimental data in Ref.~\cite{Thomas:1973uh}. This reaction was found to put weight on the $\pi\Sigma\to\pi\Sigma$ amplitude and the importance of the $N(1710)$ production in the initial stage was pointed out. The $K^*$ photoproduction reaction $\gamma p \to K^{*+}\Lambda(1405)\to K\pi \pi\Sigma$ was studied in Ref.~\cite{Hyodo:2004vt}. In this case, the correlation of the polarization of the photon beam and the polarization of the final $K^*$ filters the parity of the exchanged particle, which reduces the uncertainty of the reaction mechanism. Reference~\cite{Magas:2005vu} analyzed the $K^-p\to\pi^0\pi^0\Sigma^0$ and found that this reaction is dominated by the $\bar{K}N\to\pi\Sigma$ amplitude. Since the higher energy pole is strongly couples to the $\bar{K}N$ channel, the peak of the $\pi\Sigma$ spectrum will be shifted upward in comparison with the $\pi^-$ induced reaction. This is indeed the case in the experimental data reported in Ref.~\cite{Prakhov:2004an}. The $pp\to pK^+\Lambda(1405)\to pK^+\pi^0\Sigma^0$ reaction was analyzed in Ref.~\cite{Geng:2007vm}. The calculated spectrum was consistent with that obtained by COSY~\cite{Zychor:2007gf}. 

\begin{figure}[tb]
\begin{center}
  \includegraphics[width=0.5\textwidth,bb=0 0 400 180]{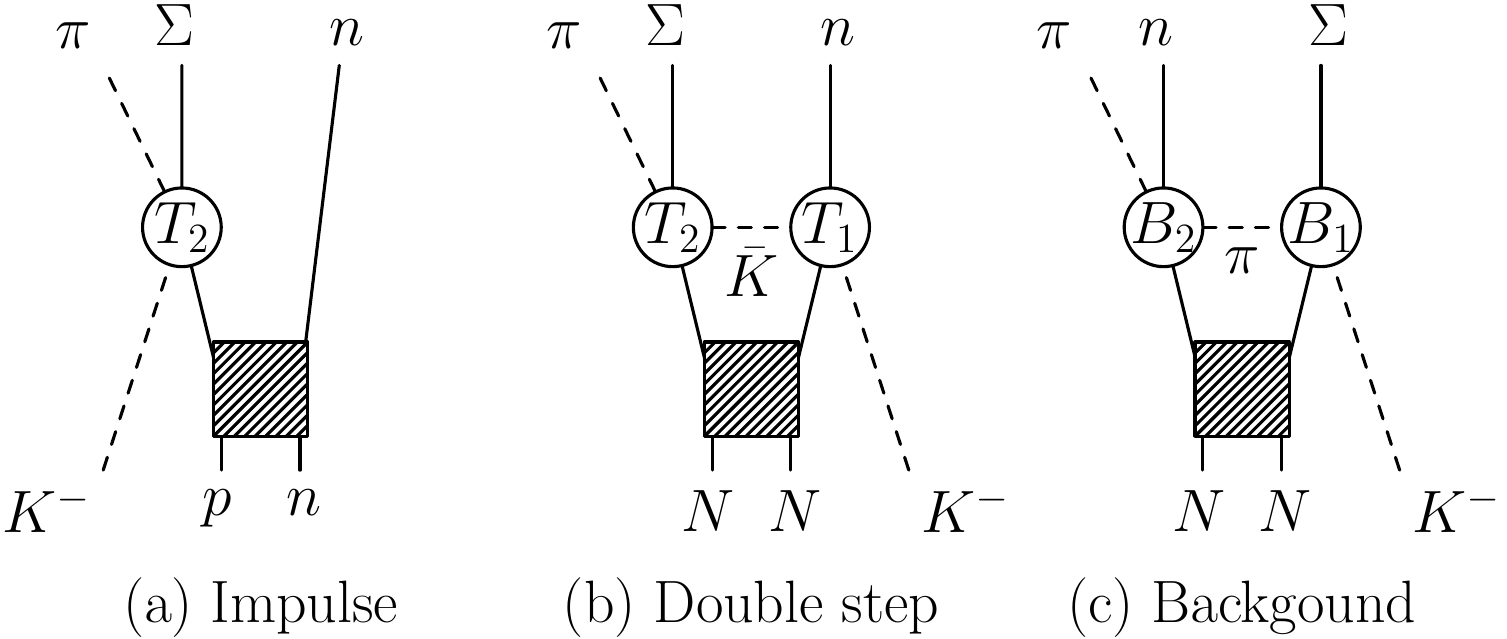}
\begin{minipage}[t]{16.5 cm}
\caption{Feynman diagrams for the $K^{-}d \to \pi\Sigma n$ reaction to observe the $\Lambda(1405)$ resonance in the $\pi\Sigma$ invariant mass spectrum~\cite{Jido:2009jf}. The diagrams (a) and (b) correspond to the impulse approximation and the double step process, respectively, in which $T_{2}$ is the scattering amplitude for $\bar KN \to \pi\Sigma$ involving the $\Lambda(1405)$ resonance and $T_{1}$ is for $\bar KN \to \bar KN$. The diagram (c) is a possible background contribution with one $\pi$ exchange, in which the final $\pi\Sigma$ does not come from $\Lambda(1405)$.
\label{fig:KdDia}}
\end{minipage}
\end{center}
\end{figure}

Among several reactions, it is constructive to find out the reaction in which $\Lambda(1405)$ is selectively initiated by the $\bar{K}N$ or $\pi\Sigma$ channel. Especially in the $\bar KN \to \pi\Sigma$ process the $\Lambda(1405)$ peak could be shifted upward from the nominal $\Lambda(1405)$ and observed around 1420 MeV. To observe the $\Lambda(1405)$ production initiated by the $\bar KN$ channel, the kaon induced reaction on the deuteron target,  $K^{-} d \to \Lambda(1405)n$, was proposed in Refs.~\cite{Jido:2009jf,Jido:2010rx}. In this reaction, the created $\Lambda(1405)$ decays to $\pi\Sigma$ with $I=0$ and is to be identified in the $\pi\Sigma$ invariant mass. The relevant diagrams for the production of $\Lambda(1405)$ are shown in Fig.~\ref{fig:KdDia}. The diagram (a) represents the $\Lambda(1405)$ production in the impulse approximation, while (b) is for the two-step process with $\bar K$ exchange. The amplitude $T_{1}$ ($T_{2}$) stands for the $s$-wave scattering process of $\bar K N \to \bar KN$ ($\bar K N \to \pi \Sigma$). In this reaction, $\Lambda(1405)$ is produced selectively by the $\bar KN$ channel, because the strangeness is brought into the system from the outside by the initial kaon. This is a different feature from the photo- or pion-production of $\Lambda(1405)$ in which the strangeness has to be created inside of the reaction system. The diagrams (a) and (b) have different kinematical characteristics. For energetic incident $K^{-}$ with several hundreds MeV/c momentum in the laboratory frame, the contribution of diagram (a) is found to be very small, since the scattered nucleon should be far off-shell to produce $\Lambda(1405)$ and the deuteron wavefunction has tiny component of such a nucleon. In contrast, in the double scattering diagram (b), the incident $K^{-}$ energy is taken away by the final neutron going to forward directions and the exchanged kaon can have a suitable energy to produce $\Lambda(1405)$ colliding with the other nucleon in the deuteron. Consequently, in the double (single) step process, $\Lambda(1405)$ is produced dominantly in backward (forward) directions in the laboratory frame. The double scattering process with a pion exchange shown in Fig.~\ref{fig:KdDia}(c) hardly contribute to the $\Lambda(1405)$ production, since $\Sigma$ and $\pi$ are emitted separately from the $B_{1}$ and $B_{2}$ amplitudes. Such diagrams give smooth backgrounds in the $\pi \Sigma$ invariant mass spectra~\cite{Yamagata}. In the theoretical calculation of this reaction with chiral unitary approach, the $\pi\Sigma$ mass spectrum with $I=0$ was found to have a peak around 1420 MeV instead of 1405 MeV as a consequence of $\Lambda(1405)$ produced purely by the $\bar KN$ channel~\cite{Jido:2009jf,Jido:2010rx}. Experimentally, this process was already observed in an old bubble chamber experiment at $K^{-}$ momenta between 686 and 844 MeV/c and it shows also the $\Lambda(1405)$ spectrum peaking at 1420 MeV in $K^{-}d \to \pi^{+}\Sigma^{-}n$~\cite{Braun:1977wd}. Further detailed experiments are certainly desirable, for instance,  at J-PARC or DA$\Phi$NE. It is also worth to mention that, in $K^{-}d \to Y\pi n$ reactions with $Y=\Sigma$ or $\Lambda$, because the resonance position of $\Lambda(1405)$ produced by $\bar KN$ will be at 1420 MeV with a narrower width, one could have a chance to observe $\Sigma(1385)$ and $\Lambda(1405)$ as separated peaks in the missing mass spectrum of the emitted neutron~\cite{Yamagata}.

Before closing the discussion of the $\Lambda(1405)$ spectrum, we point out several cautions on the analysis of the experimental data:
\begin{itemize}

\item \textit{Isospin}: The physical $\pi\Sigma$ final states ($\pi^+\Sigma^-, \pi^-\Sigma^+, \pi^0\Sigma^0$) are not the eigenstates of the total isospin $I$ as shown in Eqs.~\eqref{eq:pipSigmam}-\eqref{eq:pi0Sigma0}. The other isospin amplitudes as well as the interference terms also modify the $\pi\Sigma$ spectrum. To extract pure $I=0$ spectrum, all three $\pi\Sigma$ states must be detected in the same experiment. There is no \textit{a priori} reason to believe that $I=0$ contribution dominates the spectrum.

\item \textit{Nonresonant contribution}: The final state meson-baryon interaction $T_{i\to \pi\Sigma} $ always contains both the resonance pole term and the nonresonant background contribution. The interference of the pole term and background term also modify the spectrum from the simple Breit-Wigner form. There is no \textit{a priori} reason to believe that the resonant contribution dominates the spectrum. 

\item \textit{Model dependence}: In the theoretical calculation of the reaction process Eq.~\eqref{eq:factorization}, there is always model dependence of the initial state interaction $T_{I\to Xi}$. The final $\pi\Sigma$ spectrum does depend on the chosen reaction model. 

\end{itemize}
In view of these uncertainties, it is not an easy task to firmly establish/exclude the existence of two poles from the experiments. Nevertheless, it is important to collect the $\pi\Sigma$ spectra from different experiments to enrich the knowledge of $\Lambda(1405)$. From the theoretical side, it is desirable to find the reaction in which the model dependence of the initial process is as small as possible.

\subsection{$\Lambda(1405)$ as a meson-baryon molecular state}
\label{subsec:Lambda1405MBstate}

In the chiral unitary approach, the scattering amplitude is obtained by solving scattering equation with the interaction kernel derived in chiral perturbation theory. Since the model space consists of meson-baryon scattering states, resonances appearing in this model are naively expected to have the meson-baryon molecular structure. As argued in Section~\ref{subsec:origin}, however, it has been revealed that the component other than the meson-baryon molecule can contribute to the resonance through the subtraction constant in the loop function~\cite{Hyodo:2008xr}. Here we discuss the origin of baryon resonances $\Lambda(1405)$ and $N(1535)$, combining the scattering amplitude obtained in the phenomenological fitting to experimental data with the natural renormalization scheme introduced in Section~\ref{subsec:origin}.

The argument given in Section~\ref{subsec:origin} for the single-channel scattering can be generalized to coupled-channel systems straightforwardly. Following Ref.~\cite{Hyodo:2008xr}, we define the natural subtraction constants in the coupled-channel case by
\begin{align}
    G_i(\mu_m)
    = &
    0, \quad \mu_m= \text{min}\{M_i\}
    \label{eq:naturalcouple} ,
\end{align}
where $i$ is the channel index and $M_{i}$ is the baryon mass in the channel $i$. With this definition, both for the $\Lambda(1405)$ and $N(1535)$ channels, the natural subtraction constants are determined by $G_{i}(M_{N})=0$ and their values are shown in Table~\ref{tbl:subtractions}. Applying the derivation of Eq.~\eqref{eq:Vnatural} to the Weinberg-Tomozawa interaction in the matrix form (\ref{eq:WTtermswave}), we obtain the effective interaction in the natural renormalization scheme for the coupled-channel case as
\begin{align}
    V_{\text{natural}}(W;a_{{\rm natural},i})
    =&V_{\text{WT}}(W)\left(\bm{1}
    -A\cdot V_{\text{WT}}(W)\right)^{-1} ,
    \nonumber
\end{align}
where the diagonal matrix $A$ is given by the difference of the subtraction constants as
\begin{equation}
    A_{ij} = \frac{2M_i\Delta a_i}{16\pi^2}\delta_{ij},
    \quad
    \Delta a_i = a_{\text{pheno}, i}-a_{\text{natural}, i} .
    \nonumber
\end{equation}
The separation of the pole term from the original WT interaction as in Eq.~\eqref{eq:Vnatural} is possible only for the single-channel case. The pole positions of the effective interaction are obtained by 
\begin{equation}
    \det\left[\bm{1}
    -A\cdot V_{\text{WT}}(W)\right] =0 .
    \label{eq:polecond}
\end{equation}
Because the WT interaction is linear in $W$, Eq.~\eqref{eq:polecond} is an $n$th order algebraic equation of $W$. There are $n$ solutions which can be complex. This means that a pole in the effective interaction can have a finite imaginary part in the coupled-channel case.

To investigate $\Lambda(1405)$ and $N(1535)$, we take the phenomenological amplitudes for the $S=-1$ and $I=0$ channel and the $S=0$ and $I=1/2$ channel from Refs.~\cite{Oset:2001cn,Inoue:2001ip,Hyodo:2002pk,Hyodo:2003qa} in which the $\bar{K}N$ and $\pi N$ scattering observables were well reproduced up to the resonance energy region. The values of the phenomenological subtraction constants are shown in Table~\ref{tbl:subtractions} where the renormalization scale $\mu$ is chosen to be $\mu=M_{i} $ as in Ref.~\cite{Hyodo:2008xr}. Using the phenomenological and natural subtraction constants in Table~\ref{tbl:subtractions}, we calculate the pole positions in the effective interaction by Eq.~\eqref{eq:polecond}. The closest pole to the physical scattering region for the $N(1535)$ resonance was found at
\begin{align}
    z_{\text{eff}}^{N^*} &= 1693 \pm 37 i \text{ MeV}  .
    \nonumber
\end{align}
Reflecting a large difference between $a_{\text{pheno}, i}$ and $a_{\text{natural},i}$ for the $N(1535)$ channel as shown in Table~\ref{tbl:subtractions}, the effective mass is found at a relevant energy to the physical $N(1535)$ resonance. This result implies that the $N(1535)$ resonance is not purely molecular state but requires some additional component represented by the pole term at 1.7 GeV\footnote{Strictly speaking, we classify the origin of resonances into ``meson-baryon molecule" and ``something else". It is not possible to pin down the physical origin of the CDD pole contribution in the present approach, because we do not have further microscopic description than meson-baryon constituents in this approach. For baryon resonances, the most probable candidate for the CDD pole is the three-quark state, but any other components which are not included in the model space can contribute as the CDD pole.}. For $\Lambda(1405)$, the pole in the effective interaction was found at 
\begin{align}
    z_{\text{eff}}^{\Lambda^*} \sim 7.9 \text{ GeV},
    \nonumber
\end{align}
This implies that the possible contribution from the pole term for $\Lambda(1405)$ is out of the relevant energy region. In this way, we conclude that the dominant component of $\Lambda(1405)$ is the meson-baryon molecule.

\begin{table}[tbp]
    \centering
    \caption{Phenomenological and natural subtraction constants for the $\Lambda(1405)$ ($S=-1$ and $I=0$) and $N(1535)$ ($S=0$ and $I=1/2$) channels. The values of the phenomenological subtraction constants obtained in Ref.~\cite{Hyodo:2003qa} are shown with the regularization scale $\mu=M_i$. The natural subtraction constants are obtained by $G_{i}(M_{N})=0$. 
    \label{tbl:subtractions}}
    \vspace{0.3cm}
    \begin{tabular}{|crrrr||crrrr|} 
    \hline
    $S=-1$ & $\bar{K}N$ & $\pi\Sigma$ 
    & $\eta\Lambda$ & $K\Xi$  &
     $S=0$ & $\pi N$ & $\eta N$ & $K\Lambda$ & $K\Sigma$  \\
    \hline
    $a_{\text{pheno},i}$ & $-1.042$ & $-0.7228$ 
    & $-1.107$ & $-1.194$  &
        $a_{\text{pheno},i}$ & 1.509$\phantom{0}$ & $-0.2920$ 
    & 1.454 & $-2.813$  \\
    $a_{\text{natural},i}$ & $-1.150$ & $-0.6995$ 
    & $-1.212$ & $-1.138$  &    
    $a_{\text{natural},i}$ & $-0.3976$ & $-1.239\phantom{0}$
    & $-1.143$ & $-1.138$  \\
    \hline
    \end{tabular}
\end{table}%

\subsection{The $N_{c}$ behavior and quark structure}
\label{subsec:Nc}

In this section, we argue the quark structure of resonances from the response to the change of the number of colors $N_c$. It has been shown that the complicated dynamics of QCD can be simplified by taking the large $N_c$ limit with $g^2N_c$ being kept fixed~\cite{Hooft:1974jz,Witten:1979kh}. At sufficiently large $N_c$, the spectrum of the theory contains narrow $\bar{q}q$ mesons and glueballs, while all other components are less important in the structure of hadrons.

By utilizing this feature, a novel method to investigate the quark structure of resonances in chiral dynamics was developed in the meson sector~\cite{Pelaez:2003dy,Pelaez:2004xp,Pelaez:2006nj}. It is plausible that the ground state mesons are dominated by the $\bar{q}q$ structure. In this case, the $N_c$ scaling rule for the hadron quantities can be determined by the general rules. For instance, at leading order of the $1/N_{c}$ expansion, the meson mass $m$ and the decay constant $f$ are counted as 
\begin{equation}
    m \sim \mathcal{O}(1)
    , 
    \quad
    f \sim \mathcal{O}(N_c^{1/2}) .
    \label{eq:Nccounting}
\end{equation}
The $N_c$ counting of the low energy coefficients in the higher order terms are also known~\cite{Gasser:1985gg}. According to these counting rules, a quantity $X$ in the model is scaled as
\begin{equation}
    X\to X\left(\frac{N_c}{3}\right)^p 
    \quad \text{for} \quad X\sim \mathcal{O}(N_c^p) ,
    \nonumber
\end{equation}
to generalize the amplitude for arbitrary $N_c$. We then read off the $N_c$ scaling of the mass and width of the resonance from the trace of the pole position along with the $N_c$ variation. If the resonance is dominated by the $\bar{q}q$ structure, the mass and the width should follow the general counting rule:
\begin{equation}
    m_{\bar{q}q} \sim \mathcal{O}(1),\quad 
    \Gamma_{\bar{q}q} \sim \mathcal{O}(N_c^{-1}) .
    \label{eq:Ncscalingqbarq}
\end{equation}
Thus, by comparing the pole behavior with the general scaling rule with $N_c$, it is possible to estimate the dominance of the $\bar{q}q$ structure of the resonance. In Refs.~\cite{Pelaez:2003dy,Pelaez:2004xp,Pelaez:2006nj}, the $\sigma$ meson and the $\rho$ meson in the $\pi\pi$ scattering were analyzed. The $\rho$ meson pole follows the scaling law~\eqref{eq:Ncscalingqbarq}, while the $\sigma$ meson disappears for large $N_c$ in disagreement with the $\bar{q}q$ interpretation. In this way, the $\rho$ meson can be regarded as the $\bar{q}q$ resonance, whereas the $\sigma$ meson is not dominated by the $\bar{q}q$ structure. This approach was applied also to the axial vector resonances~\cite{Geng:2008ag}. It is remarkable that the $N_c$ scaling method makes it possible to investigate the quark structure of resonances  in the models with hadronic degrees of freedom.

In Refs.~\cite{Hyodo:2007np,Roca:2008kr}, the baron resonances $\Lambda(1405)$ and $\Lambda(1670)$ were studied in the same strategy. The baryon mass is of the order of $N_c$ and the mass splitting due to the flavor SU(3) breaking effect is counted as $\mathcal{O}(1)$~\cite{Dashen:1993jt,Dashen:1994qi}. So we introduce the $N_c$ dependence in the baryon masses as
\begin{equation}
    M_i(N_c) = M_0\frac{N_c}{3}+\delta_i ,
    \label{eq:Ncbaryonmass}
\end{equation}
with $M_0$ being the averaged mass of the octet baryons at $N_c=3$ and $\delta_i$ the symmetry breaking term for the baryon $i$. 

An important non-trivial issue in the baryonic sector is the $N_c$ dependence of the coupling strength $C_{ij}$ in the leading order WT term~\cite{Hyodo:2006yk,Hyodo:2006kg}. This $N_c$ dependence stems from the extension of the SU(3) representation of the flavor multiplet to arbitrary $N_c$ for three-flavor baryons~\cite{Karl:1985qy,Dulinski:1988yh,Piesciuk:2007xq}, which leads to the $N_c$ dependence of the coupling strength. The generalization of the flavor multiplet is applied only to the baryons ($q^{N_c}$ system for arbitrary $N_c$), not to the mesons ($\bar{q}q$ for arbitrary $N_c$), so the $N_c$ dependence in the $C_{ij}$ coefficients does not appear in the meson sector. The $N_c$ extension of the interaction strength in the SU(3) basis~\eqref{eq:CijSU3} is given by
\begin{equation}
    C_{ij}^{\text{SU(3)}}(N_c) = 
    \begin{pmatrix}
    \dfrac{9+N_c}{2} & & & \\
    & 3 & & \\
    & & 3 & \\
    & & & \dfrac{-1-N_c}{2}\\
    \end{pmatrix}
    , 
    \nonumber
\end{equation}
with the channels $\largeN{\bm{1}}$, $\largeN{\bm{8}}$, $\largeN{\bm{8}^{\prime}}$ and $\largeN{\bm{27}}$. The notation $\largeN{\bm{R}}$ represents the multiplet which reduces to $\bm{R}$ at $N_c=3$. It is worth noting that the strengths in the $\largeN{\bm{1}}$ and $\largeN{\bm{27}}$ channels have $\mathcal{O}(N_c)$ contributions. Since the WT interaction~\eqref{eq:WTtermswave} is also proportional to $f^{-2}\sim \mathcal{O}(N_c^{-1})$, the linear $N_c$ dependence of $C$ indicates that the interaction kernel $V\sim C/f^2\sim \mathcal{O}(1)$ remains finite in the large $N_c$ limit. In general, the chiral interaction is considered to vanish in the large $N_c$ limit because of the $f^{-2}$ factor. In the baryon sector, however, the $N_c$ dependence of the coupling strength leads to the nonvanishing interaction in the large $N_c$ limit. Furthermore, the attraction in the $\largeN{\bm{1}}$ channel is shown to develop a bound state in the large $N_c$ limit~\cite{Hyodo:2007np,Roca:2008kr}.

The coupling strength in the SU(3) basis can be transformed into the isospin basis by using Clebsch-Gordan coefficients at arbitrary $N_c$~\cite{Cohen:2004ki} as
\begin{equation}
    C_{ij}^{\text{isospin}}(N_c)
    = 
    \begin{pmatrix}
    \dfrac{N_c+3}{2} 
      \vspace{0.2cm}
      & -\dfrac{\sqrt{3N_c-3}}{2} 
      & \dfrac{\sqrt{3N_c+9}}{2} & 0  \\
      \vspace{0.2cm}
      & 4 & 0 & \dfrac{\sqrt{N_c+3}}{2} \\
      \vspace{0.2cm}
      &   & 0 & -\dfrac{\sqrt{9N_c-9}}{2} \\
      &   &   & \dfrac{-N_c+9}{2} \\
    \end{pmatrix}
    , 
    \label{eq:CijisospinNc}
\end{equation}
with the channels being $\bar{K}N$, $\pi\Sigma$, $\eta\Lambda$ and $K\Xi$\footnote{To be precise, we should write $\largeN{N}$ instead of $N$ for baryons, since the quantum numbers are not exactly the same with the nucleon at arbitrary $N_c$. For more detail, see Refs.~\cite{Piesciuk:2007xq,Cohen:2004ki}.}. The coupling strength in the $\bar{K}N$ channel is proportional to $N_c$, so the attractive interaction $V$ remains finite in the large $N_c$ limit. This attraction is strong enough to form a bound state. The $\bar{K}N$ bound state in the large $N_c$ limit may be related with the kaon bound state in the Skyrmion model~\cite{Callan:1985hy}. It is also interesting to note the diagonal $K\Xi$ channel. The interaction is attractive at $N_c=3$, which turns into repulsion for $N_c>9$. Namely, the sign of the interaction strength can be flipped by the $N_c$ dependence.

Using the coupling strength~\eqref{eq:CijisospinNc} together with the scaling rules in Eqs.~\eqref{eq:Nccounting} and \eqref{eq:Ncbaryonmass}, we construct the WT interaction $V$ at arbitrary $N_c$. To complete the argument, we also need to determine the $N_c$ dependence of the loop function $G$. The particle masses follow the scalings~\eqref{eq:Nccounting} and \eqref{eq:Ncbaryonmass}, so the cutoff parameters are to be specified. Given that $\Lambda(1405)$ is reproduced by the natural renormalization condition~\eqref{eq:anatural} in the dimensional regularization scheme, it is reasonable to adopt Eq.~\eqref{eq:anatural} at arbitrary $N_c$~\cite{Hyodo:2007np}. To appreciate physical energy scale of the regularization, three-momentum cutoff scheme is suitable. There are two possible scaling rules for the cutoff parameter, and it is shown that the qualitative conclusions in both cases are not different from the result with the dimensional regularization~\cite{Roca:2008kr}. 

\begin{figure}[tb]
\begin{center}
  \includegraphics[width=0.7\textwidth,bb=0 0 700 260]{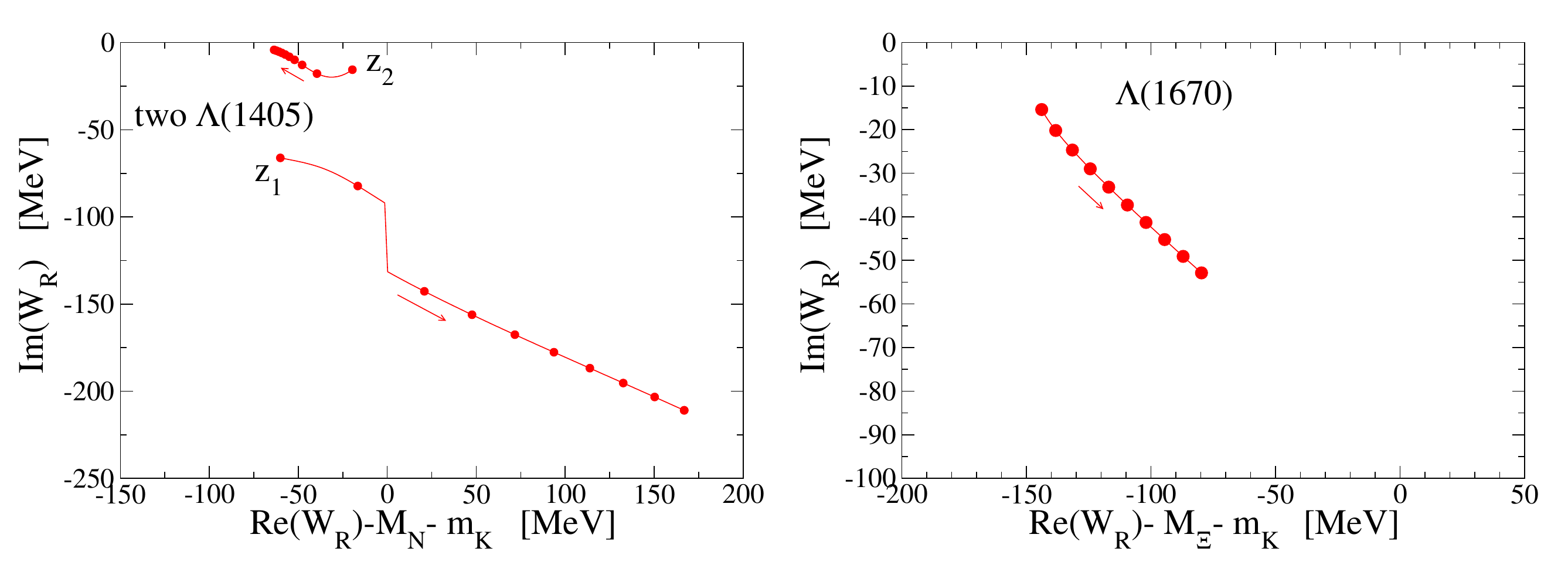}
\begin{minipage}[t]{16.5 cm}
\caption{Pole positions of the  $s$-wave meson-baryon scattering amplitudes with $I=0$ and $S=-1$ as functions of $N_c$ with the dimensional regularization scheme. The horizontal axis denotes the real part of the pole position measured from the $\bar KN$ threshold for $\Lambda(1405)$ and the $K\Xi$ threshold for $\Lambda(1670)$, and the vertical axis expresses the imaginary part of the pole position. The value of $N_{c}$ varies from 3 to 12 as indicated by arrows. Figures are taken from Ref.~\cite{Roca:2008kr}. \label{fig:Ncpole}}
\end{minipage}
\end{center}
\end{figure}

In this way we obtain the meson-baryon scattering amplitude as a function of $N_c$. General $N_c$ scaling rule of the mass $M$, the excitation energy $E$ and the width $\Gamma$ of the excited baryon with $N_c$ quarks (generalization of the three-quark state) is given by~\cite{Cohen:2003fv,Goity:2004pw}
\begin{equation}
    M_{q^{N_c}} \sim \mathcal{O}(N_c),\quad 
    E_{q^{N_c}} \sim \mathcal{O}(1),\quad 
    \Gamma_{q^{N_c}} \sim \mathcal{O}(1) .
    \label{eq:Ncscalingbaryon}
\end{equation}
We compare this general scaling rule with the behavior of the resonance poles in chiral unitary approach. The trajectories of the poles for $3\leq N_c\leq 12$ are shown in Fig.~\ref{fig:Ncpole}. To remove the trivial $N_c$ dependence of the baryon mass, the real parts for $\Lambda(1405)$ [$\Lambda(1670)$] are plotted as the excitation energy measured from the $\bar{K}N$ ($K\Xi$) threshold. It is clear that the imaginary parts of all the poles change drastically, in contrast to the leading behavior of the width of the $N_c$ quarks state~\eqref{eq:Ncscalingbaryon}. This result indicates that the three-quark component (at $N_c=3$) in $\Lambda(1405)$ and $\Lambda(1670)$ should be small. 

As $N_c$ is increased, the higher energy pole of $\Lambda(1405)$ ($z_2$ in Fig.~\ref{fig:Ncpole}) approaches the real axis and is becoming a bound state, while the imaginary parts of the other two poles grow and these resonances dissolve into continuum. This can be understood in the following way. In Section~\ref{subsec:doublepole}, we find that the higher (lower) pole of $\Lambda(1405)$ originates in the attraction in the $\bar{K}N$ ($\pi\Sigma$) channel, while $\Lambda(1670)$ is dominated by the $K\Xi$ component. The $N_c$ dependence of the relevant channels in Eq.~\eqref{eq:CijisospinNc} indicates that the $\bar{K}N$ attraction remains finite in the large $N_c$ limit, while the attractive components in the $\pi\Sigma$ and $K\Xi$ channels are gradually suppressed as $N_c$ increased. Thus, the only $\bar{K}N$ bound state remains in the large $N_c$ limit and the other states disappear. This speculation is confirmed by the properties of the residues of the poles; at large $N_c$, the would-be bound state is dominated by the $\bar{K}N$ component in isospin basis~\cite{Hyodo:2007np,Roca:2008kr}. Hence the $N_c$ behavior of these poles is indeed consistent with the expected limit of $N_c\to \infty$.

\subsection{Electromagnetic properties and the size of $\Lambda(1405)$}
\label{subsec:elemag}

The standard and traditional method to investigate the structure of a particle is to use  (virtual) external currents. For instance, the electromagnetic structure of the nucleon is known to high accuracy through the precise measurements of electron scattering and Compton scattering~\cite{ThomasWeise}. In the chiral unitary approach, $\Lambda(1405)$ is well described microscopically by meson and baryon degrees of freedom. Thus, by introducing couplings of the constituent hadrons to an external photon field, the electromagnetic properties of $\Lambda(1405)$ can be studied theoretically. A series of works on $\Lambda(1405)$ have covered the evaluation of the magnetic moments~\cite{Jido:2002yz}, the radiative decays~\cite{Geng:2007hz} and the helicity amplitudes~\cite{Doring:2010rd}. Among others, the form factors and mean squared radii provide the information of the intuitive ``size" of the particle. Knowing the size of the resonance is important for the estimation of the production yield in the heavy ion collisions in a coalescence model, which is shown to be a promising approach to extract the structure of hadrons~\cite{Cho:2010db}. The form factors of $\Lambda(1405)$ have recently been evaluated in chiral unitary approach~\cite{Sekihara:2008qk,Sekihara:2010uz}, which we discuss in this section.

The form factor of a particle is defined through the matrix element of the electromagnetic current $J_{\text{EM}}^{\mu}$. For the spin $1/2$ nucleon, the form factors are given by
\begin{equation}
    \bra{N(P^{\prime})}
    J_{\text{EM}}^{\mu}(x=0)
    \ket{N(P)}
    = 
    \bar{u}(P^{\prime})
    \left[
    \gamma^{\mu}F_1(Q^2)
    +i\sigma^{\mu\nu}\frac{q_{\nu}}{2M}
    F_2(Q^2)
    \right]
    u(P)
    , 
    \label{eq:Formfactor}
\end{equation}
where the momentum transfer is $Q^2=-(P^{\prime}-P)^2$. The Dirac ($F_1$) and Pauli ($F_2$) form factors are related to the electric ($F_E$) and magnetic ($F_M$) form factors as 
\begin{equation}
    F_E(Q^2)
    = F_1(Q^2)
    -\frac{Q^2}{4M^2}F_2(Q^2)
    , 
    \quad
    F_M(Q^2)
    = F_1(Q^2)
    +F_2(Q^2) .
    \nonumber
\end{equation}
These form factors are normalized at $Q^2=0$ as $F_E(0)=Q_N$ and $F_M(0)=\mu_N$, with the charge $Q_N$ and the magnetic moment $\mu_N$ of the nucleon $N$. In the Breit frame, Fourier transformation of the electric (magnetic) form factor represents the charge (magnetization) density distribution. In this way, we can visualize the spatial structure of the nucleon with the electromagnetic probe.

Before going to $\Lambda(1405)$, we should note that hadron resonances are unstable and decay \textit{via} strong interaction. We have already seen a consequence of the resonance nature of $\Lambda(1405)$ in Section~\ref{subsec:doublepole} where the coupling constant $g_i$ is obtained as a complex number. In the same way, for a resonance state, its form factors and mean squared radii have finite imaginary parts, whose physical interpretation is not straightforward. To start with, we should establish a reasonable definition of the form factors for an unstable particle. The detailed account on the generalization of the form factors to a resonance state is given in Ref.~\cite{Sekihara:2010uz} where the form factor reduces to the ordinary one for the stable particle~\eqref{eq:Formfactor} when the width of the resonance vanishes.

\begin{figure}[tb]
\begin{center}
  \includegraphics[width=0.7\textwidth,bb=0 0 650 120]{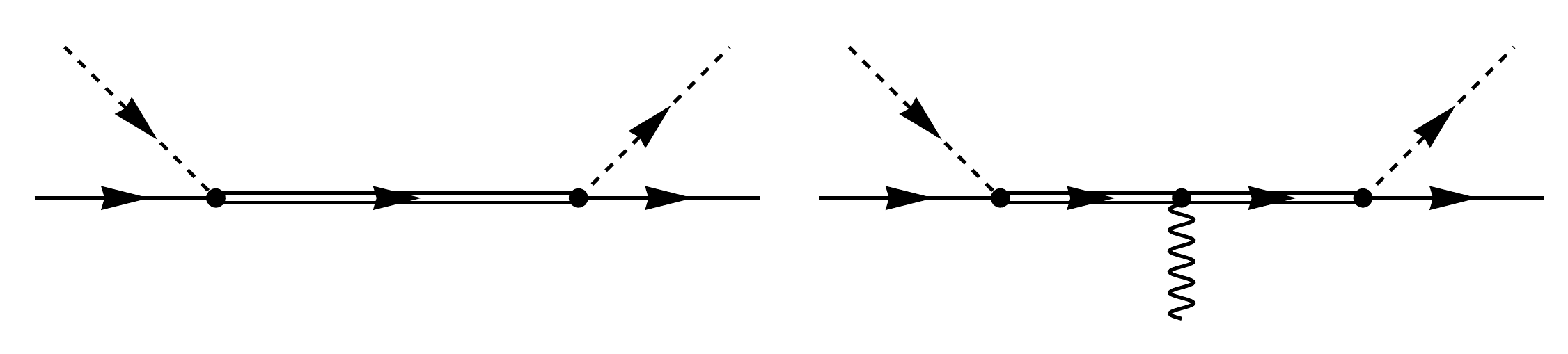}
\begin{minipage}[t]{16.5 cm}
\caption{Feynman diagrams for the meson-baryon amplitude $T_{ij}$ (left) and the current coupled meson-baryon amplitude $T_{\gamma ij}$ (right) at the pole position. The double lines correspond to the resonance state. Figures are taken from Ref.~\cite{Sekihara:2008qk}. \label{fig:diagram}}
\end{minipage}
\end{center}
\end{figure}

To evaluate the form factors, we consider two diagrams shown in Fig.~\ref{fig:diagram}. In Section~\ref{subsec:origin}, we have discussed that the resonance state can be expressed as the pole singularity of the scattering amplitude in the second Riemann sheet of the complex energy plane. When the energy $z$ is chosen at the pole position $z_R=M_R-i\Gamma_R/2$, we obtain the pole term contribution as
\begin{equation}
   \left.  T_{ij} (z) \right|_{z \to z_{R}} = 
    \frac{g_{i}g_{j}}{z -  z_{R}}  , 
   \label{eq:Tmat}
\end{equation}
which corresponds to the diagram shown in the left panel of Fig.~\ref{fig:diagram}. Next we consider the meson-baryon scattering amplitude to which an external photon is coupled as an analogy with Eq.~\eqref{eq:Formfactor}. We denote this amplitude as $T_{\gamma ij}$, whose most singular structure at the pole position is given by
\begin{equation}
   T_{\gamma ij} ( z^{\prime}, \, z; \, Q^{2} ) 
   |_{z \to z_{R},z^{\prime} \to z_{R}} 
   = - g_{i} \frac{1}{z^{\prime} - z_{R}} F (Q^{2}) 
   \frac{1}{z - z_{R}} g_{j} ,
   \label{eq:Tgammamat}
\end{equation}
where $z=\sqrt{P^2}$ and $z^{\prime}=\sqrt{P^{\prime 2}}$.
This corresponds to the diagram shown in the right panel of Fig.~\ref{fig:diagram}. It is shown that the photon-resonance vertex is given by the form factor $F(Q^2)$~\cite{Sekihara:2010uz}.
Combining Eqs.~\eqref{eq:Tmat} and \eqref{eq:Tgammamat}, we can extract the form factor as
\begin{align}
   F (Q^{2}) 
   & =  
   - \frac{(z^{\prime}-z_R)T_{\gamma ij}
   (z^{\prime}, \, z; \, Q^{2} )}{T_{ij} (z)} 
   \Bigg |_{z \to z_{R},z^{\prime} \to z_{R}}
    .
   \label{eq:Resscheme}
\end{align}
Since we have the meson-baryon amplitude $T_{ij}$ which describes $\Lambda(1405)$ microscopically well in the chiral unitary approach, our task is to calculate the current-coupled scattering amplitude $T_{\gamma ij}$. 

\begin{figure}[tb]
\begin{center}
    \includegraphics[width=0.64\textwidth,bb=0 0 600 220]{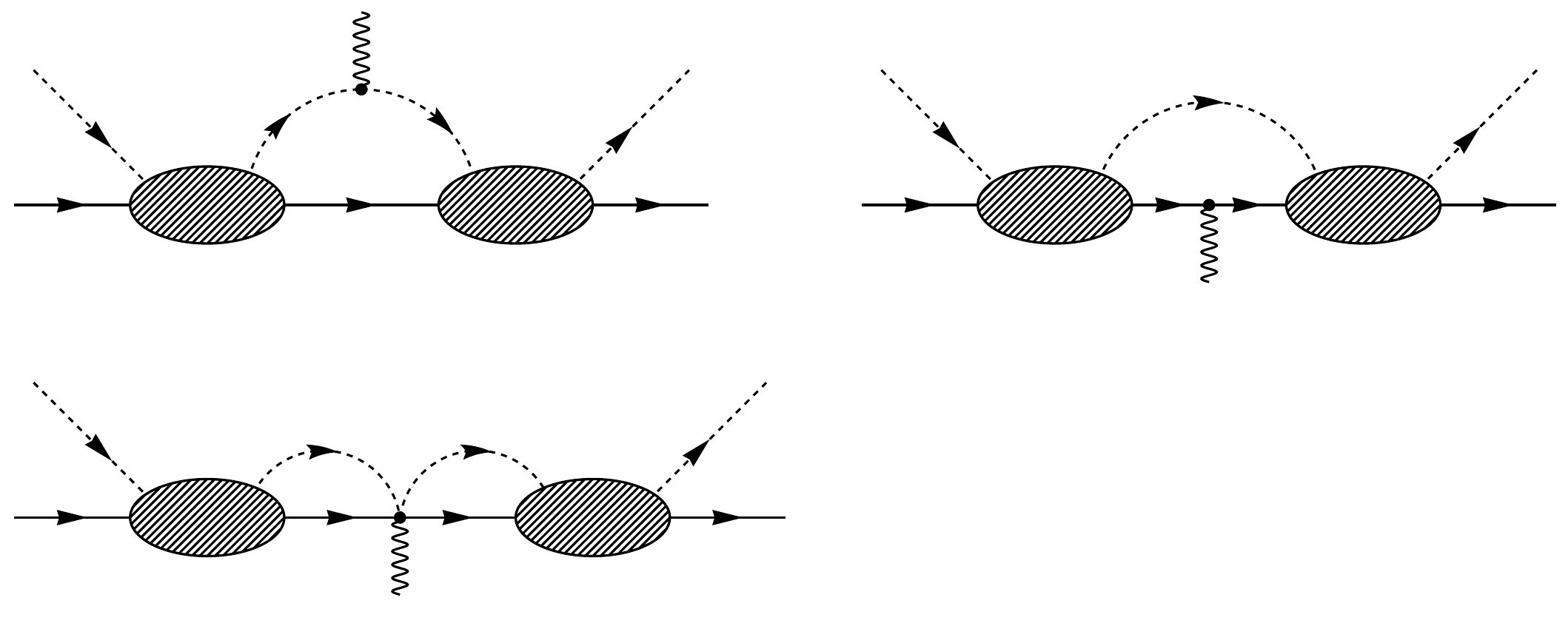}
\begin{minipage}[t]{16.5 cm}
\caption{Feynman diagrams for the current coupled meson-baryon amplitude $T_{\gamma ij}$. Figures are taken from Ref.~\cite{Sekihara:2008qk}. \label{fig:Tgamma}}
\end{minipage}
\end{center}
\end{figure}

In general, gauge invariance of the amplitude is the most fundamental constraint for the process including gauge fields. Gauge invariance in the resummation framework like the chiral unitary approach was studied in Ref.~\cite{Borasoy:2005zg}. Out of ten types of diagrams discussed in Ref.~\cite{Borasoy:2005zg}, we pick up the three diagrams shown in Fig.~\ref{fig:Tgamma} for $T_{\gamma ij}$, which are relevant to the evaluation of the form factor in Eq.~\eqref{eq:Resscheme}, because these diagrams have double-pole singularity of the resonance. In these diagrams, the photon field is attached to the constituent mesons, baryons and vertices. These photon vertices are straightforwardly given by the minimal coupling scheme together with the anomalous magnetic couplings in chiral perturbation theory\footnote{To keep gauge invariance, the minimal coupling for the meson-baryon vertices should be carefully derived according to the $s$-wave part of the interaction~\eqref{eq:WTtermswave} which is used to calculate the meson-baryon amplitude~\cite{Sekihara:2008qk,Sekihara:2010uz}.}. Therefore, in this way, the internal structure of $\Lambda(1405)$ can be probed by the external current in the chiral unitary approach. 

We calculate the $T_{\gamma ij}$ amplitude in the Breit frame and evaluate the form factors by Eq.~\eqref{eq:Resscheme}. The density distributions are obtained through Fourier transformation~\cite{Sekihara:2010uz}. Here we discuss the higher energy pole of $\Lambda(1405)$ since it gives the dominant contribution to the $\bar{K}N$ amplitude. The electric density distribution indicates that the positive (negative) charge is distributed in the inward (outward) region of $\Lambda(1405)$. This can be understood as the existence of the lighter $K^-$ in the peripheral region with the heavier proton at the central core. In addition, the decay of $\Lambda(1405)$ into $\pi\Sigma$ channel through the photon coupling is expressed by the characteristic oscillation pattern in the density distribution.

The size of $\Lambda(1405)$ can be estimated by the electric and magnetic mean squared radii~\cite{Sekihara:2010uz}
\begin{align}
   \langle r^2 \rangle_E
   & = -0.157 + 0.238i \text{ fm}^2  ,
   \quad
   \langle r^2 \rangle_M
   = 1.138 - 0.343i \text{ fm}^2 \nonumber .
\end{align}
As a representative chargeless three-quark state, we consider the neutron which has $\langle r^2 \rangle_E\sim -0.12$ fm$^2$ and $\langle r^2 \rangle_M\sim 0.66$ fm$^2$. The absolute values $|\langle r^2 \rangle_E|\sim 0.29$ fm$^2$ and $|\langle r^2 \rangle_M|\sim 1.19$ fm$^2$ for $\Lambda(1405)$ are about two times larger than those of the neutron. This result is consistent with the meson-baryon molecular picture of $\Lambda(1405)$, rather than the three-quark structure which is presumably a compact object. By assigning appropriate charges for mesons and baryons, we can construct external currents which probe the baryon number and strangeness of the system. The analysis with the baryon number and strangeness currents reinforces the meson-baryon molecular structure of $\Lambda(1405)$.

To gain physical insight into the obtained form factors, a single-channel model with a bound state is also studied. In the case of the bound state, the form factors are obtained as real numbers. The mean squared distance is calculated as a function of the binding energy $B_E$. The result for small $B_E$ is in good agreement with the behavior expected from the nonrelativistic quantum mechanics for a weakly binding system, $\langle x^2\rangle_{\text{NR}}\sim 1/4\mu B_E$ with $\mu$ being the reduced mass. This result supports that the present definition of the form factors~\eqref{eq:Resscheme} indeed reflects the size of the particle.

\subsection{Application of chiral unitary approach to other systems}
\label{subsec:application}

Since the chiral interaction given in Section~\ref{subsec:interaction} dictates the dynamics of other strangeness sectors, they can be studied by the same technique. In the $S=0$ sector, experimental data is available from the $\pi N$ scattering which contains two $S_{11}$ resonances, $N(1535)$ and $N(1650)$. The $N(1535)$ resonance was studied using the pseudo potential approach~\cite{Kaiser:1995cy}. The $S=0$ scattering amplitude, including the $N(1650)$ resonances, was further studied in the chiral unitary model~\cite{Inoue:2001ip} and in the relativistic BS equation~\cite{Nieves:2001wt}. In spite of the absence of the explicit three-body $\pi\pi N$ channel, the description of the phase shift and cross sections was excellent. In Refs.~\cite{Hyodo:2002pk,Hyodo:2003qa}, the $S=0$ and $S=-1$ sectors were studied together, from the viewpoint of the flavor SU(3) breaking effect. One of the motivations of this study was the large difference in the values of the subtraction constants in the $S=0$ and $S=-1$ sectors in Refs.~\cite{Oset:2001cn,Inoue:2001ip}. This point is related to the origin of resonances as shown in Section~\ref{subsec:origin}. The $p$-wave contribution in the $\bar KN$ scattering was discussed by introducing an explicit pole term for $\Sigma(1385)$ in Ref.~\cite{Jido:2002zk}. Turning to the $S=-2$ meson-baryon scattering, there is no experimental data. It was however shown in Ref.~\cite{Ramos:2002xh} that a $\Xi$ resonance was generated around 1.6 GeV with the natural magnitude of the subtraction constants suggested in Ref.~\cite{Oller:2000fj}. Based on this result, Ref.~\cite{Ramos:2002xh} argued that the quantum number of the observed $\Xi(1620)$ state was $1/2^-$. In this way, chiral unitary approach generates the low-lying negative parity baryon resonances in $S=0$, $S=-1$, and $S=-2$ sectors. The SU(3) extrapolation of all these states was performed in Ref.~\cite{Garcia-Recio:2003ks} with the investigation of the quark mass dependence.

Since the WT interaction is universal for the low energy NG boson scattering, the target hadron is not restricted to the octet baryons. In fact, the scattering of the NG boson off the NG boson has been studied by the similar philosophy with the chiral unitary approach~\cite{Truong:1988zp,Dobado:1990qm,Dobado:1993ha,GomezNicola:2001as}. One of the most elaborate works showed that the scalar meson nonet ($\sigma,\kappa,f_0,a_0$) and the vector mesons ($\rho,K^*,\phi$) were generated in the $s$-wave and $p$-wave nonperturbative amplitudes, respectively~\cite{Oller:1997ng,Oller:1998hw}. This result however does not directly indicate that all these mesons have the mesonic molecular structure, since the seed of vector mesons can be hidden in the low energy constant of the NLO terms in chiral perturbation theory~\cite{Ecker:1988te}. Baryon resonances with $J^P=3/2^-$ can be described, by choosing the decuplet baryons as target hadrons. In Ref.~\cite{Kolomeitsev:2003kt}, the resonances were identified by the peak of the speed plot, and the pole positions were searched for in Ref.~\cite{Sarkar:2004jh}. In contrast to the cases mentioned above, the decuplet baryons have finite widths by the strong interaction, and the $3/2^-$ resonance can also couple to the $d$-wave scattering states of the NG boson ($0^-$) and the ground state baryon ($1/2^-$). These effects may be important for the quantitative discussion of the specific resonance~\cite{Roca:2006sz,Hyodo:2006uw}. In the same way, the axial vector meson resonances were studied in the NG boson scattering off the vector mesons by the speed plot~\cite{Lutz:2003fm} and by the pole position~\cite{Roca:2005nm}. In the heavy quark sector, heavy baryons~\cite{Lutz:2003jw} and heavy mesons~\cite{Kolomeitsev:2003ac} were searched for in the speed plot by regarding the ground state heavy hadrons as target. We summarize the resonances obtained in a series of works mentioned above in Table~\ref{tab:resonances}. 

In addition to the reproduction of these resonances, various new states are predicted by the same mechanism. It is also instructive to notice the absence of the manifestly exotic states which require more than three valence quarks. Actually, there were some attempts to search for the exotic states in the chiral unitary approach~\cite{Lutz:2003jw,Kolomeitsev:2003ac,Sarkar:2004sc} but the possible signal was not clear and  very sensitive to the input parameters. The difficulty of generating exotic hadrons can be traced back to the group theoretical structure of the Weinberg-Tomozawa interaction~\cite{Hyodo:2006yk,Hyodo:2006kg}. For a general target hadron with SU(3) representation $T$, the coupling strength~\eqref{eq:CijMB} in the SU(3) basis is given by
\begin{align}
	C_{\alpha}
	=C_2(T)-C_2(\alpha)+3 ,
	\label{eq:CSU3}
\end{align}
where $\alpha$ is the representation of the combined channel and hence of the resonance state. In our convention, positive (negative) sign of $C_{\alpha}$ corresponds to the attractive (repulsive) interaction. An exotic hadron in general belongs to a larger SU(3) multiplet $\alpha$ than the ordinary hadrons. Considering the Casimir factor $C_{2}(\alpha)$ has negative sign in Eq.~\eqref{eq:CSU3}, we notice that the exotic channel is less advantageous to have an attractive interaction than the ordinary channels. To pursue this idea quantitatively, we should specify the exotic channel in which $\alpha$ is more exotic than $T$. For this purpose, the ``exoticness" quantum number $E$ is introduced as the number of valence quark-antiquark pairs to compose the given flavor multiplet $\alpha=[p,q]$ with the baryon number $B$, which is given by~\cite{Hyodo:2006yk,Hyodo:2006kg}
\begin{equation}
    E=\epsilon\theta(\epsilon)+\nu\theta(\nu) ,
    \quad
    \epsilon \equiv
    \dfrac{p+2q}{3}-B, \quad 
    \nu\equiv\dfrac{p-q}{3}-B
    \nonumber .
\end{equation}
We define the exotic channel such that the exoticness of $\alpha$ is larger than the exoticness of $T$. It is group theoretically shown that the interaction strength in the exotic channel is in most cases repulsive, and attractive interaction is only possible for the value
\begin{align}
	C_{\text{exotic}}=1 .
	\nonumber
\end{align}
This strength was shown to be not strong enough to generate a bound state~\cite{Hyodo:2006yk,Hyodo:2006kg}. Thus, it is difficult to generate the exotic hadrons in chiral dynamics.

\begin{table}
\begin{center}
\begin{minipage}[t]{16.5 cm}
\caption{Hadron resonances described in chiral unitary approaches. The results from typical works are shown. Note that different works prefer different assignments of physical resonances.
}
\label{tab:resonances}
\end{minipage}
\begin{tabular}{c|c|c|l}
\hline
 Target       & Ref. & $J^P$   & Resonances  \\
 \hline
 $J^P=1/2^+$ $\bm{8}$ baryon & & $1/2^-$ & 
   $\Lambda(1405)$, $\Lambda(1650)$, 
   $N(1535)$, $N(1670)$ \\
   & & & 
   $\Sigma(1620)$, $\Xi(1620)$      \\ 
   \hline
 $J^P=3/2^+$ $\bm{10}$ baryon &\cite{Sarkar:2004jh} & $3/2^-$ &
   $\Lambda(1520)$, $\Sigma(1670)$, $\Sigma(1940)$,   \\
   & & & 
   $N(1520)$, $\Xi(1820)$, $\Omega(2250)$     \\ 
   \hline
 heavy $\bm{3},\bm{6}$ baryon & \cite{Lutz:2003jw} & $1/2^-$ & 
   $\Lambda_c(2595)$, $\Lambda_c(2880)$, $\Xi_c(2790)$ \\
   \hline
 $J^P=0^-$ meson  & \cite{Oller:1998hw} & $0^+$ & 
   $\sigma(600)$, $\kappa(900)$, $f_0(980)$, $a_0(980)$ \\
   & & $1^-$ &  $\rho(770)$, $K^*(892)$, $\phi(1020)$ \\
   \hline
 $J^P=1^-$ meson & \cite{Roca:2005nm} & $1^+$ & 
   $b_1(1235)$, $h_1(1170)$, $h_1(1380)$, \\
   & & &
   $a_1(1260)$, $f_1(1285)$, $K_1(1270)$ \\
   \hline
 heavy $\bm{3}$ meson  & \cite{Kolomeitsev:2003ac} & $0^+$ &  $D_s(2317)$ \\
 \hline
 \end{tabular}
\end{center}
\end{table}%

We have reviewed the studies of resonances based on chiral SU(3) symmetry which are directly related to $\Lambda(1405)$. In closing, we would like to mention recent extensions of the chiral unitary approach to other sectors, with the replacement of the NG bosons by vector mesons~\cite{Molina:2008jw,Geng:2008gx,Gonzalez:2008pv,Oset:2009vf} and by heavy mesons~\cite{Hofmann:2005sw,Gamermann:2006nm,Haidenbauer:2010ch}. Strictly speaking, these sectors are not related to chiral symmetry of QCD, so the Weinberg-Tomozawa theorem cannot be applied. It is however possible to consider the vector-meson exchange mechanism as the physics behind the fundamental interaction. In this case, hidden local symmetry and flavor SU(4) symmetry lead to the similar structure of the interaction kernel with the chiral unitary approach, and these sectors are actively investigated by the same methodology. In addition, resonances in the three-body scattering has been studied using the Faddeev equation with chiral interaction~\cite{MartinezTorres:2007sr}. These works and subsequent studies have brought an interesting new perspective to hadron spectroscopy.

\section{$\Lambda(1405)$ in nuclear systems}
\label{sec:L1405many}

One of the central results in the previous section is that the structure of $\Lambda(1405)$ is dominated by the meson-baryon molecular component in the $\bar{K}N$-$\pi\Sigma$ coupled system. This conclusion indicates the important influence of $\Lambda(1405)$ to the $\bar{K}N$ interaction which is the fundamental building block to study the property of $\bar{K}$ in various many-body systems. The investigation of the antikaon nuclear systems ($\bar{K}$ nuclei) is one of the topical issues in strangeness nuclear physics. The property of $\Lambda(1405)$ in nuclear matter may be also important for the possible kaon condensation in deep interior of neutron stars. It is natural that the attractive $\bar KN$ interaction generates the bound $\bar{K}$ nuclei, but the width of the bound states can be large due to strong absorption of $\bar K$ by nucleons. This provides difficulties for direct experimental identifications of $\bar K$ bound states in nuclei. Thus, to start with, the simpler few-body systems should be studied using the two-body $\bar{K}N$ interaction. Recently, few-body hadron systems with kaons are considered as candidates of hadronic molecular states which are self-bound systems of hadrons by inter-hadron forces. In fact, $\Lambda(1405)$ itself is one of the examples of the hadronic molecules, as discussed in the previous section. This perception will help to understand more complicated many-body systems.

In this section, we first derive effective $\bar KN$ interactions by incorporating the $\pi\Sigma$ dynamics. The strength of the interaction is related with the  pole structure of $\Lambda(1405)$ discussed in Section~\ref{subsec:doublepole}. Next, we will review the present status of the $\bar{K}NN$-$\pi\Sigma N$ system as one of the examples of $\Lambda(1405)$ in few-body nuclear systems. A variational calculation shows the peculiar structure of the quasi-bound state. Then we introduce the concept of the hadronic molecular state and discuss the kaonic three-body systems along this line. We also emphasize the unique feature of the antikaon in the hadronic few-body systems. Finally, we discuss $\Lambda(1405)$ in nuclear matter and the related topics for the $\bar K$ in medium. 

\subsection{Effective $\bar KN$ interaction}
\label{subsec:KNint}

To study the property of the $\bar{K}$ in few-body systems, it is useful to construct an effective single-channel $\bar{K}N$ interaction by incorporating the dynamics of the other coupled channels. In addition, variational approach for rigorous few-body calculation favors the interaction in the nonrelativistic potential form. For instance, Ref.~\cite{Yamazaki:2007cs} constructed an energy-independent single-channel $\bar{K}N$ potential, motivated by the phenomenological interaction of Ref.~\cite{Akaishi:2002bg} in $\bar{K}N$ and $\pi\Sigma$ channels. The single-channel potential was then used to study the $\bar{K}NN$ system. Similarly, in the framework of the chiral unitary approach, an effective $\bar{K}N$ potential was derived in Ref.~\cite{Hyodo:2007jq}. The potential is not only useful for the application to the few-body systems, but also relevant to understand the property of the $\bar{K}N$ interaction along with the subthreshold extrapolation. In this section we demonstrate the construction of the effective $\bar{K}N$ potential based on the chiral unitary approach and discuss the relevance of the pole position of $\Lambda(1405)$ to the interaction strength.

We start from the scattering equation~\eqref{eq:LSEfinal} which is given in a matrix form as
\begin{align}
    T_{ij}
    =& V_{ij} + V_{ik} G_kT_{kj} .
    \label{eq:coupled}
\end{align}
The meson-baryon channels are assigned as $\bar{K}N$, $\pi\Sigma$, $\eta N$ and $K\Xi$ with $I=0$ for $i=1,\dots, 4$. Our aim is to construct an effective kernel interaction $V^{\text{eff}}$ such that the full scattering amplitude in channel 1 $(T_{11})$ is obtained by solving the single-channel scattering equation as
\begin{align}
    T_{11}
    = T^{\text{eff}}
    =& V^{\text{eff}}+V^{\text{eff}} G_1 T^{\text{eff}}
    .
    \label{eq:single}
\end{align}
Consistency with the full amplitude~\eqref{eq:coupled} requires the form of $V^{\text{eff}}$ as
\begin{align}
    V^{\text{eff}}
    =& V_{11} + \tilde{V}_{11}
    , \label{eq:effectiveint} \\
    \tilde{V}_{11} 
    =& \sum_{m=2}^{4}V_{1m}G_mV_{m1}
    +\sum_{m,l=2}^{4}V_{1m}G_m
    T^{(3)}_{ml} G_l
    V_{l1} , 
    \label{eq:Vtilde} \\
    T^{(3)}_{ml}
    =&
    V_{ml}+\sum_{k=2}^{4}V_{mk}\,G_k\,T^{(3)}_{kl} ,
    \quad
    m,l=2,3,4 ,
    \nonumber
\end{align}
where $V_{11}$ is the bare interaction in channel 1 and $\tilde{V}_{11}$ is the contribution from other channels 2-4 which consists of the iteration of the loop diagrams to all orders. As far as the two-body $\bar{K}N$ scattering amplitude is concerned, the single-channel approach~\eqref{eq:single} with the effective interaction~\eqref{eq:effectiveint} is exactly equivalent to the original coupled-channel framework of Eq.~\eqref{eq:coupled}.

As discussed in Section~\ref{subsec:interaction}, the bare $\bar{K}N$ interaction $V_{11}$ is given by the tree-level amplitude so it is a real-valued function of the energy $W$. On the other hand, $\tilde{V}_{11}$ contains the loop function $G_i$ which has an imaginary part when the energy is higher than the threshold of channel $i$. This means that the effective interaction $V^{\text{eff}}$ becomes complex when one incorporates the channel which has its threshold at lower energy in Eq.~\eqref{eq:Vtilde}. The imaginary part of the effective interaction represents the transition processes to the open channels. In the present case, the threshold of the $\pi\Sigma$ channel is lower than that of the $\bar{K}N$ channel, so the inclusion of the $\pi\Sigma$ channel generates the imaginary part of the effective interaction for energies higher than the $\pi\Sigma$ threshold. It was found in Ref.~\cite{Hyodo:2007jq} that the inclusion of the $\pi\Sigma$ channel enhances attractively the real part of the effective interaction for energies between the $\pi\Sigma$ and $\bar KN$ thresholds, but the enhancement is small in magnitude in comparison with the bare strength of $V_{11}$. The primary effect of the $\pi\Sigma$ coupled channel inclusion is to give further energy dependence to the effective interaction on top of the linear dependence in $W$ of the Weinberg-Tomozawa interaction~\eqref{eq:WTtermswave}. It was also found that the inclusion of the other coupled channels ($\eta \Lambda$ and $K\Xi$) has only minor effects in the energy region below the $\bar{K}N$ threshold. This indicates that the physics of $\Lambda(1405)$ and the subthreshold $\bar{K}N$ interaction can be well described by the $\bar{K}N$ and $\pi\Sigma$ channels.

The channel reduction~\eqref{eq:effectiveint} leads to the energy-dependent and nonlocal interaction $V^{\text{eff}}$ in the $\bar{K}N$ channel. To produce a useful input for few-body calculations, we convert the effective interaction into an equivalent local potential $U(r,E)$ in nonrelativistic quantum mechanics. The Schr\"odinger equation for the radial wave function $u(r)$ is (with $\hbar=1$)
\begin{align}
    -\frac{1}{2\mu}\frac{d^2 u(r)}{d r^2}
    +U(r,E) u(r)
    =&E u(r) ,
    \label{eq:schroedinger}
\end{align}
where the reduced mass is $\mu=M_Nm_K/(M_N+m_K)$ and the nonrelativistic energy is given by $E=W-M_N-m_K$. The energy dependence of the potential should be treated self-consistently. As explained in detail in Ref.~\cite{Hyodo:2007jq}, the local potential $U(r,E)$ has been constructed from $V^{\text{eff}}$ such that the scattering amplitude in chiral unitary approach is reproduced by Eq.~\eqref{eq:schroedinger}. This is not an exact transformation, since it is not guaranteed that a simple local potential can reproduce the complicated coupled-channel dynamics. Nevertheless, it is possible to construct a complex and energy-dependent $\bar{K}N$ potential with the gaussian form of the spatial distribution, which well reproduces the coupled-channel results in the $\bar{K}N$ channel. 

By definition, the effective interaction~\eqref{eq:effectiveint} [the equivalent local potential $U(r,E)$] reproduces the $\bar{K}N\to\bar{K}N$ amplitude in the single-channel scattering equation~\eqref{eq:single} [Schr\"odinger equation~\eqref{eq:schroedinger}]. We argue in Section~\ref{subsec:doublepole} that the resonance peak position in the $\bar{K}N\to\bar{K}N$ amplitude is observed around 1420 MeV, as a consequence of the double-pole structure. This peak position is higher than the nominal value of 1405 MeV. If we interpret the resonance peak position as the ``binding energy'' measured from the $\bar{K}N$ threshold ($\sim 1435$ MeV), we obtain $\sim 15$ MeV in chiral unitary approach. This is in contrast to the phenomenological potential in Ref.~\cite{Akaishi:2002bg} which is constructed to generate a bound state with a binding energy $\sim 30$ MeV. This difference of the binding energies leads to the weaker strength of the single-channel effective $\bar{K}N$ interaction in chiral unitary approach. In this case, strong $\pi\Sigma$ correlation supplies an additional attraction to the $\bar{K}N$ quasi-bound state to reproduce the $\Lambda(1405)$ peak in the $\pi\Sigma$ spectrum. In addition, energy dependence of the interaction is important to account for the correct behavior of the imaginary part of the amplitude which should vanish below the threshold of the lowest energy channel. As a consequence of the difference of the $\pi\Sigma$ diagonal interaction and the energy dependence of the potential, the deviation of the $\bar{K}N\to\bar{K}N$ amplitude in the chiral potential from that in the single-channel phenomenological potential becomes large at lower energy region, although they agree with each other around the $\bar{K}N$ threshold. In a recent version of the phenomenological potential~\cite{Esmaili:2009iq}, the $\pi\Sigma$ diagonal coupling is included, while the pole position of $\Lambda(1405)$ is kept fixed. The subthreshold extrapolation of the coupled-channel amplitude is found to be similar to that in the chiral unitary approach with an energy independent potential, if the pole position is set to be the nominal value~\cite{Ikeda:2011dx}. In this respect, determination of the threshold quantities of the $\pi\Sigma$ channel~\cite{Lambdacdecay,IkedaLattice} are highly demanded to reduce the uncertainty of the subthreshold extrapolation of the $\bar{K}N$ interaction.

\subsection{$\Lambda(1405)$ in few-body nuclear systems}
\label{subsec:KNN}

\begin{table}[tbp]
\begin{center}
\begin{minipage}[t]{16.5 cm}
\caption{Summary of theoretical studies on the $\bar{K}NN$-$\pi\Sigma N$ system. We denote the mass of the states as the ``binding energy" $B_{\bar{K}NN}$ measured from the $\bar{K}NN$ threshold. $\Gamma_m$ represents the width of the mesonic decay into $\pi\Sigma N$ and $\pi\Lambda N$ channels. Ref.~\cite{Ikeda:2010tk} found additional state with $B=67$-$89$ MeV and $\Gamma_{m}=244$-$320$ MeV.
}
\label{tab:KNN}
\end{minipage}
\begin{tabular}{l|ccccc}
\hline
 Refs. & \cite{Shevchenko:2006xy,Shevchenko:2007zz} 
 & \cite{Yamazaki:2007cs}
 & \cite{Ikeda:2007nz} 
 & \cite{Dote:2008in,Dote:2008hw} 
 & \cite{Ikeda:2010tk} \\
 \hline 
 interaction & \multicolumn{3}{c}{Energy independent} 
   & \multicolumn{2}{c}{Energy dependent} \\
 & pheno. & pheno. & chiral 
   & chiral & chiral \\
 method & Faddeev & variational & Faddeev
   & variational & Faddeev   \\
 $\pi\Sigma N$ dynamics & explicit & effective & explicit
   & effective & explicit  \\
 \hline 
 $B_{\bar{K}NN}$ [MeV] & 50-70 & 48 & 60-95
   & 17-23 & 9-16  \\
 $\Gamma_m$ [MeV] & 90-110 & 60 & 45-80
   & 40-70 & 34-46   \\
 \hline
 \end{tabular}
\end{center}
\end{table}%

The idea of $\Lambda(1405)$ as a $\bar KN$ quasi-bound state was developed to propose a $\bar KNN$ three-body quasi-bound state in Ref.~\cite{PL7.288} and with a modern approach in Ref.~\cite{Akaishi:2002bg}. Recent theoretical investigations of the $\bar{K}NN$ system with $I=1/2$ has been studied in various theoretical approaches~\cite{Shevchenko:2006xy,Shevchenko:2007zz,Yamazaki:2007cs,Ikeda:2007nz,Ikeda:2008ub,Dote:2008in,Dote:2008hw,Ikeda:2010tk} as summarized in Table~\ref{tab:KNN}. We find that the $\bar KNN$ system is bound below the $\bar KNN$ break-up threshold with a large width in spite of quantitative discrepancy among theoretical predictions. These works rigorously solve the three-body problem, adopting the $\bar{K}N$ and $NN$ interactions which are constrained by experimental data. The theoretical models can be characterized by the choice of the $\bar{K}N$ interaction and the method to solve the three-body system. The $\bar{K}N$ interactions can be divided into two classes: those constructed phenomenologically~\cite{Shevchenko:2006xy,Shevchenko:2007zz,Yamazaki:2007cs} and those derived from chiral low energy theorem~\cite{Ikeda:2007nz,Dote:2008in,Dote:2008hw,Ikeda:2008ub,Ikeda:2010tk}. The chiral interaction can be further classified in terms of the treatment of the energy dependence. The few-body calculation is performed by either the variational approach~\cite{Yamazaki:2007cs,Dote:2008in,Dote:2008hw} or the Faddeev approach~\cite{Shevchenko:2006xy,Shevchenko:2007zz,Ikeda:2007nz,Ikeda:2008ub,Ikeda:2010tk}. In the variational approach, it is easy to extract the structure of the bound state from the obtained wavefunction, but the dynamics of the $\pi\Sigma N$ is only effectively incorporated in the imaginary part of the $\bar{K}N$ interaction. Faddeev approach, on the other hand, can treat the $\pi\Sigma N$ channel explicitly, but the separable form factors are introduced to make the problem tractable. In view of these results, it is clear that the $\bar{K}NN$ system forms a quasi-bound state. Yet the quantitative estimation of the binding energy and the width has not  converged. The discrepancy is partly due to the different treatment of the interaction and few-body technique, but the main reason stems from the lack of information of the $\bar{K}N$ interaction far below the threshold~\cite{Ikeda:2011dx}.

The investigation of the $\bar KNN$ system in Refs.~\cite{Dote:2008in,Dote:2008hw} uses the effective $\bar{K}N$ interaction derived in Section~\ref{subsec:KNint}, together with the realistic $NN$ interaction with repulsive core. The binding energy is small compared with the other works, due to the weaker attraction of the effective $\bar{K}N$ potential. As discussed in Section~\ref{subsec:KNint}, chiral symmetry requires the strong $\pi\Sigma$ dynamics, and the attractive force to form $\Lambda(1405)$ is divided into the $\bar{K}N$ and $\pi\Sigma$ channels. As a consequence, the allotment of the $\bar{K}N$ attraction is reduced. A similar conclusion was drawn using the Faddeev equation with fixed center approximation~\cite{Bayar:2011qj}. The obtained wavefunction shows that the mean distance of two nucleons ($\bar{K}$ and a nucleon) in this system is about 2.1-2.2 fm (1.8-2.0 fm). The bound state has a smaller size than the deuteron, but the mean distance is comparable with the distance between two nucleons at normal nuclear density. This result shows that the $\bar{K}NN$ system can exist as a hadronic molecular structure.

An interesting observation was given in Ref.~\cite{Ikeda:2010tk} where the coupled-channel Faddeev calculation of the $\bar{K}NN$-$\pi\Sigma N$ system was performed with the energy dependence of the chiral interaction. In this case, the $\bar{K}N$-$\pi\Sigma$ amplitude has two poles as described in Section~\ref{subsec:doublepole}, and it is found that the solution of the three-body equation also has two poles. One pole appears at higher energy with narrow width, and the other locates at lower energy and with broad width more than 200 MeV. The higher energy pole of the three-body system is close to the result in the variational approach~\cite{Dote:2008in,Dote:2008hw}. This work suggests the possibility of the double-pole structure in the three-body $\bar{K}NN$-$\pi\Sigma N$ system. 

There have been several experimental hints on this issue. 
A peak structure in the $\Lambda N$ spectrum was observed in the stopped $K^-$ reactions with nuclear targets by FINUDA collaboration at DA$\Phi$NE and was claimed to be an evidence for the dibaryon ($\bar{K}NN$) system~\cite{Agnello:2005qj}.
The mass of the peak locates below the $\pi\Sigma N$ threshold. When measured from the $\bar{K}NN$ threshold, the central value of the peak position corresponds to the binding energy of about $B=115$ MeV and the width of $\Gamma = 67$ MeV, but the interpretation of the peak is not clear~\cite{Magas:2006fn}. Another signal was found by the reanalysis of the DISTO experiment at Saclay~\cite{Yamazaki:2010mu}. In the $pp\to K^+\Lambda p$ reaction at 2.85 GeV, a broad peak of the $\Lambda p$ spectrum was found when the final proton has high transverse momentum. The observed peak is at $M=2267$ MeV which corresponds to $B = 103$ MeV. The width is given as $\Gamma = 118$ MeV. 
The same reaction but with lower incident energy of 2.50 GeV was studied in Ref.~\cite{Kienle:2011mi}. The peak was not found in this lower energy data, which could be interpreted as the dominance of the $\Lambda(1405)$ doorway process for the production of the peak structure at 2.85 GeV. The investigation for the $\bar{K}NN$ bound state will be further explored through $K^-$ incident reactions by E15 experiment at J-PARC~\cite{Hiraiwa:2011zz}, AMADEUS project at DA$\Phi$NE~\cite{Zmeskal:2010zz}, $\pi$ incident reaction by E27 experiment at J-PARC~\cite{JPARCE27}, and through $pp$ collisions at higher energy by FOPI collaboration at GSI~\cite{Suzuki:2009zze}.

Experimental studies for narrow tribaryon systems ($\bar{K}$ and three nucleons) in the missing mass spectroscopy of the stopped $K^-$ at KEK were reported and a signal for the tribaryon states was found~\cite{Iwasaki:2003kj,Suzuki:2004ep}.
However, the new experiment of the same reaction with higher statistics found no significant structure~\cite{Sato:2007sb}. Some appreciable strength of the $\Lambda d$ correlation was also reported~\cite{Agnello:2007ph,Suzuki:2007kn} (see also discussion in Ref.~\cite{Magas:2008bp}). In any event, the precise knowledge of two- and three-nucleon absorption process of the antikaon is important to extract the possible signal of the $\bar{K}$ nuclei in the stopped $K^-$ reaction~\cite{Suzuki:2007kn,Suzuki:2007pd}.

\begin{figure}
\begin{center}
  \includegraphics[width=0.5\textwidth,bb=0 0 550 350]{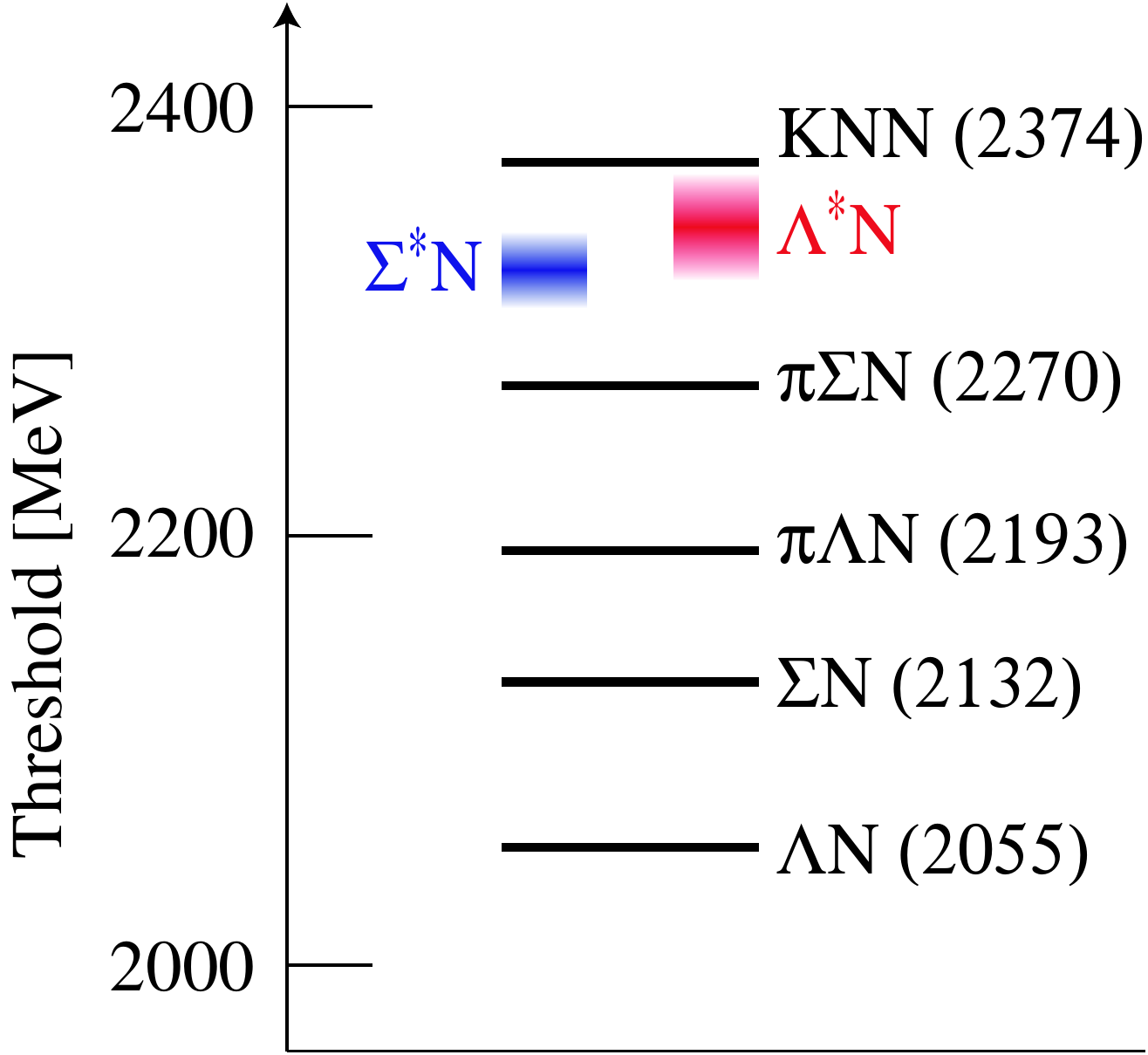}
\begin{minipage}[t]{16.5 cm}
\caption{Threshold energies of multihadron channels below the $\bar{K}NN$ threshold. $\Lambda^*$ and $\Sigma^*$ represent $\Lambda(1405)$ ($J^P=1/2^-$) and $\Sigma(1385)$ ($J^P=3/2^+$), respectively. \label{fig:KNN}}
\end{minipage}
\end{center}
\end{figure}

The signals found in FINUDA and DISTO experiments have strangeness $S=-1$ and baryon number $B=2$, so the $\bar{K}NN$ bound state is one of the possible candidates. The interpretation is however not straightforward, since there are many open channels with the same quantum numbers as shown in Fig.~\ref{fig:KNN}. The peak of FINUDA (DISTO) experiment locates slightly below (on top of) the $\pi\Sigma N$ threshold, so the substantial amount of the $\pi\Sigma N$ component should be expected. In addition, because the decay into the $\pi\Sigma N$ channel is kinematically forbidden, the width of about 100 MeV should be attributed to the decay into two-body $YN$ final states and the $\pi\Lambda N$ final state. 

Careful theoretical analyses are necessary to interpret the observed peak structure in FINUDA and DISTO experiments. All the calculations in Table~\ref{tab:KNN} did not explicitly include the two-body channels such as $\Lambda N$ where signals in FINUDA and DISTO experiments were observed. The inclusion of the $\Lambda N$ channel is mandatory, in order to show how the bound state found in the few-body calculation affects the $\Lambda N$ spectrum. Although it is difficult to study the coupled-channel problem with different number of particles (such as $\bar{K}NN$-$\Lambda N$), one promising approach was proposed in Ref.~\cite{Arai:2007qj} as the ``$\Lambda^*$-hypernuclei" picture\footnote{The word $\Lambda^*$-hypernuclei has been used in an earlier work~\cite{Fink:1990hf} to refer to the $\bar{K}$ nuclei.}. There $\Lambda(1405)$ is regarded as a fundamental degree of freedom, which may be supported by the observation in Refs.~\cite{Yamazaki:2007cs,Dote:2008in,Dote:2008hw} that the relative wavefunction of the $\bar{K}N$ pair in the $\bar{K}NN$ system is similar to the wavefunction of $\Lambda(1405)$ in vacuum. In this approach, inclusion of the $\Lambda N$ channel is rather straightforward, once the appropriate transition potential is given as in Ref.~\cite{Sekihara:2009yk}. A two-body $\Lambda^*N$ system was studied by constructing the $\Lambda^*N$ potential based on chiral unitary approach~\cite{Uchino}. The bound state was found in spin $S=0$ channel in accordance with the few-body $\bar{K}NN$-$\pi\Sigma N$ calculation, with a small binding energy as in the case of the energy-dependent chiral interaction~\cite{Ikeda:2010tk}.

\subsection{Hadronic molecular states with kaons}
\label{subsec:KKN}

The $\Lambda(1405)$ resonance as a dynamically generated state in the $\bar KN$ system can be a building-block of $\bar K$ nuclear few-body systems. This is based on the fact that the binding energy ($10\sim30$ MeV) of the $\bar KN$ quasi-bound state is not so large in comparison with typical hadron energy scale\footnote{It is also interesting to mention that the $\bar KN$ binding energy is much larger than the $NN$ bound state (deuteron).}. In other words, the kaon kinetic energy in the $\bar KN$ bound system is much smaller than the kaon mass. Such quasi-bound states are called as hadronic molecular states. The basic feature of the hadronic molecular state is that it is a (quasi)-bound state composed of hadronic constituents with keeping their identity as they are in isolated systems and appears just below the threshold of break-up into the constituent hadrons. Driving force to make hadronic molecular states is hadronic interaction rather than inter-quark dynamics and confinement force. Consequently, inter-hadron distances inside the hadronic molecular states are larger than the typical size of the low-lying hadrons which may be characterized by the quark confinement range. Well known examples of the hadronic molecules are the nuclei (hypernuclei) which are the composite systems of protons and neutrons (and $\Lambda$). For hadronic molecular states, pion has a too light mass to form bound states with other hadrons by hadronic interaction, since the pion kinetic energy in the hadronic confined system overcomes attractive potential energy. In contrast, various hadronic molecules are expected in the heavy quark sector, such as the $X(3872)$ as a $D\bar{D}^*$ bound state~\cite{Voloshin:2003nt,Tornqvist:2004qy}, $\bar{D}N$ bound state~\cite{Yasui:2009bz}, and $\Lambda_{c}N$ bound state~\cite{Liu:2011xc}.

\begin{figure}
\begin{center}
  \includegraphics[width=0.75\textwidth,bb=0 0 794 340]{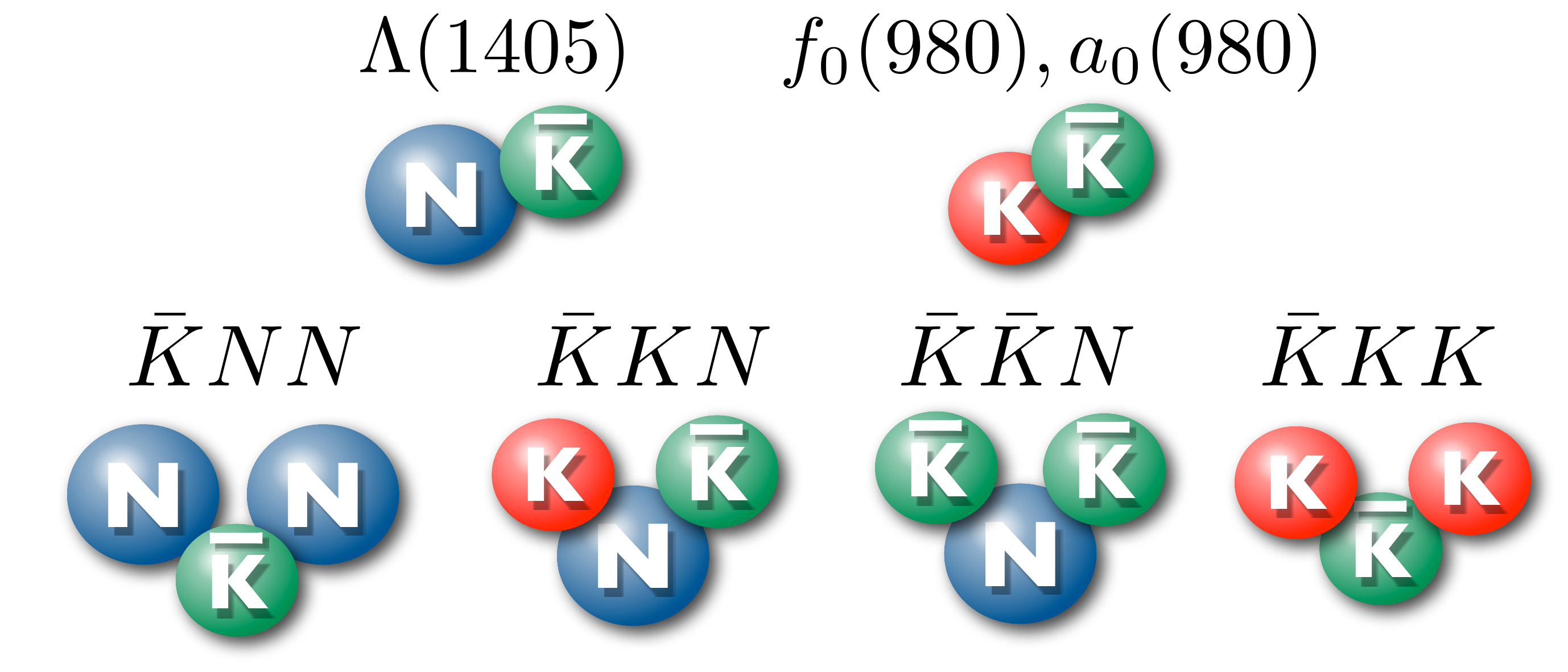}
  \caption{Family of kaonic few-body states. }
  \label{fig:family}
\end{center}
\end{figure}

The anti-kaon $\bar K$ has another interesting feature. The strength of the Weinberg-Tomozawa interaction, which is the driving force to generate $s$-wave hadronic molecular systems, is given by the flavor SU(3) symmetry in which $K$ and $N$ are classified into the same state vector in the octet representation. Therefore, considering also that $K$ and $N$ have a similarly heavy mass, one finds that the fundamental interactions in $s$ wave are very similar in the $\bar KK$ and $\bar KN$ channel. In fact, the scalar meson $f_{0}(980)$ could be explained by the $\bar KK$ quasi-bound state~\cite{Weinstein:1983gd,Baru:2003qq}, and it can be dynamically generated in the $\bar KK$-$\pi\pi$ coupled system also with the chiral unitary model, as $\Lambda(1405)$ is, and its pole appears around the $\bar KK$ threshold~\cite{Oller:1998hw,Oller:1998zr}. Consequently in the $\bar KN$ and $\bar KK$ channels with $I=0$, there are quasi-bound states of $\bar KK$ and $\bar KN$ with a dozen MeV binding energy. This similarity between $K$ and $N$ is responsible for systematics of three-body kaonic systems, $\bar KNN$, $\bar KKN$, $\bar K \bar KN$ and $\bar KKK$, as shown in Fig.~\ref{fig:family}. It is also important to emphasize that the $\bar K$ few-body systems have unavoidably larger widths due to $\bar K$ absorptions into pionic and nonmesonic modes. This is a large difference from the fermionic nuclear systems. 

The $\bar KKN$ system was investigated in Ref.~\cite{Jido:2008kp} first with a variational calculation of the three-body Schr\"odinger equation with an effective potential in which the scalar mesons $f_{0}(980)$ and $a_{0}(980)$ are reproduced as quasi-bound states of $K \bar K$. The model is based on the two-body effective potential of the $\bar KN$ system discussed in Section~\ref{subsec:KNint}, and the range parameter of the $\bar KK$ interaction was chosen to be the same. This calculation treats the $ \bar KKN$ single channel and minimal coupled-channel effects are introduced as the imaginary parts of the effective potentials which represent transition to the open channels, ($\pi \Lambda$, $\pi \Sigma$) for $\bar K N$ and ($\pi \pi$, $\pi\eta$) for $K\bar K$. In this calculation, a quasi-bound state with $I=1/2$ and $J^{P}=1/2^{+}$, which has the same quantum number as excited nucleon ($N^{*}$), was found with a mass 1910 MeV and a width 90 MeV below all of the meson-baryon decay threshold energies of the $\Lambda(1405)+K$, $f_0(980)+N$ and $a_0(980)+N$ states. This quasi-bound state was confirmed also by more sophisticated three-body calculations of the $\bar KKN$-$\pi\pi N$-$\eta \pi N$ coupled channels~\cite{MartinezTorres:2008kh} and the $K \bar KN$-$K \pi\Sigma$-$K\pi\Lambda$ channels~\cite{MartinezTorres:2010zv}. These calculations were based on a relativistic Faddeev approach developed in Ref.~\cite{MartinezTorres:2007sr}, in which two-body interactions are calculated in the chiral unitary approach.  It was also found in the Faddeev approach~\cite{MartinezTorres:2010zv} that this quasi-bound state can be essentially described by the $K\bar K N$ single channel and the rest of the channels are less relevant for the formation of the quasi-bound state. The $\bar KKN$ state was found also in a fixed center approximation of three-body Faddeev calculation~\cite{Xie:2010ig}.  A discussion on experimental observation of this $N^{*}$ can be found in Ref.~\cite{MartinezTorres:2009cw}.

\begin{figure}
\begin{center}
  \includegraphics[height=5cm,bb=0 0 369 284]{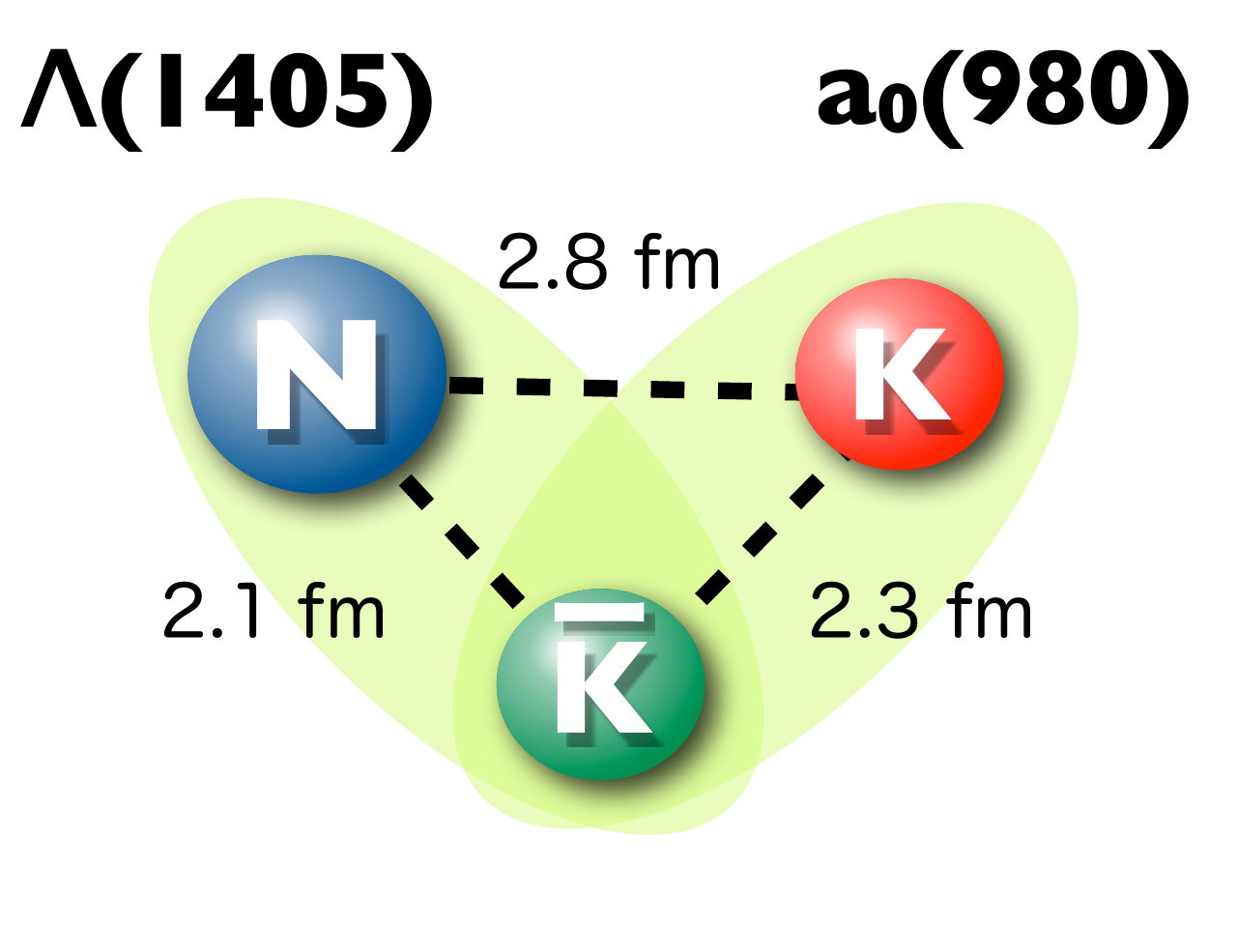} 
  \caption{Schematic structure of the $\bar KKN$ quasi-bound state with the inter-hadron distances. }
  \label{fig:bond}
\end{center}
\end{figure}

The $\bar KKN$ quasi-bound state is a loosely bound system having a 20 MeV binding energy. It was found in Ref.~\cite{Jido:2008kp} using the nonrelativistic potential model that the quasi-bound state has a spatially larger size than typical baryon resonances by showing that the root mean squared radius is as large as 1.7 fm, which is larger than the radius of the $^{4}$He nucleus. The inter-hadron distances were also calculated as 2.1 fm, 2.3 fm and 2.8 fm for the $\bar KN$, $\bar KK$ and $KN$ subsystems, respectively. These values are comparable with the inter-nucleon distance in the normal nuclei as in the case of the $\bar{K}NN$ system~\cite{Dote:2008in,Dote:2008hw}. It is also interesting that the values of the $\bar KN$ and $\bar KK$ distances are very similar to those of the inter-hadron distance in the quasi-bound states formed in the isolated two-body systems, and that the $KN$ distance is larger than the others due to the $KN$ repulsive interaction. This indicates that two subsystems, $\bar K N$ and $K\bar K$, are as loosely bound in the three-body system as they are in two-body system. For the isospin configuration of the $K\bar{K}N$ state, it was found that the $\bar K N$ subsystem has a dominant $I=0$ component thanks to the $\Lambda(1405)$ resonance in the $\bar KN$ subsystem. The $K\bar K$ subsystem has dominant contribution from the $I=1$ configuration, because, in both $I=0$ and $I=1$ channels, $K\bar K$ has attraction with very similar strength enough to provide the quasi-bound $K\bar K$ states at almost the same positions, and the $I=1$ configuration of $\bar K K$ is favorable to have total isospin 1/2 for the $K \bar K N$ with the $\bar KN$ subsystem with $I=0$. From the above discussions of the spatial and isospin structure, it was concluded~\cite{Jido:2008kp} that the $K \bar KN$ state can be understood by the structure of simultaneous coexistence of $\Lambda(1405)$ and $a_{0}(980)$ clusters as shown in Fig.~\ref{fig:bond}. This does not mean that it would be described as superposition of the $\Lambda(1405)+K$ and $a_{0}(980)+N$ wavefunctions, because they are not orthogonal to each other. The probabilities for the $K \bar K N$ system to have these states are 90\%. It means that $\bar K$ is shared by both $\Lambda(1405)$ and $a_{0}$ at the same time. 

The $\bar K \bar K N$ system with $J^{p}=(1/2)^{+}$ and $I=1/2$ also develops a bound state around 1.93 GeV as a $\Xi^*$ resonance and the binding energy from the $\Lambda(1405)+\bar K$ threshold was found to be as small as a few MeV due to the strong repulsion $\bar K \bar K$ with $I=1$~\cite{KanadaEnyo:2008wm}. In the $\bar K \bar KN$ system with $I=1/2$, once $\Lambda(1405)$ forms in a $\bar KN$ pair with $I=0$, another $\bar KN$ should have dominantly the $I=1$ configuration. This channel has weak attraction, but it is not enough to compensate the repulsive $\bar K \bar K$ interaction. Therefore, the $\bar K \bar K N$ system is very weak binding.

The $\bar KKK$ system with $I=1/2$ and $J^{p}= 0^{-}$, being quantum number of an excited state of kaon, was studied in two-body $f_{0}K$ and $a_{0}K$ dynamics~\cite{Albaladejo:2010tj}, in the three-body Faddeev calculation~\cite{Torres:2011jt} and in the non-relativistic potential model~\cite{Torres:2011jt}. Experimentally, Particle Data Group tells that there is a excited kaon around 1460 MeV observed in $K\pi\pi$ partial-wave analysis, although it is omitted from the summary table. In Ref.~\cite{Albaladejo:2010tj}, effective interactions between kaon and the $f_{0}$ and $a_{0}$ scalar mesons were derived and by solving the two-body $f_{0}K$ and $a_{0}K$ coupled channels problem with these interactions a resonance was found at around 1460 MeV. In Ref.~\cite{Torres:2011jt}, the three-body Faddeev calculation based on Ref.~\cite{MartinezTorres:2007sr} was performed in the three-body coupled-channels of $K \bar KK$, $K \pi\pi$ and $K\pi\eta$ and a similar resonance state was found at 1420 MeV, while the potential model suggested a quasi-bound state with a binding energy 20 MeV, or a 1467 MeV mass. This state is essentially described by the $\bar KKK$ single channel and its configuration is found to be mostly $f_{0}K$. In the potential model, the internal structure of the $\bar K KK$ quasi-bound state was calculated and it was found that the root mean-squared radius is as large as 1.6 fm, and the inter-hadron distances of $K$-$K$ and $(\bar KK)$-$K$ are 2.8 fm and 1.7 fm, respectively. The $K$-$K$ distance is larger due to the repulsive interaction. The distances of $K_{2}$-$\bar K$ and $K_{1}$-$(\bar K_{2}\bar K)$ before performing symmetrization of two $K$'s were also calculated and found to be 1.6 fm and 2.6 fm. The $K_{2}$-$\bar K$ distance is very similar with the $\bar KK$ distance of the isolated $f_{0}(980)$ quasi-bound state. 

It is important to emphasize a significant role of the repulsive configurations in the quasi-bound systems of $\bar KKN$ and $\bar KKK$ which are essential for the hadronic molecular state to be described by hadronic degrees of freedom. The repulsive interactions ensure such a loosely bound system that the constituent hadrons are separated well and keep their identities inside the bound state. For deeply bound states, such as states bound by hundreds MeV, the constituent hadrons come close and may be overlapped. In such a case, the hadronic molecular picture may be broken down and quark degrees of freedom should be essentially taken into account. In the present $\bar KKN$ and $\bar KKK$ systems, the repulsive interactions in $KK$ and $KN$ with $I=1$ make the $\bar KKN$ and $\bar KKK$ systems loosely bound with moderate binding energies. When taking as strong artificial attraction for $KK$ and $KN$ as $\bar KK$ and $\bar KN$, one obtains very deeply bound states with hundreds MeV binding energies~\cite{Torres:2011jt}. Therefore for the realization of the hadronic molecular state near the break-up threshold, one of the pairs should have a repulsive interaction or, at least, sufficiently weak attraction. 

\subsection{$\Lambda(1405)$ in nuclear matter}
\label{subsec:matter}

Kaon is one of the Nambu-Goldstone bosons associated with spontaneous breaking of chiral symmetry. Although the connection of the in-medium properties of kaon to the fate of the \chiral chiral symmetry in nuclear medium is not clear due to the heavy strange quark mass, in-medium properties of kaons are very interesting issues both theoretically and experimentally, giving fundamental information of kaon condensation in highly dense matter~\cite{Kaplan:1986yq}. Here we concentrate on the discussion of $\Lambda(1405)$ in nuclear medium. For the detailed discussion on the $\bar K$ in nuclear medium, for instance, see Refs.~\cite{Friedman:2007zza}.

The $\Lambda(1405)$ resonance existing in the $\bar KN$ subthreshold plays an important role for the $\bar K$ properties in nuclear matter. The importance was pointed out in the study of kaonic atoms, in which the $K^{-}$ absorption process is strongly influenced by the presence of the $\Lambda(1405)$ resonance in the $K^{-}p$ channel, because the $K^{-}p$ energies available for absorption are close to the resonance energy due to nucleon binding energy in nuclei and $K^{-}$ captures take place selectively by a proton thanks to the $I=0$ resonance~\cite{Bloom:1969as,Bardeen:1971js,Bardeen:1972ti}. The repulsive $K^{-}p$ scattering length in free space, despite the attractive effective $K^{-}p$ scattering lengths in nuclei deduced from kaonic atom data, is also explained by the presence of the $\Lambda(1405)$ resonance below the $\bar KN$ threshold. 

In Ref.~\cite{Weise:1976fe}, an explicit $\Lambda(1405)$ contribution was taken into account in the calculation of the $K^{-}$ optical potential in nuclei by considering an intermediate $\Lambda(1405)$ (particle) - proton (hole) state into the kaon self-energy together with a nonresonant background part.  In the $\Lambda(1405)$-hole state, the proton was considered to be bound in nuclei, and the mass, width and coupling to $K^{-}p$ of $\Lambda(1405)$ were fixed in free space. A parameter for the $\Lambda(1405)$ mass shift in nuclei was also introduced and determined so as to reproduce the observed level shifts and widths of kaonic atom states in $^{12}$C and $^{32}$S nuclei, and the mass shift was found phenomenologically to be of the order of 20 MeV upward in relevant densities for kaon-nucleus dynamics in atomic states, which is confined to the nuclear surface. Many-body corrections to the $\Lambda(1405)$ propagation were estimated by calculating the in-medium $\Lambda(1405)$ self-energy coming from a $\pi\Sigma$ loop with pion rescattering~\cite{Eisenberg:1976ft} and $\bar KN$ loops with the Pauli blocking effect of the intermediate nucleons and kaon rescattering~\cite{Brockmann:1978sm}. It was found that the Pauli blocking correction induced a repulsive mass shift being typically of the order of 30 MeV~\cite{Brockmann:1978sm} and the rescattering effects gave a 10 MeV repulsive mass shift and an increase of the width by 20--30\% over its free space value at the nuclear surface $\rho \sim \rho_{0}/8 $~\cite{Eisenberg:1976ft,Brockmann:1978sm}, though quantitative estimates of the mass shift and width broadening were difficult due to lack of phenomenological inputs to fix unknown cut-off factors for the $\Lambda(1405)$ couplings to $\bar KN$ and $\pi\Sigma$ appearing in the rescattering diagrams. Along the line of discussion on kaon condensation~\cite{Kaplan:1986yq} in dense nuclear matter, Ref.~\cite{Lee:1994jj} also considered an explicit $\Lambda(1405)$ contribution for the perturbative calculation of the in-medium kaon self-energy up to the next-to-next leading order (N$^{2}$LO) in chiral effective Lagrangian approach, and it was found that $\Lambda(1405)$-nucleon interactions could be important to induce kaon condensation but the critical density is insensitive to the strengths of the $\Lambda^{*}$-$N$ interactions once the kaonic atom data are reproduced within the model. 

The first investigation of in-medium properties of a dynamically generated $\Lambda(1405)$ in the $\bar KN$-$\pi\Sigma$ coupled channels was performed in Ref.~\cite{Koch:1994mj}.  Dynamical description of $\Lambda(1405)$ with its intrinsic structure enables us to perform microscopic calculations of nuclear medium corrections by following conventional many-body technique developed in nuclear physics.  In the calculation of Ref.~\cite{Koch:1994mj}, $\Lambda(1405)$ was introduced as a quasi-bound state of $\bar KN$ induced by a separable potential of the $s$-wave $\bar KN$ and $\pi\Sigma$ interactions whose strengths were fixed by the SU(3) flavor symmetry of vector-meson exchange model~\cite{Siegel:1988rq}. The cut-off parameters in the form factors of the $\bar KN$ and $\pi\Sigma$ couplings were determined phenomenologically so as to reproduce the scattering amplitude extracted by Martin in Ref.~\cite{Martin:1980qe}. The medium effects were simply taken into account by the Pauli blocking which restricts the momentum of the intermediate nucleon propagation in the $\bar KN$ loops. It turned out that the $\bar KN$ bound state feels effectively a repulsive interaction by the truncation of  momentum space due to the Pauli blocking effect, and consequently the mass of $\Lambda(1405)$ is pushed up by 65 MeV at nuclear saturation density and the $\Lambda(1405)$ width increases by about 20\% simply because of the enlarged phase space available for the decay to $\pi\Sigma$. In Ref.~\cite{Waas:1996xh}, a similar calculation was done by using a more refined $\bar KN$ scattering amplitude in free space obtained in the coupled-channels approach for the six channels ($K^{-}p$, $\bar K^{0}n$, $\pi^{+}\Sigma^{-}$, $\pi^{0}\Sigma^{0}$, $\pi^{-}\Sigma^{+}$, $\pi^{0}\Lambda$) based on chiral effective Lagrangian, which reproduces available low energy data of $K^{-}p$ elastic and inelastic scattering, the $K^{-}p$ threshold blanching ratio and the $\Lambda(1405)$ spectrum~\cite{Kaiser:1995eg}. The medium corrections from Fermi motion and nucleon binding were included in addition to the Pauli blocking and found to give opposing effects. It was also pointed out that the presence of $\Lambda(1405)$ just below the $\bar KN$ threshold in free space is responsible for the breakdown of the low density description for the $K^{-}$ optical potential $V_{\rm opt} \propto a_{K^{-}N} \rho$ at $\rho\sim\rho_{0}/10$. 

The Pauli blocking effect in nuclear matter induces repulsive contribution to the mass of the $\bar KN$ quasi-bound state which pushes up $\Lambda(1405)$ above the free $K^{-}p$ threshold~\cite{Koch:1994mj,Waas:1996xh,Ramos:1999ku}, and at further higher density $\Lambda(1405)$ could dissolve into the nuclear matter~\cite{Waas:1996xh,Waas:1996fy}. While the quasi-bound state feels repulsion as density increases, the $\bar K$ feels an enhanced attraction and its effective mass decreases in the nuclear medium~\cite{Waas:1996fy}. This is because $\Lambda(1405)$ is pushed up at higher density and does not give relevant influence to the $\bar KN$ interaction any more, and consequently the effective $\bar K$ mass is primarily determined by the leading attractive Weinberg-Tomozawa interaction~\cite{Waas:1996fy}. With the large attraction in the in-medium $\bar K$ self-energy, the modification of the $\bar K$ propagation in nuclear medium should be incorporated in the calculation of the in-medium properties of $\Lambda(1405)$. This self-consistent treatment was done in Ref.~\cite{Lutz:1997wt}, and it was found that the in-medium corrections on the $\bar K$ propagation affect the $\Lambda(1405)$ mass downward in opposite to the Pauli blocking, and thus the in-medium $\Lambda(1405)$ mass is determined by these two competing effects. After all, the $\Lambda(1405)$ mass stays almost at its free space value with an increased decay width~\cite{Lutz:1997wt}. In Ref.~\cite{Ramos:1999ku}, medium corrections to the pion self-energy were also considered together with the Pauli blocking acting on the intermediate nucleon, mean-field binding potential of the baryons and the $\bar K$ self-energy in the $\bar K$ propagation. It turned out that the inclusion of the dressing pions in the intermediate states enhances the in-medium $\Lambda(1405)$ width further and $\Lambda(1405)$ staying almost at its free space position tends to dissolves with increasing density~\cite{Ramos:1999ku}. The in-medium $\bar KN$ scattering amplitude obtained by the chiral unitary approach in these ways was applied for the calculation of the $K^{-}$ optical potential for kaonic atoms, and with the optical potential the observed energy shifts and widths of kaonic atom states were successfully reproduced in a wide mass range\cite{Hirenzaki:2000da} (see also Refs.~\cite{Friedman:2007zza}). The nonmesonic decay of $\Lambda(1405)$ in nuclear matter, $\Lambda^{*}N \to Y N$ was investigated with one-meson exchange model between $\Lambda^{*}$ and nucleon in Ref.~\cite{Sekihara:2009yk}, though the medium modification on $\Lambda(1405)$ was not taken into account. This work was motivated by the finding that $\Lambda(1405)$ keeps its properties in few-body systems and the $\Lambda(1405)$ resonance can be a doorway state of $\bar K$ absorption to nuclear systems. It was turned out that, according to the SU(3) flavor relation of the meson-baryon couplings, the $\bar K$ ($\pi$) exchange is dominated in the $\Lambda^{*} N \to \Lambda N$ ($\Lambda^{*} N \to \Sigma N$) transition, and thus the ratio of the transition rates of $\Lambda^{*}N$ to $\Lambda N$ and $\Sigma N$ is strongly sensitive to the $\Lambda^{*}$ coupling strengths to $\bar KN$ and $\pi\Sigma$. The partial decay width of $\Lambda(1405)$ in the non-mesonic decay was found to be around 20 MeV at the nuclear saturation density. 

\section{Conclusion and future perspective}

The understanding of $\Lambda(1405)$ is an interesting and challenging subject in hadron physics after fifty years since its discovery. In this article, we have summarized the status of  experimental and theoretical investigations on the $\Lambda(1405)$ resonance. Among others, we present a detailed introduction of chiral unitary approach as a successful and promising theoretical framework to describe $\Lambda(1405)$. The meson-baryon scattering amplitude is formulated with the basic principles of hadron scattering: chiral symmetry of QCD for the low energy interaction and unitarity of the S-matrix. Excited baryons are described as pole singularities in the scattering amplitude, whose physical origins are discussed at length. 

The observables in $S=-1$ meson-baryon scattering as well as the properties of the $\Lambda(1405)$ resonance are well described in the chiral unitary approach. The structure of $\Lambda(1405)$ has been studied from various aspects in the chiral unitary approach. It is shown that the $\Lambda(1405)$ resonance is associated with two poles in the complex energy plane as a consequence of the two attractive components of the chiral low energy interaction. We summarize how the mass spectra of the $\pi\Sigma$ channel in various reactions are reflected by the pole structure together with possible contamination which modifies the experimental spectra. We then focus on the internal structure of $\Lambda(1405)$. Through the analysis of the renormalization procedure, we show that the possible seed of the resonance is hidden in the loop function which can be extracted by the use of the natural renormalization scheme. In the case of $\Lambda(1405)$, such nontrivial contribution is found to be small at the energy region of the resonance, implying the dominance of the meson-baryon structure in $\Lambda(1405)$. The analysis of the $N_c$ scaling shows that the three-quark component of $\Lambda(1405)$ is small. The electromagnetic size of $\Lambda(1405)$ turns out to be larger than the typical three-quark hadrons. All these findings consistently indicate the meson-baryon molecular structure of $\Lambda(1405)$. We have also briefly reviewed the application of the chiral unitary approach to various hadron resonances in different flavor-spin sectors.

Finally we have also discussed the properties of $\Lambda(1405)$ in various environments, especially in few-body systems. Thanks to the $\bar{K}N$ quasi-bound component of $\Lambda(1405)$, we expect further various hadronic molecular systems with kaons and nucleons systematically. This novel structure is driven by the two-sided nature of the antikaon; the $\bar K$ meson has the nature of the Nambu-Goldstone particle associated with the spontaneous breaking of chiral symmetry, which deduces $s$-wave attractions in $\bar KN$ and $\bar KK$, but at the same time $\bar K$ is moderately heavy due to the strange quark mass. In the nuclear matter, $\Lambda(1405)$ exhibits complicated dynamics due to the strong coupling to $\bar{K}N$ and $\pi\Sigma$ channels. In summary, the large meson-baryon component in $\Lambda(1405)$ is responsible for the peculiar features of $\Lambda(1405)$ in many-body systems.

At present, there are several directions of studies of $\Lambda(1405)$ to be pursued. In closing this paper, we summarize future perspectives below.
\begin{itemize}
 
\item \textit{Exploration of the structure}: It has been shown that $\Lambda(1405)$ is dominated by meson-baryon molecular structure, based on several theoretical arguments. Although this is a reasonable and plausible scenario, it is important to examine this picture in experiments, by relating the internal structure with experimental observables. An attempt in this direction has been recently presented in Ref.~\cite{Cho:2010db}; it is shown that the production yield of a particle in heavy ion collisions can be used to investigate its internal structure. It is also desirable to establish a model-independent and quantitative argument to assess the structure of resonances. For this purpose, compositeness of bound states defined in Refs.~\cite{Weinberg:1962hj,Weinberg:1965zz} may be used as a baseline of the discussion~\cite{Baru:2003qq,Hyodo:2010uh,Compositeness}.

\item \textit{Theoretical foundation of resonances}: We have discussed that some quantities concerning $\Lambda(1405)$, such as coupling constants and the form factors, are obtained as complex numbers in the dynamical framework. This is because $\Lambda(1405)$ is not a stable particle but an unstable resonance state. In order to give an interpretation to these results, resonance states should be formulated on a firm theoretical ground.

\item \textit{Realistic $\bar{K}N$-$\pi\Sigma$ interaction}: Quantitatively refined theoretical models for the $\bar{K}N$-$\pi\Sigma$ amplitudes are essential both for the study of the structure of $\Lambda(1405)$ and for the applications to the $\bar{K}$ few-body systems. One of the key quantities is the binding energy of the quasi-bound $\bar{K}N$ system (position of the higher energy pole), which is closely related to the $\bar{K}N$ scattering length and the $\pi\Sigma$ mass spectrum. The present analysis can be systematically improved by including new experimental data and higher order terms of the chiral Lagrangian. Determination of the $\pi\Sigma$ threshold behavior is also essential to constrain the $\bar{K}N$-$\pi\Sigma$ amplitude~\cite{Ikeda:2011dx}. This allows one to extrapolate the amplitude down to the lower energy region.

\item \textit{Experiments and Lattice QCD}: To pin down the property and structure of the $\Lambda(1405)$ resonance, experimental studies are indispensable. As mentioned above, an urgent issue is to accumulate precise data of the $\bar{K}N$ scattering length and the $\pi\Sigma$ spectrum. The interpretation of experimental results should be guided by theoretical investigations of the reaction mechanism such as Ref.~\cite{Jido:2009jf}. In the longer term, the improvement of the cross sections of low energy $K^{-}p$ scattering and the threshold information of the $\pi\Sigma$ channel will help the calibration of the theoretical models. Experimental searches for the $\bar{K}$ few-body systems will also complement the constraints. Currently, direct application of the lattice QCD technique to the $\Lambda(1405)$ resonance is no simple task. It is however becoming possible to extract the meson-baryon interaction in lattice QCD~\cite{Fukugita:1994ve,Torok:2009dg,Ikeda:2010zz,IkedaLattice}. This will assist the determination of the observables for which experimental studies are difficult. The experimental activities and lattice QCD simulations, together with theoretical studies of the $\Lambda(1405)$ resonance will open new paradigm for the hadron and nuclear physics.

\end{itemize}

\section*{Acknowledgements}

The authors are grateful to the collaborators in a series of works,
Akinobu Dote,
Atsushi Hosaka,
Yoshiko Kanada-En'yo,
Alberto Martinez Torres,
Ulf-G. Meissner,
Jose Antonio Oller,
Eulogio Oset,
Angels Ramos,
Luis Roca,
Takayasu Sekihara
and
Wolfram Weise.
They also acknowledge the fruitful discussions with
Michael D\"oring,
Hiroyuki Fujioka,
Avraham Gal,
Yoichi Ikeda,
Mattias F. M. Lutz,
Kei Moriya,
Masayuki Niiyama,
Manolo Jose Vicente Vacas,
and 
Koichi Yazaki.
T.H.\ thanks the support from the Global Center of Excellence Program by MEXT, Japan through the Nanoscience and Quantum Physics Project of the Tokyo Institute 
of Technology. 
This work was partly supported by the Grant-in-Aid for Scientific Research from MEXT and JSPS (Nos.\
  22740161, 22105507, 
  21840026), 
the Grant-in-Aid for the Global COE Program ``The Next Generation of Physics, Spun from Universality and Emergence'' from MEXT of Japan.
This work was done in part under the Yukawa International Program for Quark-hadron Sciences (YIPQS).

\appendix
\section{Conventions}

Here we summarize conventions of the theoretical formulation used in this article. The total cross section $\sigma$ at momentum $q$ is given by the nonrelativistic scattering amplitude $f(q,\theta)$ as 
\begin{align}
    \sigma (q)
    =&
    \int d\Omega
    |f(q,\theta)|^{2} ,
    \nonumber
\end{align}
where $\theta$ is the scattering angle. The scattering amplitude can be decomposed into partial waves as 
\begin{align}
    f(q,\theta)
    =
    \sum_{l}(2l+1)f_{l}(q) P_{l}(\cos\theta) ,
    \nonumber
\end{align}
with the Legendre polynomials $P_{l}(\cos\theta)$. The total cross section is given as the sum of the contributions from partial waves due to the orthogonality of the Legendre polynomials. The $s$-wave amplitude gives the dominant contribution at low energy, so we obtain
\begin{align}
    \sigma(q)
    \sim 
    4\pi |f_{l=0}(q)|^{2} 
    \quad \text{for } q\to 0.
    \nonumber
\end{align}
For the coupled-channel case, the cross section from channel $j$ to $i$ is given by the $s$-wave scattering amplitude $f_{ij}$ as a function of the total energy $E$ as
\begin{align}
    \sigma_{ij}(E)
    \sim
    4\pi \frac{q_{i}}{q_{j}}|f_{ij}(E)|^{2} .
    \label{eq:crosssection}
\end{align}
where $q_{i}$ is the three-momentum in channel $i$.

Next we consider the scattering amplitude $T$ in the relativistic field theory, and relate it with the nonrelativistic amplitude $f$. Normalization of the Dirac spinor is given as $\bar{u}(\bm{p},r)u(\bm{p},s)=\delta_{rs}$ and $u^{\dag}(\bm{p},r)u(\bm{p},s)=\delta_{rs}E/M$ where $E=\sqrt{M^{2}+\bm{p}^{2}}$ and $M$ are the energy and mass of the fermion. To calculate the scattering amplitude, we define the $S$-operator and the $R$-operator as
\begin{align} 
    S=& 1+R.  \nonumber
\end{align}
We denote the $n$-particle scattering state as $\ket{k}$ with the label $k$ representing the momenta of the particles $k_a(a=1,...,n)$ and other labels collectively. The matrix element of the $R$-operator is related to the scattering amplitude $T$ as
\begin{align}
    \bra{k^{\prime}}R\ket{k}
    =&-i(2\pi)^4\delta^{(4)}
    (k^{\prime}-k)T_{k^{\prime}k}
    . \nonumber
\end{align}
The unitarity of the S-matrix $S^{\dag}S= 1$ leads to the relation $R^{\dag}+R=-R^{\dag}R$. Taking the matrix element, and inserting the complete set of the sum of the $n$-body intermediate states
\begin{equation}
    1
    = \sum_n\prod_{a=1}^{n} \int\frac{d^3q_a}{(2\pi)^3}
    N_a
    \ket{q_a}\bra{q_a},
    \nonumber
\end{equation}
with the normalization factor being $N_a=1/(2\omega_a)$ for bosons and $N_a=M_a/E_a$ for fermions, we obtain the optical theorem:
\begin{align}
    T_{kk^{\prime}}^{*}
    -T_{k^{\prime}k}
    =&i
    \sum_n\int d\Pi_{q}^{(n)}(k)
    T_{qk^{\prime}}^{*}T_{qk} ,
    \label{eq:optical}     
\end{align}
where the $n$-body phase space is defined as
\begin{equation}
    d\Pi_{q}^{(n)}(k)
    =\prod_{a=1}^{n} \frac{d^3q_a}{(2\pi)^3}
    N_a
    (2\pi)^4\delta^{(4)}(k-q). \label{eq:phasespace}
\end{equation}
In the case of the forward scattering $k^{\prime}=k=P$, we obtain \begin{align}
    \im T_{PP}
    =&-\frac{1}{2}\sum_n\int d\Pi_{q}^{(n)}(P)
    |T_{qP}|^2 . \label{eq:imT}
\end{align}
When we multiply Eq.~\eqref{eq:optical} by $(T^{\dag})^{-1}$ from the left and $T^{-1}$ from the right and take the forward scattering $k^{\prime}=k=P$, we obtain the imaginary part of the inverse of the amplitude as
\begin{align}
    \im T_{PP}^{-1}
    =&
    \frac{1}{2}
    \sum_n\int d\Pi_{q}^{(n)}(P) . \nonumber
\end{align}

Let us now restrict the intermediate states to be the two-body meson-baryon channel. In this case, the amplitude can be written as a function of the total energy square $s=P^{2}$ and the inverse amplitude is given by
\begin{align}
    \im T^{-1}(s)
    =&
    \frac{\rho(s)}{2} ,\quad
    \rho(s) = 
    \int d\Pi_{q}^{(2)}(P)
    = \frac{M\sqrt{(s-s_-)(s-s_+)}}{4\pi s},
    \nonumber
\end{align}
where $s_{\pm}=(M\pm m)^2$. In terms of the total energy $W=\sqrt{s}$, it is also written as
\begin{align}
    \im T^{-1}(W)
    = \frac{2M\bar{q}(W)}{4\pi W} ,
    \nonumber
\end{align}
with the momentum variable
\begin{align}
    \bar{q}(W)
    =&\frac{\sqrt{[W^2-(M-m)^2][W^2-(M+m)^2]}}{2W} 
    = \frac{\sqrt{(s-s_-)(s-s_+)}}{2\sqrt{s}} 
    = \frac{\sqrt{\lambda(s,M^{2},m^{2})}}{2\sqrt{s}} ,
    \nonumber
\end{align}
where the K\"allen function is defined as $\lambda(x,y,z)=x^{2}+y^{2}+z^{2}-2xy-2yz-2zx$. The $s$-wave $T$ matrix is related to the corresponding nonrelativistic scattering amplitude $f_0$ as 
\begin{align}
    T(W)
    =
    -\frac{4\pi \sqrt{s}}{M }\frac{q}{\bar{q}} f_0(q)
    , \label{eq:amplitudetrans}
\end{align}
with the nonrelativistic momentum $q=\sqrt{2\mu (W-M-m)}$ and the reduced mass $\mu=Mm/(M+m)$. At low energy $q\sim 0$, the deviation of the momentum is small and $q= \bar{q}=0$ at the threshold. With Eqs.~\eqref{eq:crosssection} and \eqref{eq:amplitudetrans}, the total cross section can be calculated in the chiral unitary approach. The scattering length $a$ and the effective range $r_{e}$ are defined as
\begin{align}
    f_{0}(q)
    =
    \frac{1}{q\cot \delta_{0}-qi},
    \quad
    q\cot \delta_{0}
    =\frac{1}{a}+r_{e}\frac{q^{2}}{2}+\dots ,
    \nonumber
\end{align}
where $\delta_{0}$ is the $s$-wave phase shift. The scattering length can be obtained by the amplitude at the threshold $q=0$, 
\begin{align}
    a
    =&f_{0}(q=0)
    =
    -\frac{M}{4\pi(M+m)}T(W=M+m) . \nonumber
\end{align}
In this convention, the positive (negative) scattering length represents attractive (repulsive) interaction.

Phase convention for the isospin states $\ket{I,I_{3}}$ is given by~\cite{Oset:1998it}
\begin{align}
    \ket{\pi^{+}}=&-\ket{1,1},  \quad
    \ket{K^{-}}=-\ket{1/2,-1/2},\quad
    \ket{\Sigma^{+}}=-\ket{1,1},\quad
    \ket{\Xi^{-}}=-\ket{1/2,-1/2},
    \nonumber
\end{align}
for both initial and final states. The SU(2) Clebsch-Gordan coefficients in PDG~\cite{Nakamura:2010zzi} is consistent with this convention.
  

\end{document}